\documentclass[11pt]{article}
\usepackage{amssymb}
\usepackage{epsfig}

\newcommand{\BE}{\begin{equation}}
\newcommand{\EE}{\end{equation}}
\newcommand{\BA}{\begin{eqnarray}}
\newcommand{\EA}{\end{eqnarray}}

\begin{document}
\begin{titlepage}

\vspace*{1mm}
\begin{center}

\vskip 1 pt

   {\LARGE{\bf The classical ether-drift experiments:\\ a modern re-interpretation }}

\vspace*{14mm} {\Large  M. Consoli$^{(a)}$, C. Matheson$^{(b)}$ and
A. Pluchino $^{(a,c)}$}
\vspace*{4mm}\\
{
a) Istituto Nazionale di Fisica Nucleare, Sezione di Catania, Italy ~~~~~~~~~~\\
b) Selwyn College, Cambridge, United Kingdom~~~~~~~~~~~~~~~~~~~~~~~~~~~~~~~~~\\
c) Dipartimento di Fisica e Astronomia dell'Universit\`a di Catania,
Italy }
\end{center}
\begin{center}
{\bf Abstract}
\end{center}
The condensation of elementary quanta and their macroscopic
occupation of the same quantum state, say ${\bf k}=0$ in some
reference frame $\Sigma$, is the essential ingredient of the
degenerate vacuum of present-day elementary particle physics. This
represents a sort of `quantum ether' which characterizes the
physically realized form of relativity and could play the role of
preferred reference frame in a modern re-formulation of the
Lorentzian approach. In spite of this, the so called `null results'
of the classical ether-drift experiments, traditionally interpreted
as confirmations of Special Relativity, have so deeply influenced
scientific thought as to prevent a critical discussion on the real
reasons underlying its alleged supremacy. In this paper, we argue
that this traditional null interpretation is far from obvious. In
fact, by using Lorentz transformations to connect the Earth's frame
to $\Sigma$, the small observed effects point to an average Earth's
velocity of about 300 km/s, as in most cosmic motions. A common
feature is the irregular behaviour of the data. While this has
motivated, so far, their standard interpretation as instrumental
artifacts, our new re-analysis of the very accurate Joos experiment
gives clear indications for the type of  Earth's motion associated
with the CMB anisotropy and leaves little space for this traditional
interpretation. The new explanation requires instead a view of the
vacuum as a stochastic medium, similar to a fluid in a turbulent
state of motion, in agreement with basic foundational aspects of
both quantum physics and relativity. The overall consistency of this
picture with the present experiments with vacuum optical resonators
and the need for a new generation of dedicated ether-drift
experiments are also emphasized.

\end{titlepage}

\section{Introduction}

An analysis of the ether-drift experiments, starting from the
original Michelson-Morley experiment of 1887, should be suitably
framed within a general discussion of the basic differences between
Einstein's Special Relativity \cite{einstein} and the Lorentzian
point of view \cite{lorentz,poincare,electron}. There is no doubt
that the former interpretation is today widely accepted. However, in
spite of the deep conceptual differences, it is not obvious how to
distinguish experimentally between the two formulations. This type
of conclusion was, for instance, already clearly expressed by
Ehrenfest in his lecture `On the crisis of the light ether
hypothesis' (Leyden, December 1912) as follows: ``So, we see that
the ether-less theory of Einstein demands exactly the same here as
the ether theory of Lorentz. It is, in fact, because of this
circumstance, that according to Einstein's theory an observer must
observe exactly the same contractions, changes of rate, etc. in the
measuring rods, clocks, etc. moving with respect to him as in the
Lorentzian theory. And let it be said here right away and in all
generality. As a matter of principle, there is no experimentum
crucis between the two theories". This can be understood since,
independently of all interpretative aspects, the basic quantitative
ingredients, namely Lorentz transformations, are the same in both
formulations. Their validity will be assumed in the following to
discuss the possible existence of a preferred reference frame.

For a modern presentation of the Lorentzian philosophy one can then
refer to Bell \cite{bell,brown,pla}. In this alternative approach,
differently from the usual derivations, one starts from physical
modifications of matter (namely Larmor's time dilation and
Lorentz-Fitzgerald length contraction in the direction of motion) to
deduce Lorentz transformations. In this way, due to the fundamental
group properties, the relation between two observers $S'$ and $S''$,
individually related to the preferred frame $\Sigma$ by Lorentz
transformations with dimensionless parameters $\beta'=v'/c$ and
$\beta''=v''/c$, is also a Lorentz transformation with relative
velocity parameter $\beta_{\rm rel}$ fixed by the relativistic
composition rule \BE
          \beta_{\rm rel}= {{\beta' - \beta''}\over{ 1- \beta' \beta''}}
\EE (for simplicity we restrict to the case of one-dimensional
motion). This produces a substantial quantitative equivalence with
Einstein's formulation for most standard experimental tests where
one just compares the relative measurements of a pair of observers.
Hence the importance of the ether-drift experiments where one
attempts to measure an absolute velocity.

At the same time, if the velocity of light $c_\gamma$ propagating in
the various interferometers coincides with the basic parameter $c$
entering Lorentz transformations, relativistic effects conspire to
make undetectable the individual $\beta'$, $\beta''$,...This means
that a null result of the ether-drift experiments should {\it not}
be automatically interpreted as a confirmation of Special
Relativity. As stressed by Ehrenfest, the motion with respect to
$\Sigma$ might remain unobservable, yet one could interpret
relativity ` \`a la Lorentz'. This could be crucial, for instance,
to reconcile faster-than-light signals with causality \cite{annals}
and thus provide a different view of the apparent non-local aspects
of the quantum theory \cite{scarani}.

However, to a closer look, is it really impossible to detect the
motion with respect to $\Sigma$? This possibility, which was
implicit in Lorentz' words \cite{electron} ``...it seems natural not
to assume at starting that it can never make any difference whether
a body moves through the ether or not..", may induce one to
re-analyze the classical ether-drift experiments.  Let us first give
some general theoretical arguments that could motivate this
apparently startling idea.

A possible observation is that Lorentz symmetry might not be an
exact symmetry. In this case, one could conceivably detect the
effects of absolute motion. For instance Lorentz symmetry could
represent an `emergent' phenomenon and thus reflect the existence of
some underlying form of ether.  This is an interesting conceptual
possibility which, in many different forms, objectively reflects the
fast growing interest of part of the physics community, a partial
list including i) the idea of the vacuum as a quantum liquid
\cite{volo,volo1} (which can explain in a natural way the huge
difference between the typical vacuum-energy scales of modern
particle physics and the cosmological term needed in Einstein's
equations to fit the observations) ii) the idea of Lorentz symmetry
as associated with an infrared fixed point \cite{chadha,nielsen} in
non-symmetric quantum field theories iii) the quantum-gravity
literature which, by starting from the original concept
\cite{wheeler1} of `space-time foam', explicitly models the vacuum
as a turbulent fluid \cite{migdal,ng2,ng3} iv) the idea of
deformations of Lorentz symmetry in a theoretical scheme (`Doubly
Special Relativity') \cite{amelino1,amelino2,amelino3} where besides
an invariant speed there is also an invariant length associated with
the Planck scale v) the representation of relativistic particle
propagation from the superposition, at very short time scales, of
non-relativistic particle paths with different Newtonian mass
\cite{kleinert}.

Here, however, we shall adopt a different perspective and
concentrate our analysis on a peculiar aspect of today's quantum
field theories: the representation of the vacuum as a `condensate'
of elementary quanta. These condense because their trivially empty
vacuum is a meta-stable state and not the true ground state of the
theory. In the physically relevant case of the Standard Model of
electroweak interactions, this situation can be summarized by saying
\cite{thooft} that ``What we experience as empty space is nothing
but the configuration of the Higgs field that has the lowest
possible energy. If we move from field jargon to particle jargon,
this means that empty space is actually filled with Higgs particles.
They have Bose condensed". The explicit translation from field
jargon to particle jargon, with the substantial equivalence between
the effective potential of quantum field theory and the energy
density of a dilute particle condensate, can be found for instance
in ref.\cite{mech}.

The trivial empty vacuum will eventually be re-established by
heating the system above a critical temperature $T=T_c$ where the
condensate `evaporates'. This temperature in the Standard Model is
so high that one can safely approximate the ordinary vacuum as a
zero-temperature system (think of $^4$He at a temperature $10^{-12}$
$^{o}$K). This observation allows one to view the physical vacuum as
a superfluid medium \cite{volo} where bodies can flow without any
apparent friction, consistently with the experimental results.
Clearly, this form of quantum vacuum is not the kind of ether
imagined by Lorentz. However, if possible, this modern view of the
vacuum state is even more different from the empty space-time of
Special Relativity that Einstein had in mind in 1905. Therefore, one
might ask \cite{pagano} if Bose condensation, i.e. the macroscopic
occupation of the same quantum state, say ${\bf{k}}=0$ in some
reference frame $\Sigma$, can represent the operative construction
of a `quantum ether'. This characterizes the {\it physically
realized form of relativity} and could play the role of the
preferred reference frame in a modern Lorentzian approach.

Usually this possibility is not considered with the motivation,
perhaps, that the average properties of the condensed phase are
summarized into a single quantity which transforms as a world scalar
under the Lorentz group, for instance, in the Standard Model, the
vacuum expectation value $\langle\Phi\rangle$ of the Higgs field.
However, this does not imply that the vacuum state itself has to be
{\it Lorentz invariant}. Namely, Lorentz transformation operators
$\hat{U}'$, $\hat{U}''$,..might transform non trivially the
reference vacuum state $|\Psi^{(0)}\rangle$ (appropriate to an
observer at rest in $\Sigma$) into $| \Psi'\rangle$, $|
\Psi''\rangle$,.. (appropriate to moving observers $S'$, $S''$,..)
and still, for any Lorentz-invariant operator $\hat{G}$, one would
find \BE \langle \hat{G}\rangle_{\Psi^{(0)}}=\langle
\hat{G}\rangle_{\Psi'}=\langle
\hat{G}\rangle_{\Psi''}=..\end{equation} Here, we are assuming the
existence of a suitable operatorial representation of the Poincar\'e
algebra for the quantum theory in terms of 10 generators $P_\alpha$,
$M_{\alpha,\beta}$ ( $\alpha$ ,$\beta$=0, 1, 2, 3) where $P_\alpha$
are the 4 generators of the space-time translations and
$M_{\alpha\beta}=-M_{\beta\alpha}$ are the  6 generators of the
Lorentzian rotations with commutation relations
\begin{equation} \label{tras1} [P_\alpha,P_\beta]=0 \end{equation}
\begin{equation} \label{boost} [M_{\alpha\beta}, P_\gamma]=
\eta_{\beta\gamma}P_\alpha - \eta_{\alpha\gamma}P_\beta
\end{equation} \begin{equation} \label{tras2} [M_{\alpha\beta},
M_{\gamma\delta}]= \eta_{\alpha\gamma}M_{\beta\delta}+
\eta_{\beta\delta}M_{\alpha\gamma}
-\eta_{\beta\gamma}M_{\alpha\delta}-\eta_{\alpha\delta}M_{\beta\gamma}
\end{equation} where $\eta_{\alpha\beta}={\rm diag}(1,-1,-1,-1)$.

With these premises, the possibility of a Lorentz-non-invariant
vacuum state was addressed in refs.\cite{epjc,dedicated} by
comparing two basically different approaches. In the first
description, as in the axiomatic approach to quantum field theory
\cite{cpt}, one could describe the physical vacuum as an eigenstate
of the energy-momentum vector. This physical vacuum state
$|\Psi^{(0)}\rangle$ \footnote{ We ignore here the problem of vacuum
degeneracy by assuming that any overlapping among equivalent vacua
vanishes in the infinite-volume limit of quantum field theory (see
e.g. S. Weinberg, {\it The Quantum Theory of Fields}, Cambridge
University press, Vol.II, pp. 163-167).} would maintain both zero
momentum and zero angular momentum, i.e. (i,j=1,2,3)\begin{equation}
\label{groundi}
\hat{P}_i|\Psi^{(0)}\rangle=\hat{M}_{ij}|\Psi^{(0)}\rangle=0
\end{equation} but, at the same time, be characterized by a
non-vanishing energy
\begin{equation} \label{ground}
\hat{P}_0|\Psi^{(0)}\rangle=E_0|\Psi^{(0)}\rangle \end{equation}
This vacuum energy might have different explanations. Here, we shall
limit ourselves to exploring the physical implications of its
existence by just observing that, in interacting quantum field
theories, there is no known way to ensure consistently the condition
$E_0=0$ without imposing an unbroken supersymmetry, which is not
phenomenologically acceptable. In this framework, by using the
Poincar\'e algebra of the boost and energy-momentum operators, one
then deduces that the physical vacuum cannot be a Lorentz-invariant
state and that, in any moving frame, there should be a non-zero
vacuum spatial momentum $\langle {\hat{P}_i}\rangle_{\Psi'}\neq 0$
along the direction of motion. In this way, for a moving observer
$S'$ the physical vacuum would look like some kind of ethereal
medium for which, in general, one can introduce a momentum density
$\langle \hat{W}_{0i}\rangle_{\Psi'}$ through the relation (i=1,2,3)
\begin{equation} \label{density} \langle {\hat{P}_i}\rangle_{\Psi'}\equiv \int
d^3x~\langle \hat{W}_{0i}\rangle_{\Psi'}  \neq 0 \end{equation} On
the other hand, there is an alternative approach where one tends to
consider the vacuum energy $E_0$ as a spurious concept and only
concentrate on an energy-momentum tensor of the following form
\cite{zeldovich,weinberg}
\begin{equation}\label{zeld} \langle \hat{W}_{\mu\nu}\rangle_
{\Psi^{(0)}}=\rho_v ~\eta_{\mu\nu}\end{equation} ($\rho_v$ being a
space-time independent constant). In this case, one is driven to
completely different conclusions since, by introducing the Lorentz
transformation matrices $\Lambda^\mu_\nu$ to any moving frame $S'$,
defining $\langle \hat{W}_{\mu\nu}\rangle_{\Psi'}$ through the
relation
\begin{equation}\langle \label{cov}
\hat{W}_{\mu\nu}\rangle_{\Psi'}=\Lambda^{\sigma}{_\mu}\Lambda^{\rho}{_\nu}
~\langle\hat{W}_{\sigma\rho}\rangle_{\Psi^{(0)}}\end{equation} and
using Eq.(\ref{zeld}), it follows that the expectation value of
$\hat{W}_{0i}$ in any
 boosted vacuum state $| \Psi'\rangle$ vanishes, just as it vanishes
in $|\Psi^{(0)}\rangle$, i.e. \begin{equation} \label{density1} \int
d^3x~ \langle \hat{W}_{0i}\rangle_{\Psi'} \equiv \langle
{\hat{P}_i}\rangle_{\Psi'}= 0 \end{equation} As discussed in
ref.\cite{epjc}, both alternatives have their own good motivations
and it is not so obvious how to decide between Eq.(\ref{density})
and Eq.(\ref{density1}) on purely theoretical grounds. For instance,
in a second-quantized formalism, single-particle energies
$E_1({\bf{p}})$ are defined as the energies of the corresponding
one-particle states $|{\bf{p}}\rangle$ minus the energy of the
zero-particle, vacuum state. If $E_0$ is considered a spurious
concept,  $E_1({\bf{p}})$ will also become an ill-defined quantity.
At a deeper level, one should also realize that in an approach based
solely on Eq.(\ref{zeld}) the properties of $|\Psi^{(0)}\rangle$
under a Lorentz transformation are not well defined. In fact, a
transformed vacuum state $| \Psi'\rangle$ is obtained, for instance,
by acting on $|\Psi^{(0)}\rangle$ with the boost generator
$\hat{M}_{01}$. Once $|\Psi^{(0)}\rangle$ is considered an
eigenstate of the energy-momentum operator, one can definitely show
\cite{epjc} that, for $E_0\neq 0$, $| \Psi'\rangle$ and
$|\Psi^{(0)}\rangle$ differ non-trivially. On the other hand, if
$E_0=0$ there are only two alternatives: either
$\hat{M}_{01}|\Psi^{(0)}\rangle=0$, so that
$|\Psi'\rangle=|\Psi^{(0)}\rangle$,  or
$\hat{M}_{01}|\Psi^{(0)}\rangle$ is a state vector proportional to
$|\Psi^{(0)}\rangle$, so that $| \Psi'\rangle$ and
$|\Psi^{(0)}\rangle$ differ by a phase factor.

Therefore, if the structure in Eq.(\ref{zeld}) were really
equivalent to the exact Lorentz invariance of the vacuum, it should
be possible to show similar results, for instance that such a
$|\Psi^{(0)}\rangle$ state can remain invariant under a boost, i.e.
be an eigenstate of \BE \hat{M}_{0i}=-i\int
d^3x~(x_i\hat{W}_{00}-x_0 \hat{W}_{0i}) \EE with zero eigenvalue. As
far as we can see, there is no way to obtain such a result by just
starting from Eq.(\ref{zeld}) (this only amounts to the weaker
condition $\langle \hat{M}_{0i}\rangle_ {\Psi^{(0)}}=0$). Thus,
independently of the finiteness of $E_0$, it should not come as a
surprise that one can run into contradictory statements once
$|\Psi^{(0)}\rangle$ is instead characterized by means of
Eqs.(\ref{groundi})$-$(\ref{ground}). For these reasons, it is not
obvious that the local relations (\ref{zeld}) represent a more
fundamental approach to the vacuum.

Alternatively, one could argue that a satisfactory solution of the
vacuum-energy problem lies definitely beyond flat space. A non-zero
$\rho_v$, in fact, should induce a cosmological term in Einstein's
field equations and a non-vanishing space-time curvature which
anyhow dynamically breaks global Lorentz symmetry. Nevertheless, in
our opinion, in the absence of a consistent quantum theory of
gravity, physical models of the vacuum in flat space can be useful
to clarify a crucial point that, so far, remains obscure: the huge
renormalization effect which is seen when comparing the typical
vacuum-energy scales of modern particle physics with the
experimental value of the cosmological term needed in Einstein's
equations to fit the observations. For instance, as anticipated, the
picture of the vacuum as a superfluid can explain in a natural way
why there might be no non-trivial macroscopic curvature in the
equilibrium state where any liquid is self-sustaining \cite{volo}.
In any liquid, in fact, curvature requires {\it deviations} from the
equilibrium state. The same happens for a crystal at zero
temperature where all lattice distortions vanish and electrons can
propagate freely as in a perfect vacuum.  In such representations of
the lowest energy state, where  large condensation energies (of the
liquid and of the crystal) play no observable role, one can
intuitively understand why curvature effects can be orders of
magnitude smaller than those naively expected by solving Einstein's
equations with the full $\langle \hat{W}_{\mu\nu}\rangle_
{\Psi^{(0)}}$ as a cosmological term. In this perspective,
`emergent-gravity' approaches \cite{barcelo1,barcelo2,ultraweak},
where gravity somehow arises from the same physical flat-space
vacuum, may become natural \footnote{ In this sense, by exploring
emergent-gravity approaches based on an underlying superfluid
medium, one is taking seriously Feynman's indication : "...the first
thing we should understand is how to formulate gravity so that it
doesn't interact with the energy in the vacuum" \cite{rule}.} and,
to find the effective form for the cosmological term to be inserted
in Einstein's field equations, we are lead to sharpen our
understanding of the vacuum structure and of its excitation
mechanisms by starting from the physical picture of a superfluid
medium. To decide between Eqs.(\ref{density}) and (\ref{density1}),
one could then work out the possible observable consequences and
check experimentally the existence of a fundamental energy-momentum
flow.

\section{Vacuum energy-momentum flow as an ether drift}

To explore the idea of a non-zero vacuum energy-momentum flow, one
can adopt a phenomenological model \cite{epjc} where the physical
vacuum is described as a relativistic fluid \cite{eckart}. In this
representation, a non-zero $\langle \hat{W}_{0i}\rangle_{\Psi'}$
gives rise to a tiny heat flow and an effective thermal gradient in
a moving frame $S'$. This would represent a fundamental perturbation
which, if present, is likely too small to be detectable in most
experimental conditions by standard calorimetric devices. However,
it could eventually be detected through very accurate ether-drift
experiments performed in forms of matter that react by producing
convective currents in the presence of arbitrarily small thermal
gradients, i.e. in gaseous systems.

To better explain this possibility, let us first recall that in the
modern version of these experiments one looks for a possible
anisotropy of the two-way velocity of light through the relative
frequency shift $\Delta\nu(\theta)$ of two orthogonal optical
cavities \cite{applied,lammer}. Their frequency \BE \label{nutheta0}
  \nu (\theta)= {{ \bar{c}_\gamma(\theta) m}\over{2L(\theta)}}
\EE is proportional to the two-way velocity of light
$\bar{c}_\gamma(\theta)$ within the cavity through an integer number
$m$, which fixes the cavity mode, and the length of the cavity
$L(\theta)$ as measured in the laboratory. In principle, by filling
the resonating cavities with some gaseous medium, the existence of a
vacuum energy-momentum flow could produce two basically different
effects:

~~~a) modifications of the solid parts of the apparatus. These can
change the cavity length upon active rotations of the apparatus or
under the Earth's rotation.

~~~b) convective currents of the gas molecules {\it inside} the
optical cavities. These can produce an anisotropy of the two-way
velocity of light. In this sense, the reference frame $S'$ where the
solid container of the gas is at rest would not define a true state
of rest.

Now, an anisotropy of the cavity length, in the laboratory frame,
would amount to an anisotropy of the basic atomic parameters, a
possibility which is severely limited experimentally. In fact, in
the most recent versions of the original Hughes-Drever experiment
\cite{hughes,drever}, where one measures the atomic energy levels as
a function of their orientation with respect to the fixed stars,
possible deviations from isotropy have been found below the
$10^{-20}$ level \cite{will}. This is incomparably smaller than any
other effect on the velocity of light that we are going to discuss.
Therefore, mechanism a), if present, is completely negligible and,
from now on, we shall assume $L(\theta)=L=$constant. In this way,
one re-obtains the standard relation adopted in the analysis of the
experiments
 \BE \label{bbasic2}
 {{\Delta \nu^{\rm phys}(\theta) }\over{\nu_0}}  =
     {{\bar{c}_\gamma(\pi/2 +\theta)- \bar{c}_\gamma (\theta)} \over
       {c }} \equiv {{\Delta \bar{c}_\theta } \over{c}}
 \EE where $\nu_0$ is
the reference frequency of the two optical resonators and the suffix
``${\rm phys}$" indicates a hypothetical physical part of the
frequency shift after subtraction of all spurious effects.

Let us now estimate the possible effects of mechanism b) by first
recalling that rigorous treatments of light propagation in
dielectric media are based on the extinction theory \cite{bornwolf}.
This was originally formulated for continuous media where the
inter-particle distance is smaller than the light wavelength. In the
opposite case of an isotropic, dilute random medium \cite{weber} as
a gas, it is relatively easy to compute the scattered wave in the
forward direction and obtain the refractive index. However, the
presence of convective currents would produce an anisotropy of the
velocity of refracted light.

To derive the relevant relations, let us introduce from scratch the
refractive index ${\cal N}$ of the gas. By assuming isotropy, the
time $t$ spent by refracted light to cover some given distance $L$
within the medium is $t={\cal N}L/c$. This can be expressed as the
sum of $t_0=L/c$ and $t_1=({\cal N}-1)L/c$ where $t_0$ is the same
time as in the vacuum and $t_1$ represents the additional, average
time by which refracted light is slowed down by the presence of
matter. If there are convective currents, due to the motion of the
laboratory with respect to a preferred reference frame $\Sigma$,
then $t_1$ will be different in different directions, and there will
be an anisotropy of the velocity of light proportional to $({\cal
N}-1)$. In fact, let us consider light propagating in a
2-dimensional plane and express $t_1$ as
\begin{equation} t_1={{L}\over{c}}f({\cal N}, \theta, \beta)
\end{equation} with $\beta=V/c$, $V$ being (the projection on the
considered plane of) the relevant velocity with respect to $\Sigma$
where the isotropic form
\begin{equation}
\label{boundary} f({\cal N}, \theta, 0)={\cal N}-1
\end{equation}
is assumed. By expanding around ${\cal N}=1$ where, whatever
$\beta$, $f$ vanishes by definition, one finds for gaseous systems
(where ${\cal N}-1 \ll 1$) the universal trend
\begin{equation} f({\cal N}, \theta,\beta)\sim ({\cal N}-1)F(\theta,\beta) \end{equation}
with
\begin{equation}
F(\theta,\beta)\equiv (\partial f/\partial {\cal N})|_{ {\cal N}=1}
\end{equation} and $F(\theta,0)=1$.
Therefore, by introducing the one-way velocity of light
\begin{equation} t({\cal N},\theta,\beta)= {{L}\over{c_\gamma({\cal
N},\theta,\beta)}}\sim {{L}\over{c}}+ {{L}\over{c}}({\cal
N}-1)~F(\theta,\beta)
\end{equation} one gets
\begin{equation}
c_\gamma({\cal N},\theta,\beta)\sim {{c}\over{ {\cal N} }}~
\left[1- ({\cal N}-1) ~(F(\theta,\beta) -1)\right]
\end{equation}
Analogous relations hold for the two-way velocity $
\bar{c}_\gamma({\cal N},\theta,\beta)$  \begin{equation}
\bar{c}_\gamma({\cal N},\theta,\beta)={{2~c_\gamma({\cal
N},\theta,\beta)c_\gamma({\cal N},\pi
+\theta,\beta)}\over{c_\gamma({\cal N},\theta,\beta) +c_\gamma
({\cal N},\pi +\theta,\beta)}} \sim {{c}\over{ {\cal N} }} \left[1-
({\cal N}-1)
 ~\left( {{F(\theta,\beta) + F(\pi+\theta,\beta)}\over{2}} -1\right) \right]
\end{equation} A more explicit expression can be obtained by exploring some general
properties of the function $F(\theta,\beta)$. By expanding in powers
of $\beta$
\begin{equation}
F(\theta,\beta)-1 = \beta F_1(\theta) + \beta^2 F_2(\theta)+...
\end{equation}
and taking into account that, by the very definition of two-way
velocity, $\bar{c}_\gamma({\cal N},\theta,\beta)=
\bar{c}_\gamma({\cal N},\theta,-\beta)$, it follows that
$F_1(\theta)=-F_1(\pi + \theta)$. Therefore, to ${\cal O}(\beta^2)$,
we get the general structure \cite{dedicated}
\begin{equation}
\label{legendre} \bar{c}_\gamma({\cal N},\theta,\beta) \sim
{{c}\over{ {\cal N} }} \left[1- ({\cal N}-1)~\beta^2
\sum^\infty_{n=0}\zeta_{2n}P_{2n}(\cos\theta)
  \right]
\end{equation}
in which we have expressed the combination $F_2(\theta) + F_2(\pi
+\theta)$ as an infinite expansion of even-order Legendre
polynomials with unknown coefficients $\zeta_{2n}={\cal O}(1)$ which
depend on the characteristics of the induced convective motion of
the gas molecules inside the cavities.

Eq.(\ref{legendre}), in principle, is exact to the given accuracy
but it is of limited utility if one wants to compare with real
experiments. In fact, it would require the complete control of all
possible mechanisms that can produce the gas convective currents by
starting from scratch with the macroscopic Earth's motion in the
physical vacuum. This general structure can, however, be compared
with the particular form (see Eq.(\ref{twoway}) of the Appendix)
obtained by using Lorentz transformations to connect $S'$ to the
preferred frame
\begin{equation} \label{twoway1} \bar{c}_\gamma({\cal N},\theta,\beta)\sim {{c}\over{ {\cal N}
}}~[1-\beta^2~({\cal N}-1)(A+B\sin^2\theta)] \end{equation} with $
A=2$ and $B= -1$  which corresponds to setting $\zeta_0=4/3$,
$\zeta_{2}= 2/3$ and all $\zeta_{2n}=0$ for $n
> 1$ in
Eq.(\ref{legendre}). Eq.(\ref{twoway1}) represents a definite
realization of the general structure in (\ref{legendre}) and a
particular case of the Robertson-Mansouri-Sexl (RMS) scheme
\cite{robertson,mansouri} for anisotropy parameter $|{\cal B}|={\cal
N} -1$ (see the Appendix). In this sense, it provides a partial
answer to the problems posed by our limited knowledge of the
electromagnetic properties of gaseous systems and will be adopted in
the following as a tentative model for the two-way velocity of light
\footnote{One conceptual detail concerns the gas refractive index
whose reported values are experimentally measured on the Earth by
two-way measurements. For instance for air, the most precise
determinations are at the level $10^{-7}$, say ${\cal N}_{\rm
air}=1.0002926..$ at STP (Standard Temperature and Pressure). By
assuming a non-zero anisotropy in the Earth's frame, one should
interpret the isotropic value $c/{\cal N_{\rm air}}$ as an angular
average of Eq.(\ref{twoway1}), i.e.
\begin{equation} \label{nair} {{c}\over{ {\cal N}_{\rm air} }}\equiv
\langle\bar{c}_\gamma(\bar{\cal N}_{\rm air},\theta,\beta)\rangle_\theta=
{{c}\over{ \bar {\cal N} _{\rm air} }} ~[1-{{3}\over{2}} (\bar{\cal
N}_{\rm air} -1)\beta^2]
\end{equation} From this relation, one can determine in principle
the unknown value $\bar {\cal N} _{\rm air} \equiv {\cal N}(\Sigma)$
(as if the gas were at rest in $\Sigma$), in terms of the
experimentally known quantity ${\cal N}_{\rm air}\equiv{\cal
N}(Earth)$ and of $V$. In practice, for the standard velocity values
involved in most cosmic motions, say $ V \sim $ 300 km/s, the
difference between ${\cal N}(\Sigma)$ and ${\cal N}(Earth)$ is at
the level $10^{-9}$ and thus completely negligible. The same holds
true for the other gaseous systems at STP (say nitrogen, carbon
dioxide, helium,..) for which the present experimental accuracy in
the refractive index is, at best, at the level $10^{-6}$. Finally,
the isotropic two-way speed of light is better determined in the
low-pressure limit where $({\cal N}-1)\to 0$. In the same limit, for
any given value of $V$, the approximation ${\cal N}(\Sigma)={\cal
N}(Earth)$ becomes better and better.}.

Summarizing: in this scheme, the theoretical estimate for a possible
anisotropy of the two-way velocity of light is \BE
\label{anigas}\left[{{\Delta\nu }\over{\nu_0}}\right]^{\rm
Theor}_{\rm gas} =\left[{{\Delta\bar{c}_\theta}\over{c}}\right]^{\rm
Theor}_{\rm gas}\sim ({\cal N}_{\rm gas} -1)~{{V^2}\over{c^2}} \EE
Then, by assuming the typical velocity of most Earth's cosmic
motions $V\sim$ 300 km/s, one would expect
${{\Delta\bar{c}_\theta}\over{c}}\lesssim 10^{-9}$ for experiments
performed in air at atmospheric pressure, where ${\cal N}\sim
1.00029$, or ${{\Delta\bar{c}_\theta}\over{c}}\lesssim 10^{-10}$ for
experiments performed in helium at atmospheric pressure, where
${\cal N}\sim 1.000035$. Therefore these potential effects are much
larger than those possibly associated with vacuum cavities. In fact,
from experiments one finds \cite{brillet}$-$\cite{schillernew} \BE
\left[{{\Delta\nu }\over{\nu_0}}\right]^{\rm EXP}_{\rm vacuum} =
\left[{{\Delta\bar{c}_\theta}\over{c}}\right]^{\rm EXP}_{\rm vacuum}
\sim 10^{-15}\EE or smaller and thus completely negligible when
compared with those of Eq.(\ref{anigas}).

On the other hand, if one were considering light propagation in a
strongly bound system, such as a solid or liquid transparent medium,
the small energy flow generated by the motion with respect to the
vacuum condensate should mainly dissipate by heat conduction with no
appreciable particle flow and no light anisotropy in the rest frame
of the container of the medium. This conclusion is in agreement with
the experiments \cite{pla,cimento} that seem to indicate the
existence of {\it two} regimes. A former region of gaseous systems
where ${\cal N}\sim 1$ and there are small residuals which are
roughly consistent with Eq.(\ref{anigas}). A latter region  where
the difference of ${\cal N}$ from unity is substantial, (e.g. ${\cal
N}\sim 1.5$ as with perspex in the experiment by Shamir and Fox
\cite{fox}), where light propagation is seen isotropic in the rest
frame of the medium (i.e. in the Earth's frame). Although it would
be difficult to describe in a fully quantitative way the transition
between the two regimes, some simple arguments can be given along
the lines suggested by de Abreu and Guerra (see pages 165-170 of
ref.\cite{guerra}).

For this reason, it was proposed in refs.\cite{pla,epjc,dedicated}
that one should design a new class of dedicated experiments in
gaseous systems. Such a type of `non-vacuum' experiment would be
along the lines of ref.\cite{holger} where just the use of optical
cavities filled with different materials was considered as a useful
complementary tool to study deviations from exact Lorentz
invariance. In the meantime, due to the heuristic nature of our
approach, and to further motivate this new series of experiments,
one could try to obtain quantitative checks by applying the same
interpretative scheme to the classical ether-drift experiments
(Michelson-Morley, Miller, Illingworth, Joos,...). These old
experiments were performed with interferometers where light was
propagating in air or helium at atmospheric pressure. In this
regime, where $({\cal N}-1)$ is a very small number, the theoretical
fringe shifts expected on the basis of Eqs.(\ref{legendre}) and
(\ref{twoway1}) are much smaller than the classical prediction
${\cal O} (\beta^2)$ and it becomes conceivable that tiny non-zero
effects might have been erroneously interpreted as `null results'.

To make this more evident, let us adopt Eq.(\ref{twoway1}). Then, an
anisotropy of the two-way velocity of light  could be measured by
rotating a Michelson interferometer. As anticipated, in the rest
frame $S'$ of the apparatus, the length $L$ of its arms does not
depend on their orientation so that the interference pattern between
two orthogonal beams of light depends on the time difference
 \BE \label{deltaT} \Delta T(\theta)=
{{2L}\over{\bar{c}_\gamma({\cal N},\theta,\beta)}}-
{{2L}\over{\bar{c}_\gamma({\cal N},\pi/2+\theta,\beta)}} \EE
In this way, by introducing the wavelength $\lambda$ of the light
source and the projection $v$ of the relative velocity in the plane
of the interferometer, one finds to order ${{v^2}\over{c^2}}$  the
fringe shift \BE\label{fringe0} {{\Delta
\lambda(\theta)}\over{\lambda}} \sim {{c\Delta T(\theta)}\over{{\cal
N}\lambda}} \sim {{L}\over{\lambda}}{{v^2_{\rm
obs}}\over{c^2}}\cos2(\theta-\theta_0)\EE In the above equation the
angle $\theta_0=\theta_0(t)$ indicates the apparent direction of the
ether-drift in the plane of the interferometer (the `azimuth') and
the square of the  {\it observable} velocity  \BE\label{vobs}
v^2_{\rm obs}(t) \sim 2({\cal N}-1)v^2(t) \EE is re-scaled by the
tiny factor $2({\cal N}-1)$ with respect to the true {\it
kinematical } velocity $v^2(t)$. We emphasize that $v_{\rm obs}$ is
just a short-hand notation to summarize into a single quantity the
combined effects of a given kinematical $v$ and of the gas
refractive index ${\cal N}$. In this sense, one could also avoid its
introduction altogether. However, in our opinion, it is a useful,
compact parametrization since, in this way, relation (\ref{fringe0})
is formally identical to the classical prediction of a
second-harmonic effect with the only replacement $v \to v_{\rm
obs}$. For this reason, as we shall see in the following sections,
it is in terms of $ v_{\rm obs}$, rather than in terms of the true
kinematical $v$, that one can more easily compare with the original
analysis of the classical ether-drift experiments.

In conclusion, in this scheme, the interpretation of the experiments
is transparent. According to Special Relativity, there can be no
fringe shift upon rotation of the interferometer. In fact, if light
propagates in a medium, the frame of isotropic propagation is always
assumed to coincide with the laboratory frame $S'$, where the
container of the medium is at rest, and thus one has $v_{\rm
obs}=v=0$. On the other hand, if there were fringe shifts, one could
try to deduce the existence of a preferred frame $\Sigma\neq S'$
provided the following minimal requirements are fulfilled : i) the
fringe shifts exhibit an angular dependence of the type in
Eq.(\ref{fringe0}) ii) by using gaseous media with different
refractive index one gets consistency with Eq.(\ref{vobs}) in such a
way that different $v_{\rm obs}$ correspond to the same kinematical
$v$.

Before starting with the analysis of the classical experiments, one
more remark is in order. In principle, even a {\it single}
observation, within its experimental accuracy, can determine the
existence of an ether-drift. However interpretative models are
required to compare results obtained at different times and in
different places. In the scheme of Eqs.(\ref{fringe0}) and
(\ref{vobs}), the crucial information is contained in the two
time-dependent functions $v=v(t)$ and $\theta_0=\theta_0(t)$,
respectively the magnitude  of the velocity and the apparent
direction of the azimuth in the plane of the interferometer. For
their determination, the standard assumption is to consider a cosmic
Earth's velocity with well defined magnitude $V$, right ascension
$\alpha$ and angular declination $\gamma$ that can be considered
constant for short-time observations of a few days where there are
no appreciable changes due to the Earth's orbital velocity around
the Sun. In this framework, where the only time dependence is due to
the Earth's rotation, one identifies $v(t)\equiv \tilde v(t)$ and
$\theta_0(t)\equiv\tilde\theta_0(t)$ where $\tilde v(t)$ and
$\tilde\theta_0(t)$ derive from the simple application of spherical
trigonometry \cite{nassau} \BE \label{nassau1}
       \cos z(t)= \sin\gamma\sin \phi + \cos\gamma
       \cos\phi \cos(\tau-\alpha)
\EE \BE\label{nassau2}
      {{\tilde v_x(t)}\over{V}}\equiv \sin z(t)\cos\tilde\theta_0(t)= \sin\gamma\cos \phi -\cos\gamma
       \sin\phi \cos(\tau-\alpha)
\EE \BE\label{nassau3}
      {{\tilde v_y(t)}\over{V}}\equiv \sin z(t)\sin\tilde\theta_0(t)= \cos\gamma\sin(\tau-\alpha) \EE
 \BE \label{projection}
       \tilde v(t)\equiv \sqrt{ \tilde v^2_x(t) + \tilde v^2_y(t)}  =V \sin z(t) ,
\EE Here $z=z(t)$ is the zenithal distance of ${\bf{V}}$, $\phi$ is
the latitude of the observatory, $\tau=\omega_{\rm sid}t$ is the
sidereal time of the observation in degrees ($\omega_{\rm sid}\sim
{{2\pi}\over{23^{h}56'}}$) and the angle $\theta_0$ is counted
conventionally from North through East so that North is $\theta_0=0$
and East is $\theta_0=90^o$.

To explore the observable implications, let us first re-write the
basic Eq.(\ref{fringe0}) as \BE\label{fringe} {{\Delta
\lambda(\theta)}\over{\lambda}} \sim {{2L ({\cal N}-1)
}\over{\lambda}}{{v^2(t)}\over{c^2}}\cos2(\theta-\theta_0(t))\equiv
2C(t)\cos2\theta +2S(t) \sin 2\theta \EE where \BE \label{CS}
C(t)={{L ({\cal N}-1)
}\over{\lambda}}{{v^2(t)}\over{c^2}}\cos2\theta_0(t)
~~~~~~~~~~~~~~~~~~~~  S(t)={{L({\cal N}-1)
}\over{\lambda}}{{v^2(t)}\over{c^2}}\sin2\theta_0(t) \EE Then Eqs.
(\ref{nassau1})$-$(\ref{projection}) amount to the structure\BE
 \label{amorse1}
      S(t)\equiv {\tilde S}(t) =
      {S}_{s1}\sin \tau +{S}_{c1} \cos \tau
       + {S}_{s2}\sin(2\tau) +{S}_{c2} \cos(2\tau)
\EE \BE \label{amorse2}
      C(t)\equiv {\tilde C}(t) = {C}_0 +
      {C}_{s1}\sin \tau +{C}_{c1} \cos \tau
       + {C}_{s2}\sin(2 \tau) +{C}_{c2} \cos(2 \tau)
\EE with Fourier coefficients (${\cal R}\equiv {{L ({\cal
N}-1)}\over{\lambda}} {{V^2}\over{c^2}}) $ \BE \label{C0} C_0=
-{{1}\over{4}}{\cal R}(3\cos 2\gamma -1) \cos^2\phi \EE
 \BE \label{C1} {C}_{s1}= - {{1}\over{2}}
{\cal R}\sin\alpha\sin 2\gamma\sin
2\phi~~~~~~~~~~~~~~~~~~~~~~{C}_{c1}= -{{1}\over{2}}{\cal R} \cos
\alpha\sin 2\gamma \sin 2\phi \EE \BE \label{C2} {C}_{s2}=
{{1}\over{2}}{\cal R} \sin2\alpha\cos^2\gamma(1+ \sin^2\phi)
~~~~~~~~~~~~~{C}_{c2}={{1}\over{2}} {\cal R} \cos
2\alpha\cos^2\gamma(1+ \sin^2\phi) \EE and \BE \label{S1} {S}_{s1}=
- {{ {C}_{c1}}\over{\sin \phi}}~~~~~~~~~~~~~{S}_{c1}= {{
{C}_{s1}}\over{\sin \phi}} \EE\BE \label{S2}{S}_{s2}= - {{ 2 \sin
\phi}\over{1+ \sin^2\phi}}{C}_{c2}~~~~~~~~~~~~~{S}_{c2}= {{ 2 \sin
\phi}\over{1+ \sin^2\phi}}{C}_{s2} \EE These standard forms are
nowadays adopted in the analysis of the data of the ether-drift
experiments \cite{peters}.  However, one should not forget that
Eq.(\ref{twoway1}) represents only an {\it approximation} for the
full structure Eq.(\ref{legendre}). Therefore, even for short-time
observations, one might not obtain from the data completely
consistent determinations of the kinematical parameters
$(V,\alpha,\gamma)$. In addition, by using a physical analogy, and
by representing the Earth's motion in the physical vacuum as the
motion of a body in a fluid, the scheme Eqs.(\ref{amorse1})$,
$(\ref{amorse2}) of smooth sinusoidal variations associated with the
Earth's rotation corresponds to the conditions of a pure laminar
flow associated with a simple regular motion. Instead, the physical
vacuum might behave as a turbulent fluid, where large-scale and
small-scale flows are only {\it indirectly} related.

In this modified perspective, which finds motivations in some basic
foundational aspects of both quantum physics and relativity
\cite{troshkin,puthoff,tsankov,chaos,plafluid} and in those
representations of the vacuum as a form of `space-time foam' which
indeed resembles a turbulent fluid \cite{wheeler1,migdal,ng2,ng3},
the ether-drift might exhibit forms of time modulations that do {\it
not} fit in the scheme of Eqs.(\ref{amorse1})$, $(\ref{amorse2}). To
evaluate the potential effects, and by still retaining the
functional form Eq.(\ref{fringe}), one could first re-write
Eqs.(\ref{CS}) as \BE \label{amplitude10}
       C(t)= {{L ({\cal N}-1)
}\over{\lambda}}~ {{v^2_x(t)- v^2_y(t)  }
       \over{c^2}}~~~~~~~~~~~~~~
       S(t)= {{L ({\cal N}-1)
}\over{\lambda}} ~{{2v_x(t)v_y(t)  }\over{c^2}} \EE where
$v_x(t)=v(t) \cos\theta_0(t)$ and $v_y(t)=v(t) \sin\theta_0(t)$.
Then, by exploiting the turbulence scenario, one could model the
two velocity components $v_x(t)$ and $v_y(t)$ as stochastic fluctuations.
In this different scheme, where now $v(t)\neq\tilde v(t)$ and
$\theta_0(t)\neq\tilde\theta_0(t)$, experimental results which, on
consecutive days and at the same sidereal time, deviate from
Eqs.(\ref{nassau1})$-$(\ref{projection}) do {\it not} necessarily
represent spurious effects. Equivalently, if data collected at the
same sidereal time average to zero this does {\it not} necessarily
mean that there is no ether-drift. This particular aspect will be
discussed at length in the rest of the paper.

After this important premise, we shall now proceed in Sects. 3-8
with our re-analysis of the classical experiments. In the end,
Sect.9 will contain a summary, a brief discussion of the modern experiments and
our conclusions.

\section{The original Michelson-Morley experiment}

The Michelson-Morley experiment \cite{michelson} is probably the
most celebrated experiment in the history of physics. Its result and
its interpretation have been (and are still) the subject of endless
controversies. For instance, for some time there was the idea
\cite{righi} that, by taking  into account the reflection from a
moving mirror and other effects, the predicted shifts would be
largely reduced and become unobservable. These points of view are
summarized in Hedrick's contribution to the `Conference on the
Michelson-Morley experiment' \cite{conference} (Pasadena, February
1927) which was attended by the greatest experts of the time, in
particular Lorentz and Michelson. The arguments presented by Hedrick
were, however, refuted by Kennedy \cite{kennedy} in a paper of 1935
where, by using Huygens principle, he re-obtained to order $v^2/c^2$
the classical result of Eq.(\ref{fringe0}) (with the identification
$v_{\rm obs}=v$).

\begin{figure}[ht] \psfig{figure=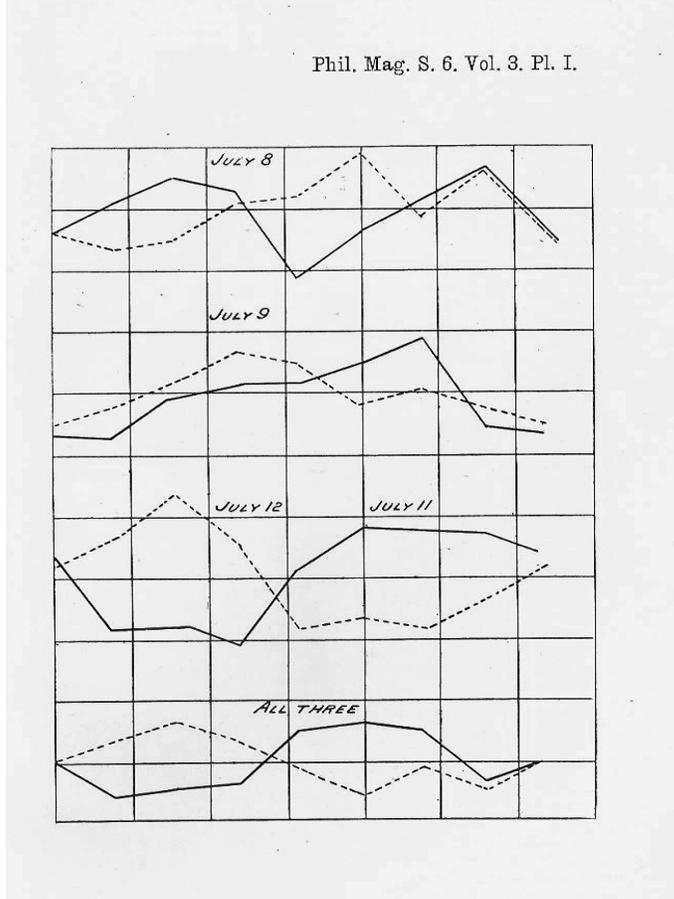,height=12 true
cm,width=9 true cm,angle=0} \caption{\it The Michelson-Morley fringe
shifts as reported by Hicks \cite{hicks}. Solid and dashed lines
refer respectively to noon and evening observations. }
\end{figure}
In this framework, the fringe shift is a second-harmonic effect,
i.e. periodic in the range $[0,\pi]$, whose amplitude $A_2$ is
predicted differently by using the classical formulas or Lorentz
transformations (\ref{fringe0})
 \BE \label{a2class}
 A^{\rm class}_2={{ L }\over{\lambda}} {{{v}^2}\over{c^2}}
~~~~~~~~~~~~~~~~~~~~~~~~~~~
  A^{\rm rel}_2= {{ L }\over{\lambda}}
{{{v_{\rm obs}}^2}\over{c^2}}\sim 2({\cal N}-1) A^{\rm class}_2 \EE
Notice also that upon rotation of $\pi/2$ with respect to
$\theta=\theta_0$ the predicted fringe shift is  $2{A}_2$. Now, for
the Michelson-Morley interferometer the whole effective optical path
was about $L=11$ meters, or about $2\cdot 10^7$ in units of light
wavelengths, so for a velocity $v\sim 30$ km/s (the Earth's orbital
velocity about the Sun, and consequently the minimum anticipated
drift velocity) the expected classical 2nd-harmonic amplitude was
${A}^{\rm class}_2\sim 0.2$. This value can thus be used as a
reference point to obtain an observable velocity, in the plane of
the interferometer, from the actual measured value of $A_2$ through
the relation \BE \label{ab2} v_{\rm obs} \sim 30 ~ \sqrt { {{A_2
}\over{ 0.2 }} }~~{\rm km/s} \EE Michelson and Morley performed
their six observations in 1887, on July 8th, 9th, 11th and 12th, at
noon and in the evening, in the basement of the Case Western
University of Cleveland.  Each experimental session consisted of six
turns of the interferometer performed in about 36 minutes. As well
summarized by Miller in 1933 \cite{miller}, ``The brief series of
observations was sufficient to show clearly that the effect did not
have the anticipated magnitude. However, and this fact must be
emphasized, {\it the indicated effect was not zero}".

The same conclusion had already been obtained by Hicks in 1902
\cite{hicks}: "..the data published by Michelson and Morley, instead
of giving a null result, show distinct evidence for an effect of the
kind to be expected". Namely, there was a second-harmonic effect.
But its amplitude was substantially smaller than the classical
expectation (see Fig.1).

Quantitatively, the situation can be summarized in Figure 2, taken
from Miller \cite{miller}, where the values of the effective
velocity measured in various ether-drift experiments are reported
and compared with a smooth curve fitted by Miller to his own results
as function of the sidereal time.
\begin{figure}[ht] \psfig{figure=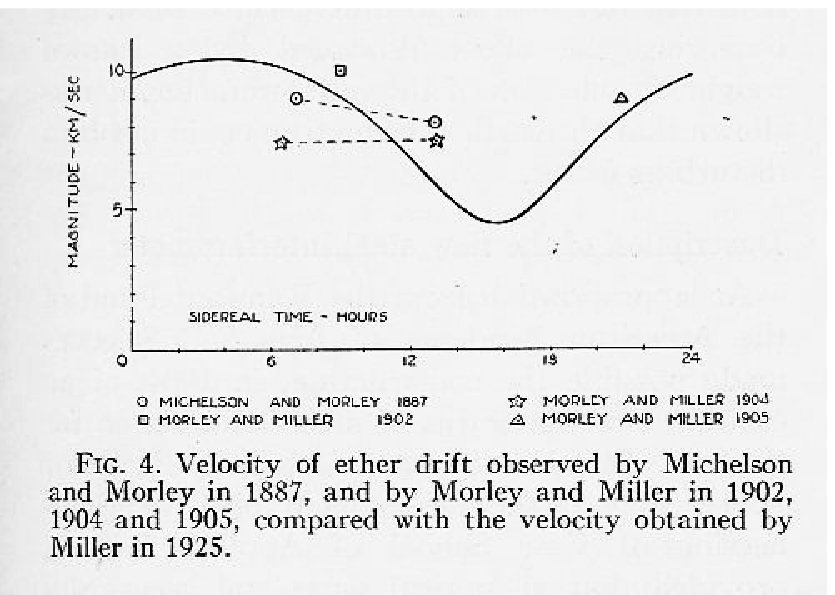,height=8 true
cm,width=11 true cm,angle=0} \caption{\it The magnitude of the
observable velocity measured in various experiments as reported by
Miller \cite{miller}.}
\end{figure}

For the Michelson-Morley experiment, the average observable velocity
reported by Miller is about 8.4 km/s. Comparing with the classical
prediction for a velocity of 30 km/s, this means an experimental
2nd- harmonic amplitude \BE \label{a2} A^{\rm EXP}_2\sim
0.2~({{8.4}\over{30}})^2\sim 0.016 \EE which is about twelve times
smaller than the expected result.

Neither Hicks nor Miller reported an estimate of the error on the
2nd harmonic extracted from the Michelson-Morley data. To understand
the precision of their readings, we can look at the original paper
\cite{michelson} where one finds the following statement : "The
readings are divisions of the screw-heads. The width of the fringes
varied from 40 to 60 divisions, the mean value being near 50, so
that one division means 0.02 wavelength". Now, in their tables
Michelson and Morley reported the readings with an accuracy of 1/10
of a division (example 44.7, 44.0, 43.5,..). This means that the
nominal accuracy of the readings was $\pm 0.002$ wavelengths. In
fact, in units of wavelengths, they reported values such as 0.862,
0.832, 0.824,.. Furthermore, this estimate of the error agrees well
with Born's book \cite{born}. In fact, Born, when discussing the
classically expected fractional fringe shift upon rotation of the
apparatus by $90^o$, about 0.37, explicitly says: ``Michelson was
certain that the one-hundredth part of this displacement would still
be observable" (i.e. 0.0037). Therefore, to be consistent with both
the original Michelson-Morley article and Born's quotation of
Michelson's thought, we shall adopt  $\pm 0.004$ as an estimate of
the error \footnote{To confirm that such estimate should not be
considered unrealistically small, we report explicitly Michelson's
words from ref.\cite{conference}:``I must say that every beginner
thinks himself lucky if he is able to observe a shift of 1/20 of a
fringe. It should be mentioned however that with some practice
shifts of 1/100 of a fringe can be measured, and that in very
favorable cases even a shift of 1/1000 of a fringe may be
observed."}.

With this premise, the Michelson-Morley data were re-analyzed in
ref.\cite{cimento}. To this end, one should first follow the well
defined procedure adopted in the classical experiments as described
in Miller's paper \cite{miller}. Namely, by starting from each set
of seventeen entries (one every $22.5^o$), say $E(i)$, one has first
to correct the data for the observed linear thermal drift. This is
responsible for the difference $E(1) -E(17)$ between the 1st entry
and the 17th entry obtained after a complete rotation of the
apparatus. In this way, by adding 15/16 of the correction to the
16th entry, 14/16 to the 15th entry and so on, one obtains a set of
16 corrected entries \BE E_{\rm corr}(i)={{i-1}\over{16}}
(E(1)-E(17)) + E(i) \EE \vskip 10 pt The fringe shifts are then
defined by the differences between each of the corrected entries
$E_{\rm corr}(i)$ and their average value $\langle E_{\rm corr}
\rangle$ as  \BE \label{corr}
 {{\Delta \lambda (i)}\over{\lambda}}=
E_{\rm corr}(i) - \langle E_{\rm corr} \rangle \EE The resulting
data are reported in Table 1.

\begin{table*} \caption{\it The fringe
shifts ${{\Delta \lambda(i)}\over{\lambda}}$ for all noon (n.) and
evening (e.) sessions of the Michelson-Morley experiment.}
\begin{center}
\begin{tabular}{clllllll}
\hline i ~~     &  July 8 (n.) & July 9 (n.) & July 11 (n.)
       &  July 8 (e.) & July 9 (e.) & July 12 (e.)     \\
\hline
1 ~~     & $-$0.001 & +0.018 &+0.016& $-$0.016& +0.007& +0.036 \\
2 ~~     & +0.024 & $-$0.004 &$-$0.034& +0.008& $-$0.015& +0.044 \\
3~~      & +0.053 & $-$0.004 &$-$0.038& $-$0.010& +0.006& +0.047 \\
4 ~~    & +0.015 & $-$0.003 &$-$0.066& +0.070& +0.004& +0.027 \\
5 ~~     & $-$0.036 & $-$0.031 &$-$0.042& +0.041& +0.027& $-$0.002 \\
6 ~~     & $-$0.007 & $-$0.020 &$-$0.014& +0.055& +0.015& $-$0.012 \\
7 ~~     & +0.024 & $-$0.025 &+0.000& +0.057& $-$0.022& +0.007 \\
8 ~~     & +0.026 & $-$0.021 &+0.028& +0.029& $-$0.036& $-$0.011 \\
9 ~~     & $-$0.021 & $-$0.049 &+0.002& $-$0.005& $-$0.033& $-$0.028 \\
10~~     & $-$0.022 & $-$0.032 &$-$0.010& +0.023& +0.001& $-$0.064 \\
11~~     & $-$0.031 & +0.001 &$-$0.004& +0.005& $-$0.008& $-$0.091 \\
12~~     & $-$0.005 & +0.012 &+0.012& $-$0.030& $-$0.014& $-$0.057 \\
13~~     & $-$0.024 & +0.041 &+0.048& $-$0.034& $-$0.007& $-$0.038 \\
14~~     & $-$0.017 & +0.042 &+0.054& $-$0.052& +0.015& +0.040 \\
15~~     & $-$0.002 & +0.070 &+0.038& $-$0.084& +0.026& +0.059 \\
16~~     & +0.022 & $-$0.005 &+0.006& $-$0.062& +0.024& +0.043 \\
\hline
\end{tabular}
\end{center}
\end{table*}
\begin{table*}
\caption{\it The amplitude of the fitted second-harmonic component
$A^{\rm EXP}_2$ for the six experimental sessions of the
Michelson-Morley experiment.}
\begin{center}
\begin{tabular}{cl}
\hline
SESSION       & ~~~~~~      $A^{\rm EXP}_2$   \\
\hline
July 8  (noon) & $0.010 \pm 0.005$  \\
July 9  (noon) & $0.015 \pm 0.005$   \\
July 11 (noon) & $0.025 \pm 0.005$    \\
July 8  (evening) & $0.014 \pm 0.005$  \\
July 9  (evening) &$0.011 \pm 0.005$   \\
July 12 (evening) & $0.024 \pm 0.005$  \\
\hline

\end{tabular}
\end{center}
\end{table*}

\begin{figure}[ht]
\psfig{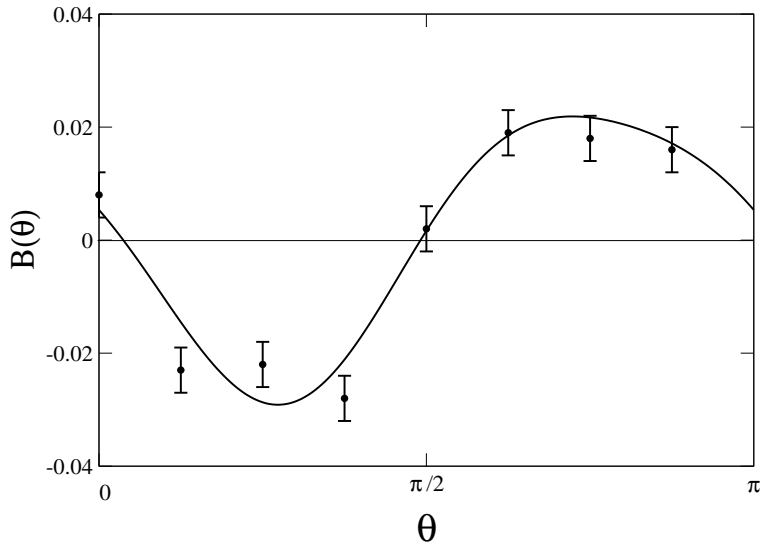}
\caption{\it A fit to the even combination $B(\theta)$
Eq.(\ref{even}). The second harmonic amplitude is ${A}^{\rm
EXP}_2=0.025 \pm 0.005$ and the fourth harmonic is ${A}^{\rm
EXP}_4=0.004 \pm 0.005$. The figure is taken from
ref.\cite{cimento}. Compare the data with the solid curve of July
11th shown in Fig.1. }
\end{figure}
With this procedure, the fringe shifts Eq.(\ref{corr}) are given as
a periodic function, with vanishing mean, in the range $0 \leq
\theta \leq 2\pi$, with $\theta={{i-1}\over{16}} ~2\pi$, so that
they can be reproduced in a Fourier expansion. Notice that in the
evening observations the apparatus was rotated in the opposite
direction to that of noon.

One can thus extract the amplitude and the phase of the 2nd-harmonic
component by fitting the even combination of fringe shifts
 \BE \label{even}  B(\theta)=
 {{\Delta \lambda (\theta) +\Delta \lambda (\pi
 +\theta)}\over{2\lambda}}\EE (see Fig.3).
 This is essential to cancel the 1st-harmonic contribution
originally pointed out by Hicks \cite{hicks}. Its theoretical
interpretation is in terms of the arrangements of the mirrors and,
as such, this effect has to show up in the outcome of real
experiments. For more details, see the discussion given by Miller,
in particular Fig.30 of ref.\cite{miller}, where it is shown that
his observations were well consistent with Hicks' theoretical study.
The observed 1st-harmonic effect is sizeable, of comparable
magnitude or even larger than the second-harmonic effect. The same
conclusion was also obtained by Shankland et {\it al.}
\cite{shankland} in their re-analysis of Miller's data. The
2nd-harmonic amplitudes from the six individual sessions are
reported in Table 2.

Due to their reasonable statistical consistency, one can compute the
mean and variance of the six determinations reported in Table 2 by
obtaining $A^{\rm EXP}_2 \sim 0.016 \pm 0.006$. This value is
consistent with an observable velocity \BE \label{vobsmm} v_{\rm
obs} \sim 8.4^{+1.5}_{-1.7} ~~~{\rm km/s} \EE Then, by using
Eq.(\ref{vobs}), which connects the observable velocity to the
projection of the kinematical velocity in the plane of the
interferometer through the refractive index of the medium where
light propagation takes place (in our case air where ${\cal N}\sim
1.00029$), we can deduce the average value \BE \label{vearth} v \sim
349^{+62}_{-70}~{\rm km/s}\EE

While the individual values of $A_2$ show a reasonable consistency,
there are substantial changes in the apparent direction $\theta_0$
of the ether-drift effect in the plane of the interferometer. This
is the reason for the strong cancelations obtained when fitting
together all noon sessions or all evening sessions \cite{handshy}.
For instance, for the noon sessions, by taking into account that the
azimuth is always defined up to $\pm 180^o$, one choice for the
experimental azimuths is $357^o \pm 14^o$, $285^o \pm 10^o$ and
$317^o \pm 8^o$ respectively for July 8th, 9th and 11th. For this
assignment, the individual velocity vectors $v_{\rm
obs}(\cos\theta_0,-\sin\theta_0)$ and their mean are shown in Fig.4.
\begin{figure}[ht]
\psfig{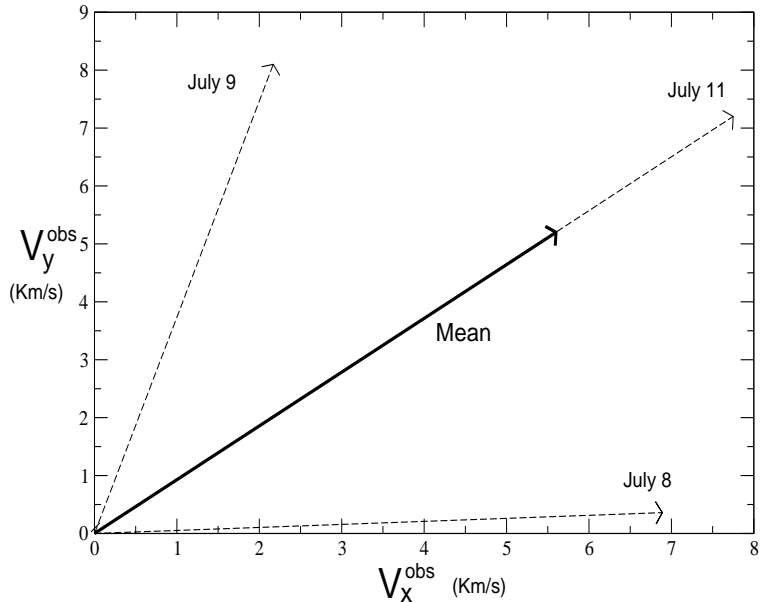}
\caption{\it The observable velocities for the three noon sessions
and their mean. The x-axis corresponds to $\theta_0=0^o\equiv 360^o$
and the y-axis to $\theta_0=270^o$. Statistical uncertainties of the
various determinations are ignored. All individual directions could
also be reversed by 180$^o$.}
\end{figure}
According to the usual interpretation, the large spread of the
azimuths is taken as indication that any non-zero fringe shift is
due to pure instrumental effects. However, as anticipated in Sect.2,
this type of discrepancy could also indicate an unconventional form
of ether-drift where there are substantial deviations from
Eq.(\ref{twoway1}) and/or from the smooth trend in
Eqs.(\ref{nassau1})$-$(\ref{projection}). For instance, in agreement
with the general structure Eq.(\ref{legendre}), and differently from
July 11 noon, which represents a very clean indication, there are
sizeable 4th- harmonic contributions (here ${A}^{\rm EXP}_4=0.019
\pm 0.005$ and ${A}^{\rm EXP}_4=0.008 \pm 0.005$ for the noon
sessions of July 8 and July 9 respectively). In any case, the
observed strong variations of $\theta_0$ are in qualitative
agreement with the analogous values reported by Miller. To this end,
compare with Fig.22 of ref.\cite{miller} and in particular with the
large scatter of the data taken around August 1st, as this
represents the epoch of the year which is closer to the period of
July when the Michelson-Morley observations were actually performed.
Thus one could also conclude that individual experimental sessions
indicate a definite non-zero ether-drift but the azimuth does not
exhibit the smooth trend expected from the conventional picture
Eqs.(\ref{nassau1})$-$(\ref{projection}).

For completeness, we add that the large spread of the
$\theta_0-$values might also reflect a particular systematic effect
pointed out by Hicks \cite{hicks}. As described by Miller
\cite{miller}, `` before beginning observations the end mirror on
the telescope arm is very carefully adjusted to secure vertical
fringes of suitable width. There are two adjustments of the angle of
this mirror which will give fringes of the same width but which
produce opposite displacements of the fringes for the same change in
one of the light-paths''. Since the relevant shifts are extremely
small, ``...the adjustments of the mirrors can easily change from
one type to the other on consecutive days. It follows that averaging
the results of different days in the usual manner is not allowable
unless the types are all the same. If this is not attended to, the
average displacement may be expected to come out zero $-$ at least
if a large number are averaged" \cite{hicks}. Therefore averaging
the fringe shifts from various sessions represents a delicate issue
and can introduce uncontrolled errors.  Clearly, this relative sign
does not affect the values of $A_2$  and this is why averaging the
2nd-harmonic amplitudes is a safer procedure. However, it can
introduce spurious changes in the apparent direction $\theta_0$ of
the ether-drift. In fact, an overall change of sign of the fringe
shifts at all $\theta-$values is equivalent to replacing $\theta_0
\to \theta_0 \pm \pi/2$. As a matter of fact, Hicks concluded that
the fringes of July 8th were of different type from those of the
remaining days. Thus for his averages (in our Fig.1) ``the values of
the ordinates are one-third of July 9 + July 11 $-$ July 8 and
one-third of July 9 + July 12 $-$ July 8'' \cite{hicks} for noon and
evening sessions respectively. If this were true, one choice for the
azimuth of July 8th could now be $\theta^{\rm EXP}_0=267^o \pm
14^o$. This would orient the arrow of July 8th in Fig.4 in the
direction of the y$-$axis and change the average azimuth from
$\langle \theta^{\rm EXP}_0\rangle \sim 317^o$ to $\langle
\theta^{\rm EXP}_0\rangle\sim 290^o$. We'll return to this
particular aspect in our Appendix II.

Let us finally compare with the interpretation that Michelson and
Morley gave of their data. They start from the observation that
"...the displacement to be expected was 0.4 fringe" while "...the
actual displacement was certainly less than the twentieth part of
this". In this way, since the displacement is proportional to the
square of the velocity, "...the relative velocity of the earth and
the ether is... certainly less than one-fourth of the orbital
earth's velocity". The straightforward translation of this upper
bound is $v_{\rm obs} < $ 7.5 km/s. However, this estimate is likely
affected by a theoretical uncertainty. In fact, in their Fig.6,
Michelson and Morley reported their measured fringe shifts together
with the plot of a theoretical second-harmonic component. In doing
so, they plotted a wave of amplitude $A_2=0.05$, that they interpret
as {\it one-eight} of the theoretical displacement expected on the
base of classical physics, thus implicitly assuming $A^{\rm
class}_2$=0.4. As discussed above, the amplitude of the classically
expected second-harmonic component is {\it not} 0.4 but is just
one-half of that, {\it i.e.} 0.2. Therefore, their experimental
upper bound $A^{\rm exp}_2 < {{0.4}\over{20}}=$0.02 is actually
equivalent to $v_{\rm obs} <$ 9.5 km/s. If we now consider that
their estimates were obtained after superimposing the fringe shifts
obtained from various sessions (where the overall effect is reduced,
see our Fig.1), we deduce a substantial agreement with our result
Eq. (\ref{vobsmm}).

\section{ Morley-Miller}

After the original 1887 experiment, there was much interest in the
Michelson-Morley result that, being too small to meet any classical
prediction, was apparently contradicting two cornerstones of
physics: Galilei's transformations and/or the existence of the
ether. For this reason, one of the most influential physicists of
the time, Lord Kelvin, after his conference at the 1900 Paris Expo,
induced Morley and his young collaborator Dayton Miller to design a
new interferometer (where the effective optical path was increased
up to 32 meters) to improve the accuracy of the measurement over the
1887 result.

It must be emphasized that Morley and Miller \cite{morley}, in their
observations of 1905, superimposed the data of the morning with
those of the evening. As explained by Miller \cite{conference}, the
two physicists were assuming that the ether drift had to be obtained
by combining the motion of the solar system relative to nearby
stars, i.e. toward the constellation of Hercules with a velocity of
about 19 km/s, with the annual orbital motion (``We now computed the
direction and the velocity of the motion of the centre of the
apparatus by compounding the annual motion in the orbit of the earth
with the motion of the solar system toward a certain point in the
heavens...There are two hours in each day when the motion is in the
desired plane of the interferometer" \cite{morley}).  The
observations at the two times (about 11:30 a.m. and 9:00 p.m.) were,
therefore, combined in such a way that the presumed azimuth for the
morning observations coincided with that for the evening (``The
direction of the motion with reference to a fixed line on the floor
of the room being computed for the two hours, we were able to
superimpose those observations which coincided with the line of
drift for the two hours of observation" \cite{morley}). However, the
observations for the two times of the day gave results having nearly
opposite phases. When these were combined, the result was nearly
zero. For this reason, the value then reported of an observable
velocity of 3.5 km/s is incorrect and does {\it not} correspond to
the actual results of the basic observations. The error was later
understood and corrected by Miller who found that the two sets of
data were each indicating an effective velocity of about 7.5 km/s
(see Figure 11 of Miller's paper \cite{miller}). For this reason,
the correct average observable velocities for the entire period
1902-1905 are those shown in our Figure 2 between 7 and 10 km/s or
\BE \label{vobst}
       v_{\rm obs} \sim (8.5 \pm 1.5)~{\rm km/s}
\EE By using Eq.(\ref{vobs}), we then deduce the average value \BE
\label{vearthm} v \sim (353\pm 62)~{\rm km/s}\EE

\section {Kennedy-Illingworth}

An interesting development was proposed by Kennedy in 1926. As
summarized in his contribution to the previously mentioned
Conference on the Michelson-Morley experiment \cite{conference}, his
small optical system was enclosed in an effectively insulated,
sealed metal case containing helium at atmospheric pressure. Because
of its small size, "...circulation and variation in density of the
gas in the light paths were nearly eliminated. Furthermore, since
the value of ${\cal N}-1$ is only about 1/10 that for the air at the
same pressure, the disturbing changes in density of the gas
correspond to those in air to only 1/10 of the atmospheric
pressure". The essential ingredient of Kennedy's apparatus consisted
in the introduction of a small step, 1/20 of wavelength thick, in
one of the total reflecting mirrors of the interferometer allowing,
in principle, for an ultimate fringe shift accuracy $1\cdot
10^{-4}$. To take full advantage of this possibility, Kennedy should
have disposed of perfect mirrors and of a suitable (hotter) source
of light. In the original version of the experiment, these
refinements were not implemented giving an actual fringe shift
accuracy of $2\cdot 10^{-3}$. In these conditions, as Kennedy
explicitly says\cite{conference}, "...the velocity of 10 km/s found
by Prof. Miller would produce a fringe shift corresponding to
$8\cdot 10^{-3}$", four times larger than the experimental
resolution. Since the effect is quadratic in the velocity, Kennedy's
result, fringe shifts $<2\cdot 10^{-3}$, can then be summarized as
\BE v_{\rm obs} <  5 ~{\rm km/s} \EE By using Eq.(\ref{vobs}), for
helium at atmospheric pressure where ${\cal N}\sim 1.000035$, this
bound amounts to restrict the kinematical value by $v < $ 600 km/s.
\begin{table*}
\caption{\it The infra-session averages $\langle D_A\rangle$ and
$\langle D_B\rangle$ obtained from the 10 sets of rotations in each
of the 32 sessions of Illingworth's experiment. These values have
been obtained from the weights of Illingworth's Table III by
applying the conversion factor 0.002.}
\begin{center}
\begin{tabular}{llllllll}
\hline~5~A.M.&~5~A.M.&~11~A.M.&~11~A.M.&~~5~P.M.&~~5~P.M.&~~11~P.M.&~~11~P.M. \\
 ~~~~$\rm \langle D_A\rangle$&~~~~$\rm\langle D_B\rangle$&
~~~~~$\rm \langle D_A\rangle $&~~~~~$\rm \langle D_B\rangle $&
~~~~~$\rm \langle D_A\rangle $&~~~~~$\rm \langle D_B\rangle
$&~~~~~$\rm \langle D_A\rangle $&~~~~~$\rm \langle D_B\rangle
$\\\hline
$+0.00024$ & $-0.00066$ & $+0.00070$ &  $-0.00022$ & $+0.00024$ & $+0.00044$ & $-0.00010$ &$+0.00024$\\
$+0.00114$ & $+0.00024$ & $-0.00042$ &  $-0.00036$ & $-0.00056$ & $-0.00046$ & $+0.00018$ &$+0.00018$\\
$+0.00000$ & $+0.00000$ & $-0.00006$ &  $-0.00052$ & $-0.00144$ & $-0.00080$ & $-0.00126$ &$-0.00006$\\
$+0.00020$ & $-0.00044$ & $-0.00030$ &  $+0.00012$ & $-0.00016$ & $+0.00004$ & $-0.00044$ &$-0.00026$\\
$+0.00064$ & $+0.00000$ & $-0.00022$ &  $+0.00038$ & $+0.00018$ & $+0.00016$ & $+0.00000$ &$+0.00024$\\
$-0.00002$ & $-0.00010$ & $+0.00048$ &  $+0.00020$ & $+0.00030$ & $+0.00030$ & $-0.00040$ &$-0.00004$\\
           &            & $-0.00014$ &  $-0.00006$ & $+0.00030$ & $+0.00014$ &            &        \\
           &            & $-0.00006$ &  $+0.00004$ & $+0.00036$ & $-0.00036$ &            &        \\
           &            & $-0.00006$ &  $+0.00016$ & $+0.00006$ & $-0.00006$ &            &        \\
           &            & $+0.00000$ &  $+0.00024$ & $-0.00010$ & $+0.00010$ &            &        \\
\hline
\end{tabular}
\end{center}
\end{table*}

Kennedy's apparatus was further refined by Illingworth in 1927
\cite{illingworth}. Besides improving the quality of the mirrors and
of the source, Illingworth's data taking was also designed to reduce
the presence of steady thermal drift and of odd harmonics. Looking
at Illingworth's paper, one finds that his refinements reached
indeed the nominal ${\cal O} (10^{-4})$ accuracy mentioned by
Kennedy, namely about 1/1500 of wavelength for the individual
readings and $(1\div 2)\cdot 10^{-4}$ at the level of average
values.

Let us now analyze Illingworth's results. He performed four series
of observations in the first ten days of July 1927. These consisted
of 32 experimental sessions, conducted daily at 5 A.M. (6), 11 A.M.
(10), 5 P.M. (10) and 11 P.M.(6), in which he was measuring the
fringe displacement caused by  a rotation through a right angle of
the apparatus. To take into account $90^o$ rotations let us first
re-write Eq.(\ref{fringe0}) as \BE\label{fringe2} {{\Delta
\lambda(\theta)}\over{\lambda}}= A_2 \cos2(\theta-\theta_0)\EE
Therefore Illingworth, in his first set (set A) of 10 rotations,
North, East, South, West and back to North, was actually  measuring
$D_A\equiv 2A_2\cos2\theta_0$. In a second set (set B), North-East,
North-West, South-West, South-East and back to North-East, performed
immediately after the set A, he was then measuring $D_B\equiv 2A_2
\sin2\theta_0$. Notice that both $D_A$ and $D_B$ differ from the
positive-definite quantity $D\equiv 2A_2$ that should be inserted in
Illingworth's numerical relation for his apparatus $v_{\rm
obs}=112\sqrt{D}$. Therefore, the reported values for the two
velocities $v_A=112\sqrt{|D_A|}$ and $v_B=112\sqrt{|D_B|}$ should
only be taken as {\it lower} bounds for the true $v_{\rm obs}$. The
mean values $\langle D_A\rangle$ and $\langle D_B\rangle$ obtained
from the 10 sets of rotations in the 32 individual sessions can be
obtained from Illingworth's Table III and, for the convenience of
the reader, are reported in our Table 3.

From Table 3, one finds that the quantity $\sqrt{\langle
D_A\rangle^2+\langle D_B\rangle^2}$ has a mean value of about
0.00045, which corresponds to $v_{\rm obs}\sim 2.4$ km/s. Thus, by
using Eq.(\ref{vobs}) for helium at atmospheric pressure, we would
tentatively deduce an average value $v\sim 284$ km/s.

However, this is only a very partial view. To go deeper into
Illingworth's experiment we have to consider his basic measurements,
i.e. the individual turns of his interferometer. In this case, the
only known basic set of data reported by Illingworth is set A of
July 9th, 11 A.M. This set has been re-analyzed by M\'unera
\cite{munera} and  his values for the fringe shifts are reported in
our Table 4.

\begin{table*}
\caption{\it Illingworth's set A of July 9th, 11 A.M. as re-analyzed
by M\'unera \cite{munera}.}
\begin{center}
\begin{tabular}{clll}
\hline Rotation & ~~~~~~$ \rm D_A $
&~~ $\rm |D_A| $&$\rm v_A$[km/s]\\
\hline
1   &  $-0.00100$ &  $+0.00100$& 3.54\\
2  &  $+0.00066$& $+0.00066 $ &2.89\\
3   &  $-0.00066$  &  $+0.00066$&2.89\\
4 &  $-0.00066 $ &  $ +0.00066$ &2.89\\
5   &  $-0.00166$ &  $+0.00166$ &4.57 \\
6  &  $+0.00234$& $+0.00234$ &5.41\\
7   &  $+0.00100 $  &  $+0.00100$ &3.54\\
8 &  $+0.00034 $ &  $ +0.00034$ & 2.04\\
9   &  $+0.00000$ &  $+0.00000$ & 0.00\\
10  &  $-0.00100$& $+0.00100$ & 3.54\\
\hline

\end{tabular}
\end{center}
\end{table*}

As one can see, the fringe shifts are not small and correspond to an
observable velocity in the range 2-5 km/s. However, their sign seems
to change randomly. Therefore, if one attempts to extract the
observable velocity from the mean of the 10 determinations, $\langle
D_A\rangle\sim -0.00006$, the resulting value 0.9 km/s is much
smaller than all individual determinations. The basis of M\'unera's
analysis was instead to estimate $v_{\rm obs}$ from $\langle |D_A|
\rangle$, from which he obtained an average velocity $v_{\rm obs}=
3.13 \pm 1.04$ km/s.

Now, the standard interpretation of such apparently random changes
of sign is in terms of typical instrumental effects and the standard
method for eliminating these is the original averaging procedure as
employed by Illingworth. But we will now show that they could also
indicate an unconventional form of stochastic drift, of the type
already mentioned in the previous sections, and in which M\'unera's
re-estimate has a definite significance.  To this end, we shall
first use the relations \BE \label{DAB} D_A(t) = 4
C(t)~~~~~~~~~~~~~~~~~~~~~D_B(t)=4S(t) \EE where the two functions
$C(t)$ and $S(t)$ have been introduced in Eqs.(\ref{CS}) and
(\ref{amplitude10}). Thus Eqs.(\ref{DAB}) can be re-written as \BE
\label{amplitude100}
       D_A(t)= {{8L ({\cal N}-1)
}\over{\lambda}}~ {{v^2_x(t)- v^2_y(t)  }
       \over{2c^2}}~~~~~~~~~~~~~~
       D_B(t)= {{8L ({\cal N}-1)
}\over{\lambda}} ~{{v_x(t)~v_y(t)  }\over{c^2}} \EE where
$v_x(t)=v(t) \cos\theta_0(t)$ and $v_y(t)=v(t) \sin\theta_0(t)$. In
this way, by using the numerical relation for Illingworth's
experiment ${{L}\over{\lambda}} ~{{(30 {\rm km/s})^2
}\over{c^2}}\sim 0.035$ and the value of the helium refractive
index, we obtain \label{amplitude1000}
      \BE  D_A(t) \sim  {{v^2_x(t)- v^2_y(t)}
       \over{\rm 2\cdot (300~km/s)^2 }}\cdot 10^{-3}~~~~~~~~~~~~~~
       D_B(t) \sim {{v_x(t)~v_y(t)}\over{\rm (300~km/s)^2}}\cdot 10^{-3} ~ \EE

The required random ingredient can then be introduced by
characterizing the two velocity components $v_x(t)$ and $v_y(t)$ as
turbulent fluctuations. To this end, there can be several ways. Here
we shall restrict to the simplest choice of a turbulence which, at
small scales, appears statistically isotropic and homogeneous
\footnote{This picture reflects the basic Kolmogorov theory
\cite{kolmo} of a fluid with vanishingly small viscosity.}. This
represents a zeroth-order approximation which is motivated by the
substantial reading error of the Illingworth measurements (it turns
out to be comparable to the effects of turbulence). However, it is a
useful example to illustrate basic phenomenological features
associated with an underlying stochastic vacuum. To explore the
resulting temporal pattern of the data, we have followed
refs.\cite{landau,fung} where velocity flows, in statistically
isotropic and homogeneous 3-dimensional turbulence, are generated by
unsteady random Fourier series. The perspective is that of an
observer moving in the turbulent fluid who wants to simulate the two
components of the velocity in his x-y plane at a given fixed
location in his laboratory. This leads to the general expressions
\BE \label{vx} v_x(t)= \sum^{\infty}_{n=1}\left[
       x_n(1)\cos \omega_n t + x_n(2)\sin \omega_n t \right] \EE
\BE \label{vy} v_y(t)= \sum^{\infty}_{n=1}\left[
       y_n(1)\cos \omega_n t + y_n(2)\sin \omega_n t \right] \EE
where $\omega_n=2n\pi/T$, T being a time scale which represents a
common period of all stochastic components. We have adopted the
typical value $T=T_{\rm day}$= 24 hours. However, we have also
checked with a few runs that the statistical distributions of the
various quantities do not change substantially by varying $T$ in the
rather wide range $0.1~T_{\rm day}\leq T \leq 10~T_{\rm day}$.

The coefficients $x_n(i=1,2)$ and $y_n(i=1,2)$ are random variables
with zero mean. They have the physical dimension of a velocity and
we shall denote by $[-\tilde v,\tilde v]$ the common interval for
these four parameters. In terms of $\tilde v$ the statistical average of
the quadratic
values can be expressed as \BE \label{quadratic} \langle x^2_n(i=1,2)\rangle_{\rm stat}=\langle
y^2_n(i=1,2)\rangle_{\rm stat}={{{\tilde v}^2 }\over{3 ~n^{2\eta}}} \EE  for
the uniform probability model (within the interval $[-\tilde
v,\tilde v]$) which we have chosen for our simulations. Finally, the
exponent $\eta$ controls the power spectrum of the fluctuating
components. For the simulations, between the two values $\eta=5/6$
and $\eta=1$ reported in ref.\cite{fung}, we have chosen $\eta=1$
which corresponds to the point of view of an observer moving in the
fluid.

Thus, within this simple model for $D_A(t)$ and $D_B(t)$, $\tilde v$
is the only parameter whose numerical value could reflect the
properties of a large-scale motion, for instance of the Earth's
motion with respect to the Cosmic Microwave Background (CMB).  For
this reason, here, we have adopted the fixed value $\tilde{v}=V_{\rm
CMB}= $ 370 km/s.
\begin{figure}[ht]
\begin{center}
\psfig{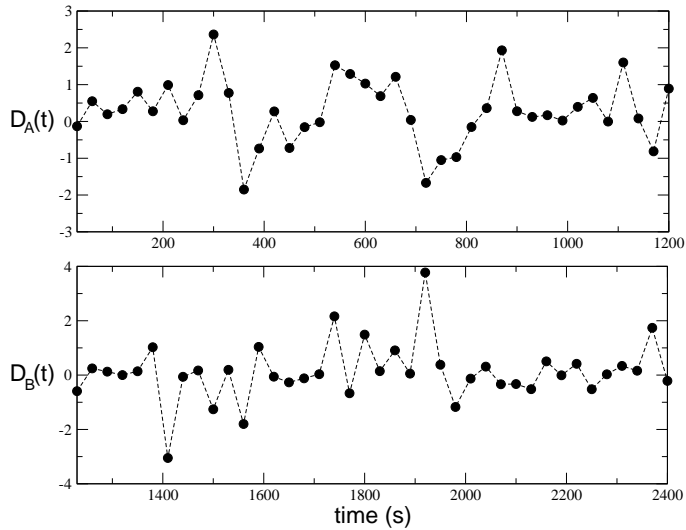}
\end{center}
\caption{ {\it A simulation of $\rm D_A(t)$ and $\rm D_B(t)$, in
units $10^{-3}$ and every 30 seconds, from typical sequences of 1200
seconds. The average values are $\langle D_A \rangle=0.00028$ and
$\langle D_B \rangle= 0.00011$. The velocity parameter is
$\tilde{v}=V_{\rm CMB}= $ 370 km/s.}} \label{Fig.5}
\end{figure}
With these premises, our results can be illustrated by first
considering the basic set of 10 complete rotations of the apparatus
during which Illingworth's fringe shifts (produced by $90^o$
rotations) were recorded every 30 seconds. Therefore, this type of
simulations consists in generating 40 values during a total time of
1200 seconds. As an illustration, two typical sequences of $\rm
D_A(t)$ and $\rm D_B(t)$, in units $10^{-3}$, are shown in Fig.5.

\begin{figure}
\begin{center}
\psfig{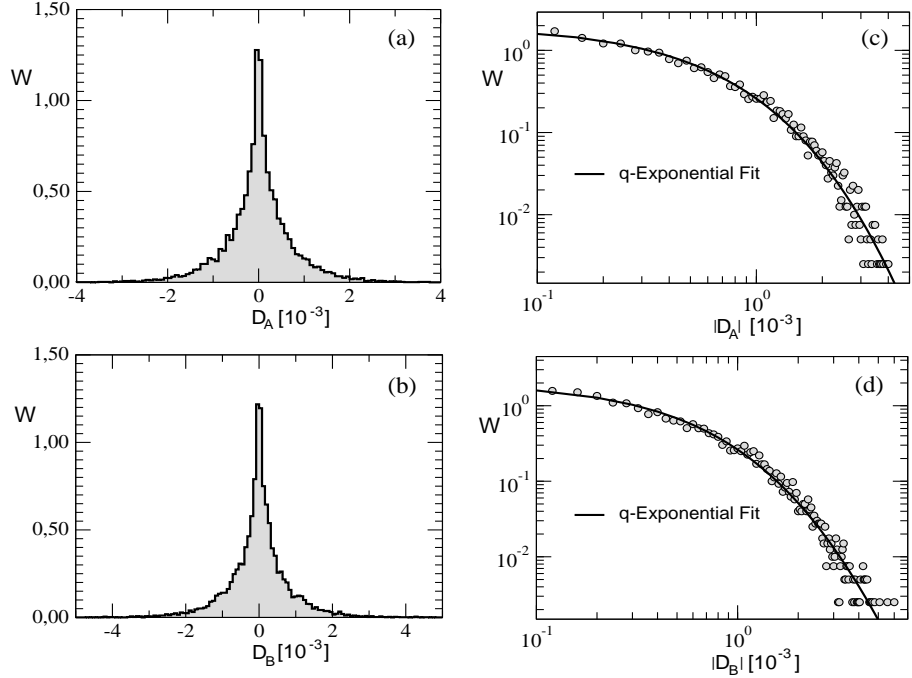}
\end{center}
\caption{ {\it We show, see (a) and (b), the histograms $W$ obtained
from a simulation for $\rm D_A=D_A(t)$ and $\rm D_B=D_B(t)$. The
vertical normalization is to a unit area. The mean values are $\rm
\langle D_A\rangle =0.75 \cdot 10^{-5}$, $\rm \langle D_B\rangle
=-1.1 \cdot 10^{-5}$ and the standard deviations $\rm
\sigma(D_A)=0.75 \cdot 10^{-3}$, $\rm \sigma(D_B)=0.83\cdot
10^{-3}$. We also show, see (c) and (d), the corresponding plots in
logarithmic scale and the fits with Eq.(\ref{qexp}). The parameters
of the fit are q=1.07, a=2 and b=2.2 for $\rm D_A$ and q=1.12, a=2
and b=2.3 for $\rm D_B$. The total statistics correspond to 10.000
values generated at steps of 30 seconds. The velocity parameter is
$\tilde{v}=V_{\rm CMB}= $ 370 km/s}. } \label{Fig.6}
\end{figure}
As one can see, the magnitude ${\cal O}(10^{-3})$ and the random
nature of the instantaneous values is completely consistent with the
entries of Table 4. Also the resulting infra-session averages
$\langle D_A \rangle=0.00028$ and $\langle D_B \rangle= 0.00011$ are
completely consistent with the typical entries of Table 3.

To obtain further insight, we have then performed extensive
simulations for large sequences of measurements. The histograms of a
set of 10000 determinations of $D_A(t)$ and $D_B(t)$ (again
generated every 30 seconds) are reported in panels (a) and (b) of
Fig.6.

Notice that these distributions are clearly ``fat-tailed'' and very
different from a Gaussian shape. This kind of behavior is
characteristic of probability distributions for instantaneous data
in turbulent flows (see e.g. \cite{sreenivasan,beck}). To better
appreciate the deviation from Gaussian behavior, in panels (c) and
(d) we plot the same data in a log$-$log scale. The resulting
distributions are well fitted by the so-called $q-$exponential
function \cite{tsallis} \BE \label{qexp} f_q(x) = a (1 - (1 - q) x
b)^{1 / (1 - q)}\EE  with entropic index $q\sim 1.1$. For such large
samples of data, the statistical averages $ \langle D_A\rangle$ and
$ \langle D_B\rangle$ are vanishingly small in units of the typical
instantaneous values ${\cal O}(10^{-3})$ and any non-zero average
has to be considered as statistical fluctuation. On the other hand,
the standard deviations $\sigma(D_A)$ and $\sigma(D_B)$ have
definite non-zero values which reflect the magnitude of the scale
parameter $\tilde v$. By keeping
 $\tilde v$ fixed at 370 km/s, we have found
\BE \label{sigmas} \sigma(D_A)\sim
(0.74 \pm 0.05)\cdot 10^{-3}
 ~~~~~~~~~~~~~~~~~~ \sigma(D_B)\sim(0.83 \pm 0.06)\cdot 10^{-3}\EE
whose uncertainties reflect the observed variations due to the
truncation of the Fourier modes in Eqs.(\ref{vx}), (\ref{vy}) and to
the dependence on the random sequence. Taking this calculation into
account gives a mean spread slightly less, about $0.65 \cdot
10^{-3}$, for the effect of stochastic drift in Illingworth's
measurements. This is comparable to the uncertainty of the
individual readings which, in the best case, was of 1/1500
wavelengths, i.e. $\pm 0.7 \cdot 10^{-3}$. By combining in
quadrature the two uncertainties, one gets a good agreement with our
Table 4 where the variance of the mean is about $\pm 1\cdot
10^{-3}$. Finally, the simulation is also useful to get indications
on the expected value of the observable velocity. In fact, with
vanishingly small values of $ \langle D_A\rangle$ and $ \langle
D_B\rangle$ one gets $\langle D^2_A\rangle\sim \sigma^2(D_A)$ and
$\langle D^2_B\rangle\sim \sigma^2(D_B)$. Therefore one obtains the
following two average estimates of $v_{\rm obs}$ \BE v_{\rm obs}\sim
112 \sqrt{\sigma(D_A)}\sim 3.05 ~{\rm km/s}
~~~~~~~~~~~~~~~~~~~v_{\rm obs}\sim 112 \sqrt{\sigma(D_B)}\sim 3.23
~{\rm km/s} \EE with a mean value of 3.14 km/s which is very close
to M\'unera's determination
 $v_{\rm obs}= 3.13 \pm 1.04$ km/s.

We emphasize that one could further improve the stochastic model by
introducing time modulations and/or slight deviations from isotropy.
For instance, $\tilde v$ could become a function of time $\tilde
v=\tilde v(t)$. By still retaining statistical isotropy, this could
be used to simulate the possible modulations of the projection of
the Earth's velocity in the plane of the interferometer. Or, one
could fix a range, say $[-\tilde v_x,\tilde v_x]$, for the two
random parameters $x_n(1)$ and $x_n(2)$, which is different from the
range $[-\tilde v_y,\tilde v_y]$ for the other two parameters
$y_n(1)$ and $y_n(2)$. Finally, $\tilde v_x$ and $\tilde v_y$ could
also become given functions of time, for instance $\tilde
v_x(t)\equiv \tilde v(t) \cos \tilde\theta_0(t)$ $\tilde
v_y(t)\equiv \tilde v(t) \sin \tilde\theta_0(t)$, $\tilde v(t)$ and
$\tilde\theta_0(t)$ being defined in Eqs.
(\ref{nassau1})$-$(\ref{projection}). We shall discuss this other
alternative later on, in connection with the much more accurate Joos
1930 experiment.

In any case, by accepting this type of picture of the ether-drift,
it is clear that further reduction of the data by performing
inter-session averages ($\langle\langle...\rangle\rangle$) among the
various sessions, can wash out completely the physical information
contained in the original observations. In Table 5, we report the
final inter-session averages $\langle\langle D_A\rangle\rangle $ and
$\langle\langle D_B\rangle\rangle $ obtained by Illingworth for the
various observation times.

\begin{table*}
\caption{\it Illingworth's final inter-session averages.}
\begin{center}
\begin{tabular}{cll}
\hline Observations  & ~~~~~~~~~~~$\langle \langle D_A\rangle\rangle
$
&~~~~~~~~~~ $ \langle \langle D_B\rangle \rangle $\\
\hline
5 A.M.   &  $+0.00036\pm 0.00012$ &  $-0.00016 \pm 0.00009$ \\
11 A.M.  &  $-0.00001\pm 0.00007$& $-0.00000 \pm 0.00006$ \\
5 P.M.   &  $-0.00008 \pm 0.00012$  &  $-0.00005 \pm 0.00008$\\
11 P.M. &  $-0.00034 \pm 0.00014$ &  $ +0.00005\pm 0.00006$ \\

\hline
\end{tabular}
\end{center}
\end{table*}

Nevertheless, in spite of the strong cancelations expected from the
averaging reduction process mentioned above, some non-zero value is
still surviving. Therefore, regardless of our simulations, one could
draw the following conclusions. Traditionally, from these final
averages for $\langle\langle D_A\rangle \rangle $ at 5 A.M. and at
11 P.M. one has been deducing the values $v_A \sim 2.12 $ km/s and
$v_A\sim 2.07$ km/s respectively. Therefore, from these two
estimates of $v_A$ that, as anticipated, represent {\it lower}
bounds for $v_{\rm obs}$, it follows that there were values of
$v_{\rm obs}$ which clearly had to be {\it larger} than both. For
this reason, this 2.1 km/s velocity value reported by Illingworth,
rather than being interpreted as an {\it upper} bound  could also be
interpreted as a {\it lower} bound placed by his experiment. In this
way, by combining with the previous Kennedy's upper bound $v_{\rm
obs} < 5$ km/s, one would deduce that these two experiments, where
light was propagating in helium at atmospheric pressure, give a
range for the observable velocity \BE{\rm (Kennedy +
Illingworth)}\hskip 80 pt 2~ {\rm km/s} \lesssim v_{\rm obs} < 5 ~
{\rm km/s}\hskip 90 pt \EE in complete agreement with M\'unera's
determination \BE v_{\rm obs}=3.1 \pm 1.0~{\rm km/s}\EE From this
last estimate, by using Eq.(\ref{vobs}) and taking into account that
for helium at atmospheric pressure the refractive index is ${\cal
N}\sim 1.000035$, one obtains a kinematical velocity \BE v \sim (370
\pm 120)~{\rm km/s}\EE consistently with the velocity values
Eqs.(\ref{vearth}) and (\ref{vearthm}) from the Michelson-Morley and
Morley-Miller experiments.

\section{Miller}

M\'unera's analysis \cite{munera} is also interesting because he
applied the same method used for Illingworth's observations to the
only known Miller set of data explicitly reported in the literature.
In this case, his value $v_{\rm obs}=8.2 \pm 1.4$ km/s,
after correcting with Eq.(\ref{vobs}), confirms
the estimate $v\sim 350$ km/s for the average velocity in
the plane of the interferometer.

This close agreement with the Michelson-Morley value 8.4 km/s is
also confirmed by the critical re-analysis of Shankland et {\it al.}
\cite{shankland}. Differently from the original Michelson-Morley
experiment, Miller's data were taken over the entire day and in four
epochs of the year. However, after the critical re-analysis of the
original raw data performed by the Shankland team, there is now an
independent estimate of the average determinations $A^{\rm EXP}_2$
for the four epochs. Their values 0.042, 0.049, 0.038 and 0.045,
respectively for April 1925, July 1925, September 1925 and February
1926 (see page 170 of ref.\cite{shankland}) are so well
statistically consistent that one can easily average them. The
overall determination from Table III of \cite{shankland} \BE  A^{\rm
EXP}_2=0.044 \pm 0.022 \EE  when compared with the equivalent
classical prediction for Miller's interferometer $A^{\rm
class}_2={{L}\over{\lambda}} {{(30 {\rm km/s})^2 }\over{c^2}}\sim
0.56$ corresponds to an average observable velocity \BE v_{\rm obs}
= 8.4^{+1.9}_{-2.5} ~{\rm km/s}\EE and, by using Eq.(\ref{vobs}), to
a true kinematical value \BE v = 349^{+79}_{-104}~{\rm km/s} \EE We
are aware that our conclusion goes against the widely spread belief,
originating precisely from the paper of Shankland et {\it al.}
ref.\cite{shankland}, that Miller's results might actually have been
due to statistical fluctuation and/or local temperature conditions.
To a closer look, however, the arguments of Shankland et {\it al.}
are not so solid as they appear when reading the Abstract of their
paper \footnote{A detailed rebuttal of the criticism raised by the
Shankland team can be found in ref.\cite{demeo}.}. In fact, within
the paper these authors say that ``...there can be little doubt that
statistical fluctuations alone cannot account for the periodic
fringe shifts observed by Miller" (see page 171 of
ref.\cite{shankland}). Further, although ``...there is obviously
considerable scatter in the data at each azimuth position,...the
average values...show a marked second harmonic effect" (see page 171
of ref.\cite{shankland}). In any case, interpreting the observed
effects on the basis of the local temperature conditions is
certainly not the only explanation since ``...we must admit that a
direct and general quantitative correlation between amplitude and
phase of the observed second harmonic on the one hand and the
thermal conditions in the observation hut on the other hand could
not be established" (see page 175 of ref.\cite{shankland}).

\begin{table*}
\caption{\it The symmetric combination of fringe shifts
$B(\theta)={{\Delta \lambda(\theta) + \Delta
\lambda(\pi+\theta)}\over{2\lambda}} $ at the various values of
$\theta$ for the set of 20 turns of the interferometer reported in
Fig.8 of ref.\cite{miller}. For our global fit, following
ref.\cite{shankland}, the nominal accuracy of each entry has been
fixed to $\pm 0.050$.}
\begin{center}
\begin{tabular}{cllllllll}
\hline Turn &~~~ $ 0^o$
&~ $ 22.5^o$&~~$45^o$&$67.5^o$&~~$90^o$&$112.5^o$&~~$135^o$&$157.5^o$\\
\hline
1 & +0.091& +0.159& +0.028& +0.047& $-$0.034& $-$0.116& $-$0.147& $-$0.028\\
2 & $-$0.025& +0.063& +0.050& +0.088& $-$0.075& $-$0.038& +0.000& $-$0.063\\
3 & +0.022& +0.103& +0.084& +0.016& $-$0.053& $-$0.072& $-$0.091& $-$0.009\\
4 & +0.034& $-$0.009& $-$0.053& $-$0.047& $-$0.041& +0.016& +0.022& +0.078\\
5 & +0.169& +0.081& +0.044& $-$0.044& $-$0.081& $-$0.169& $-$0.056& +0.056\\
6 & $-$0.025& +0.025& +0.025& +0.025& +0.025& $-$0.025& $-$0.025& $-$0.025\\
7 & +0.081& +0.094& +0.056& +0.069& $-$0.119& $-$0.106& $-$0.094& +0.019\\
8 & +0.066& +0.072& $-$0.022& $-$0.066& $-$0.059& $-$0.003& +0.003& +0.009\\
9 & +0.041& +0.084& +0.078& +0.022& $-$0.134& $-$0.141& +0.003& +0.047\\
10& +0.016& +0.072& +0.078& $-$0.016& $-$0.009& $-$0.003& $-$0.047& $-$0.091\\
11& +0.009& +0.053& +0.097& $-$0.009& $-$0.116& $-$0.072& +0.022& +0.016\\
12& +0.022& +0.016& +0.059& +0.003& $-$0.053& $-$0.009& $-$0.016& $-$0.022\\
13& +0.000& +0.063& +0.025& +0.038& +0.050& $-$0.038& $-$0.075& $-$0.063\\
14& $-$0.034& +0.047& +0.078& +0.009& $-$0.009& $-$0.028& $-$0.047& $-$0.016\\
15& +0.113& +0.125& +0.138& +0.000& $-$0.088& $-$0.125& $-$0.113& $-$0.050\\
16& +0.025& +0.050& +0.025& +0.050& $-$0.025& $-$0.050& $-$0.025& $-$0.050\\
17& +0.000& $-$0.012& $-$0.025& +0.063& +0.000& $-$0.012& $-$0.025& +0.013\\
18& +0.044& +0.050& +0.019& $-$0.019& $-$0.056& $-$0.044& $-$0.031& +0.031\\
19& +0.053& +0.059& +0.016& $-$0.028& $-$0.022& $-$0.066& $-$0.009& $-$0.003\\
20& +0.059& +0.041& +0.122& +0.003& $-$0.066& $-$0.084& $-$0.053& $-$0.022\\
\hline
\end{tabular}
\end{center}
\end{table*}

Most surprisingly, however, Shankland et {\it al.} seem not to
realize that Miller's average value $ A^{\rm EXP}_2\sim 0.044 $,
obtained after {\it their own} re-analysis of his observations at
Mt.Wilson, when compared with the reference classical value $ A^{\rm
class}_2=0.56$ for his apparatus, was giving the same observable
velocity $v_{\rm obs}\sim 8.4$ km/s obtained from Miller's
re-analysis of the Michelson-Morley experiment in Cleveland.
Conceivably, their emphasis on the role of temperature effects would
have been re-considered had they realized the perfect identity of
two determinations obtained in completely different experimental
conditions. In this sense, an interpretation in terms of a
temperature gradient is only acceptable provided this gradient
represents a {\it non-local} effect, as in our model of the ether
drift from a fundamental vacuum energy-momentum flow.

Another criticism of Miller's work was recently presented by Roberts
\cite{roberts}. This author, using the set of data reported in Fig.8
of ref.\cite{miller}, raises several objections to the validity of
Miller's observations. The two main objections concern i) the
subtraction of the steady thermal drift, which was approximated by
Miller as a pure linear effect, and ii) the statistical significance
of the measurements. Concerning remark i), Roberts reports in his
Fig.3 a broken line that reproduces the expected linear trend. He
also reports some chosen points (differing from the corners of the
broken line by 180 degrees) that, due to the 2nd-harmonic nature of
the ether-drift effect, should lie on the line. However, this
expectation ignores that, as already pointed out for the
Michelson-Morley experiment, real measurements contains large
first-harmonic effects. These only cancel when taking symmetric
combinations of data at the various angles $\theta$ and $\pi +
\theta$. As a matter of fact, the autocorrelative methods and
further tests applied by the Shankland team over all of Miller's
data confirmed the linear drift approximation as remarkably good
(see their footnote 21 on page 177 of \cite{shankland}).

Concerning remark ii), according to Roberts, the experimental
uncertainties are so large that the observed 2nd-harmonic effect has
no statistical significance. To check this point we have re-computed
ourselves the fringe shifts for the set of 20 turns of the
interferometers (reported in Fig.8 of ref.\cite{miller}) considered
by Roberts, by following the same procedure explained in Sect.3. The
resulting symmetric combinations of fringe shifts \BE
B(\theta)={{\Delta \lambda(\theta) + \Delta
\lambda(\pi+\theta)}\over{2\lambda}} \EE are reported in our Table
6.

We have then fitted these data by including both 2nd and 4th
harmonic terms. Notice that, differently from Roberts' analysis, we
do not perform any averaging of data obtained from different turns
of the interferometer. For our global fit, to estimate the accuracy
of the various determinations, we have followed ref.\cite{shankland}
and adopted a nominal uncertainty $\pm 0.050$ for each entry of
Table 6. From the fit, where the 4th harmonic is completely
consistent with the background ($ A^{\rm EXP}_4=0.004\pm 0.012$), we
have obtained a chi-square of 130 for $157$ degrees of
freedom and the following values \BE \label{small} A^{\rm
EXP}_2=0.061 \pm 0.012~~~~~~~~~~~~~~~~~~~~~~~~\theta^{\rm
EXP}_0=24^o\pm 7^o\EE Here errors correspond to the overall boundary
$\Delta\chi^2=+3.67$, as appropriate \footnote{This probability
content assumes a Gaussian distribution as for typical statistical
errors. } for a 70$\%$ C. L. in a 3-parameter fit \cite{minuit}.
Notice that, even though the fitted $A_2$ Eq.(\ref{small}) is only
20$\%$ larger than the nominal accuracy $\pm 0.050$ of each entry,
the data are distributed in such a way to produce a $5\sigma$
evidence for a non-zero 2nd harmonic.

As for Illingworth's experiment, we have also analyzed the results
obtained from the individual turns of the interferometer. To this
end, we report in Figs. 7 and 8 the plots of the azimuth and of the
2nd harmonic for the 20 rotations.

\begin{figure}[ht]
\begin{center}
\psfig{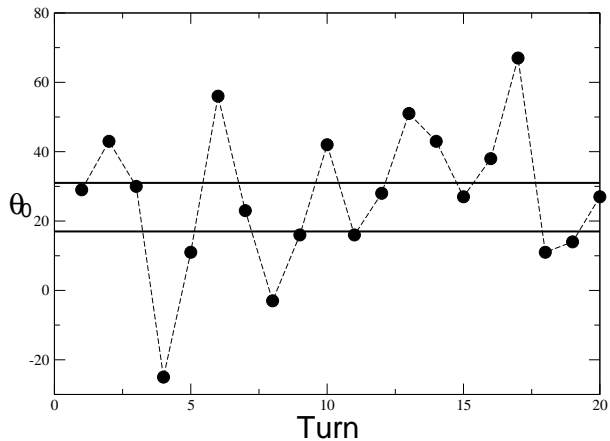}
\end{center}
\caption{ {\it The azimuth (in degrees) for the 20 individual turns
of the interferometer reported in Table 6. The average uncertainty
of each determination is about $\pm 20^o$. The band between the two
horizontal lines corresponds to the global fit $\theta_0= 24^o \pm
7^o$. Each individual value could also be reversed by 180 degrees.}}
\label{Fig.7}
\end{figure}
\begin{figure}[ht]
\begin{center}
\psfig{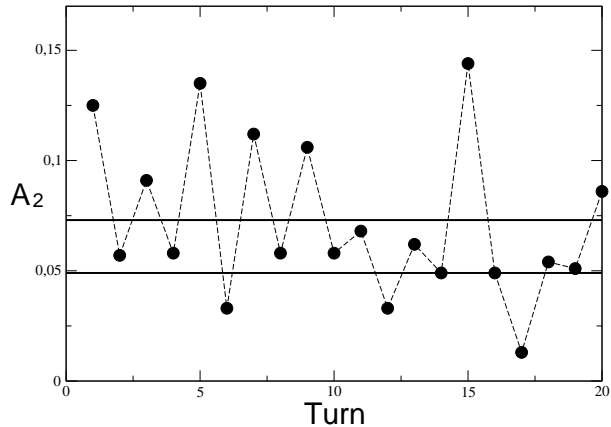}
\end{center}
\caption{ {\it The 2nd-harmonic amplitude for the 20 individual
turns of the interferometer reported in Table 6. The average
uncertainty of each determination is about $\pm 0.030$. The band
between the two horizontal lines corresponds to the global fit $A_2=
0.061 \pm 0.012$. Within their errors, these individual values
correspond to an observable velocity in the range 4$\div$14 km/s. }}
\label{Fig.8}
\end{figure}

To conclude our analysis of Miller's experiment, we want to mention
that other objections to the overall consistency of his solution for
the Earth's cosmic motion \cite{miller} were raised by von Laue
\cite{laue} and Thirring \cite{thirring}. Their argument, which
concerns the observed displacement of the maximum of the fringe
pattern {\it averaged over all sidereal times}, was also re-proposed
by Shankland et al. \cite{shankland} and amounts to the following.

By assuming relations (\ref{nassau1})$-$(\ref{S2}) and denoting by
$\langle...\rangle$ the daily average of any given quantity, one
finds, at any angle $\theta$, the daily averaged fringe shift
\BE\langle{{ \Delta\lambda(\theta)}\over{\lambda}}\rangle = 2\langle
\tilde C(t) \rangle \cos2\theta \EE since $\langle \tilde S(t)
\rangle =0$ with \BE \langle \tilde C(t) \rangle = -{{L ({\cal N}-1)
}\over{\lambda}}{{V^2}\over{c^2}}~{{1}\over{4}}(3\cos 2\gamma -1)
\cos^2\phi \EE The result can then be cast into the form
\cite{shankland} \BE \label{avfringe} \langle{{
\Delta\lambda(\theta)}\over{\lambda}}\rangle=V^2 F(\gamma,\phi) \cos
2\theta \EE Therefore, since the latitude $\phi$ is a constant and
the angular declination $\gamma$ is fixed at any specific epoch, the
daily averaged fringe shifts should all have a common maximum at the
value $\theta=0$. Only the amplitude can be different at different
epochs. Instead, in Miller's observations the location of the
maximum was differently displaced from the meridian (see Figs.25 of
ref.\cite{miller} and Fig.3 of ref.\cite{shankland}). The presence
of such effect has always represented a problem for the overall
consistency of Miller's solution for the Earth's cosmic motion
\cite{miller}.

However, in this derivation, one assumes that any physical signal
should only exhibit the smooth modulations expected from the Earth's
rotation. As anticipated in Sect.2, and discussed in connections
with the Michelson-Morley and Illingworth experiments, one might be
faced with the more general scenario where the two velocity
components $v_x(t)$ and $v_y(t)$ in Eq.(\ref{amplitude10}) are not
smooth periodic functions but exhibit stochastic behaviour. In this
different perspective, combining observations of different days and
different epochs becomes more delicate and there might be
non-trivial deviations from Eq.(\ref{avfringe}). We shall therefore
conclude our analysis of Miller's experiments by recalling the
remarkable consistency of the velocity value $v\sim 350$ km/s
(obtained from the 2nd-harmonic amplitude $A^{\rm EXP}_2 \sim 0.044$
computed by the Shankland team) with those from the
Michelson-Morley, Morley-Miller and Kennedy-Illingworth experiments.
In this sense, this bulk of Miller's work will remain.

\section{Michelson-Pease-Pearson}


Let us further compare with the experiment performed by Michelson,
Pease and Pearson \cite{mpp,mpp2}. They do not report numbers so
that we can only quote from the original article \cite{mpp2} which
reports the outcome of the measurements performed in the most
refined version of the experiment: `` In the final series of
experiments, the apparatus was transferred to a well-sheltered
basement room of the Mount Wilson Laboratory. The length of the
light path was increased to eighty-five feet, and the results showed
that the precautions taken to eliminate temperature and pressure
disturbances were effective. The results gave no displacement as
great as {\it one-fiftieth } of that to be expected on the
supposition of an effect due to a motion of the solar system of
three hundred kilometers per second". On the other hand, in
ref.\cite{mpp}, after similar comments on the length of the
apparatus and on the precautions taken to eliminate the various
disturbances, one finds this other statement ``The results gave no
displacement as great as {\it one-fifteenth} of that to be expected
on the supposition of an effect due to a motion of the solar system
of three hundred kilometers per second. These results are
differences between the displacements observed at maximum and at
minimum at sidereal times, the directions corresponding to Dr.
Str\"omberg's calculations of the supposed velocity of the solar
system". In the same paper, the authors report that, according to
Str\"omberg's calculations `` a displacement of 0.017 of the
distance between fringes should have been observed at the proper
sidereal times".

Clearly, although not explicitly stated, they were assuming that
some unknown mechanism was largely reducing the fringe shifts with
respect to the naive non-relativistic value associated with a
kinematical velocity of 300 km/s. Thus one could try to conclude
that their experiment implies fringe shifts
${{|\Delta\lambda|}\over{\lambda}} \lesssim {{1}\over{15}}~
0.017\sim 0.001$. However this is not what they say (they speak of
{\it differences} between fringe displacements) and, in any case,
this interpretation does not fit with the result reported by
Shankland et al. \cite{shankland} (see their Table I). According to
these other authors, the typical observed fringe shifts observed by
Michelson, Pease and Pearson were of the order of $\pm 0.005$.

To try to understand this intricate issue, we have been looking at
another article \cite{pease} which, surprisingly, was signed by F.
G. Pease alone. Here, one discovers that, in the first stage of the
experiment, the fringe shifts had a typical magnitude of about $\pm
0.030$. Later on, however, by reducing substantially the rotation
speed of the apparatus, the observed effects became considerably
smaller.

Pease declares that, in their experiment, to test Miller's claims,
they concentrated on a purely `differential' type of measurement.
For this reason, he only reports the difference \BE \epsilon
(\theta)= \langle {{\Delta\lambda(\theta) }\over{\lambda}}
\rangle_{5.30} - \langle {{\Delta\lambda(\theta) }\over{\lambda}}
\rangle_{17.30} \EE between the mean fringe shifts $\langle
{{\Delta\lambda }\over{\lambda}} \rangle_{5.30}$, obtained after
averaging over a large set of observations performed at sidereal
time 5.30, and the mean fringe shifts $\langle {{\Delta\lambda
}\over{\lambda}} \rangle_{17.30}$ obtained after averaging in the
same period at sidereal time 17.30. The quantity $\epsilon(\theta)$
has typical magnitude of $\pm 0.004$ or smaller. However, as already
anticipated in Sect.2, by averaging observations performed at a
given sidereal time one is assuming the smooth modulations of the
signal described by Eqs.(\ref{amorse1})$, $(\ref{amorse2}).
Otherwise, one will introduce uncontrolled errors. For instance if,
consistently with Illingworth's and Miller's data, there were
substantial stochastic components in the signal, the cancelations
introduced by a naive averaging process would become stronger and
stronger by increasing the number of observations.

Therefore, from these values, nothing can be said about the
magnitude of the fringe shifts $ {{\Delta\lambda(\theta)
}\over{\lambda}}$ obtained, before any averaging procedure and before any
subtraction, in
individual measurements at various hours of the day. Pease reports a
plot of just a single observation, performed when the length of the
optical path was still 55 feet, where the even fringe shift
combinations Eq.(\ref{even}) vary approximately in the range $\pm 0.007$. This is
equivalent to fringe shifts of about $\pm 0.011$ with a length of 85
feet and could hardly be taken as indicative of the whole sample of
measurements. In this situation, one can only adopt the estimate $A_2 \sim
0.010\pm ...$ for the value of the 2nd-harmonic amplitude, for
optical path L=85 feet, whose uncertainty cannot be estimated in the
absence of information on the other individual sessions. Then, for
this configuration, where ${{L}\over{\lambda}} ~{{(30 {\rm km/s})^2
}\over{c^2}}\sim 0.45$, this is equivalent to \BE \label{vobstmpp}
       v_{\rm obs} = (4.5\pm ...)~{\rm km/s}
\EE or, by using Eq.(\ref{vobs}), to  \BE \label{vearthmpp} v =
(185\pm ...)~{\rm km/s}\EE We emphasize that Miller's extensive
observations, as reported in Fig.22 of ref.\cite{miller} (see also
our Fig.8), gave fluctuations of the observable velocity lying,
within the errors, in the range 4$-$14 km/s which has been smoothed
in our Fig.2. For this reason, even though Miller's reconstruction
of the Earth's cosmic motion is not internally consistent, a single
observation which gives  $v_{\rm obs} \sim$ 4.5 km/s does not
represent a refutation of the whole Miller experiment. This becomes
even more true by noticing that the single session selected by
Pease, within a period of several months, was chosen to represent an
example of extremely small ether-drift effect.

\vskip 10 pt

\section {Joos}

One more classical experiment, performed by Georg Joos in 1930, has
finally to be considered. For the accuracy of the measurements (data
collected at steps of 1 hour to cover the full sidereal day that
were recorded by photocamera), this experiment cannot be compared
with the other experiments (e.g. Michelson-Morley, Illingworth)
where only observations at few selected hours were performed and for
which, in view of the strong fluctuations of the azimuth, one can
just quote the average magnitude of the observed velocity. Moreover,
differently from Miller's, the amplitudes of all basic Joos'
observations can be reconstructed from the published articles
\cite{joos,joos1}. As such, this experiment deserves a more refined
analysis and will play a central role in our work.

Joos' optical system was enclosed in a hermetic housing and,
traditionally, it was always assumed that the fringe shifts were
recorded in a partial vacuum. This is supported by several elements.
For instance, when describing his device for electromagnetic fine
movements of the mirrors, Joos explicitly refers to the condition of
an evacuated apparatus, see p.393 of \cite{joos}. This aspect is
also confirmed by Miller who, quoting Joos' experiment, explicitly
refers to an ``evacuated metal housing'' in his article
\cite{miller} of 1933. This is particularly important since later
on, in 1934, Miller and Joos had a public letter exchange
\cite{joos2} and Joos did not correct Miller's statement. On the
other hand, Swenson \cite{loyd2} explicitly reports that fringe
shifts were finally recorded with optical paths placed in a helium
bath. In spite of the fact that this important aspect is never
mentioned in Joos' papers, we shall follow Swenson and assume that
during the measurements the interferometer was filled by gaseous
helium at atmospheric pressure.
\begin{figure}[ht]
\psfig{figure=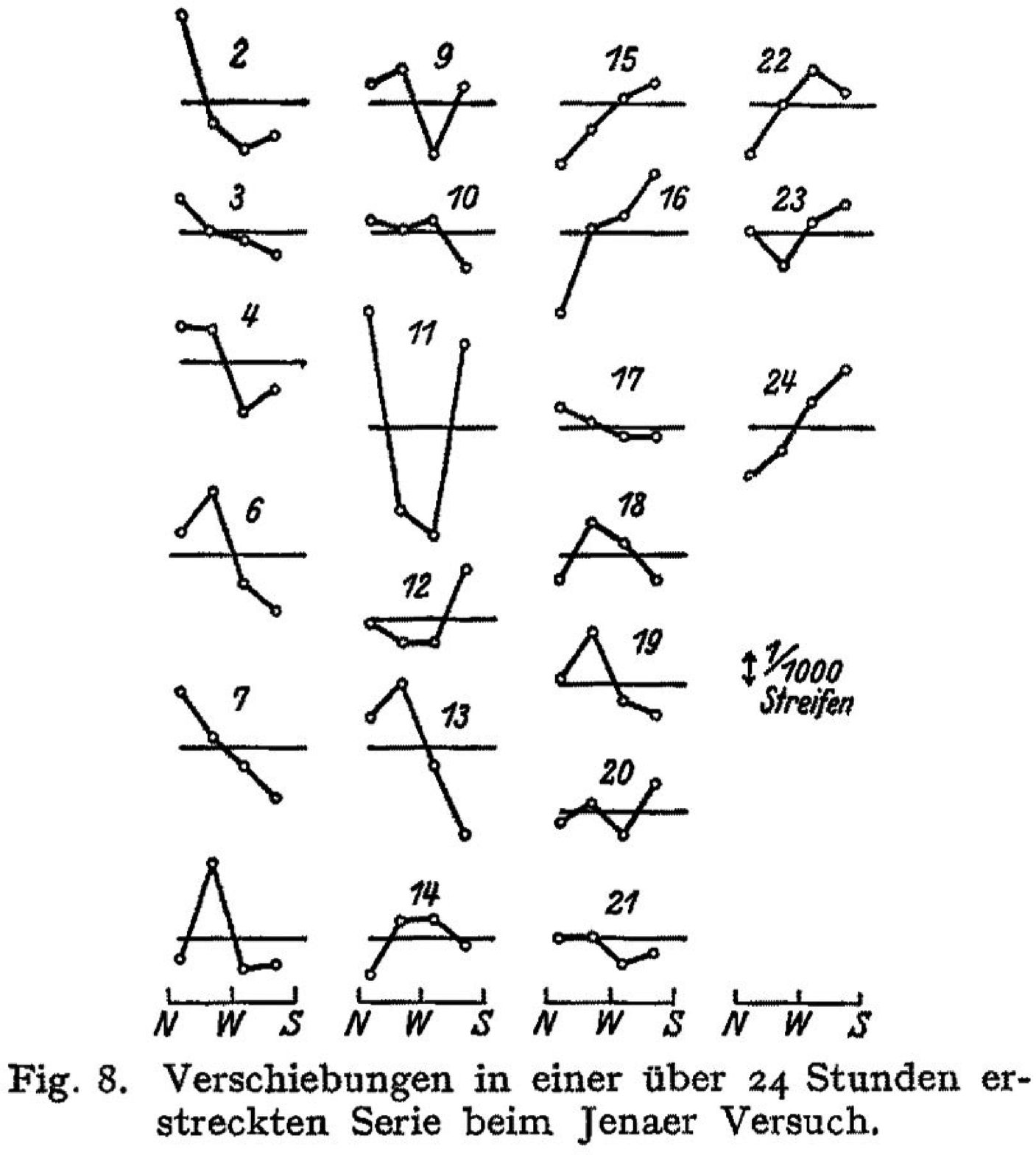,height=10 true cm,width=10 true
cm,angle=0} \caption{\it The selected set of data reported by Joos
\cite{joos,joos1}. The yardstick corresponds to 1/1000 of a
wavelength so that the experimental dots have a size of about $0.4
\cdot 10^{-3}$. This corresponds to an uncertainty $\pm 0.2 \cdot
10^{-3}$ in the extraction of the fringe shifts. }
\end{figure}

The observations were performed in Jena in 1930 starting at 2 P.M.
of May 10th and ending at 1 P.M. of May 11th. Two measurements, the
1st and the 5th, were finally deleted by Joos with the motivation
that there were spurious disturbances. The data were combined
symmetrically, in order to eliminate the presence of odd harmonics,
and the magnitude of the fringe shifts was typically of the order of
a few thousandths of a wavelength. To this end, one can look at
Fig.8 of \cite{joos1} (reported here as our Fig.9) and compare with
the shown size of 1/1000 of a wavelength. From this picture, Joos
decided to adopt 1/1000 of a wavelength as an upper limit and
deduced an observable velocity $v_{\rm obs} \lesssim 1.5$ km/s. To
derive this value, he used the fact that, for his apparatus, an
observable velocity of 30 km/s would have produced a 2nd-harmonic
amplitude of 0.375 wavelengths.

Still, since it is apparent from Fig.9 that some fringe
displacements were definitely larger than 1/1000 of a wavelength, we
have decided to extract the values of the 2nd-harmonic amplitude
$A_2$ from the 22 pictures. Differently from the values of the
azimuth, this can be done unambiguously. The point is that, due to
the camera effect, it is not clear how to fix the reference angular
values in Fig.9 for the fringe shifts. Thus, one could choose for
instance the set (k=1, 2, 3, 4) $\theta_k\equiv$($0^o$, $45^o$,
$90^o$, $135^o$) or the different set $\theta_k\equiv$($360^o$,
$315^o$, $270^o$, $225^o$). Or, by noticing that in Fig.9 there is a
small misalignment angle $\theta^*\sim 17^o$ (which actually from
\cite{joos} might instead be $22.5^o$) between the dots of Joos'
fringe shifts and the N, W, and S marks, one could also adopt other
two set of values, namely $\theta_k\equiv$($0^o +\theta^*$,
$45^o+\theta^*$, $90^o+\theta^*$, $135^o+\theta^*$) or
$\theta_k\equiv$($360^o -\theta^*$, $315^o -\theta^*$,
$270^o-\theta^*$, $225^o-\theta^*$). By fitting the fringe shifts of
Fig.9 to the 2nd-harmonic form Eq.(\ref{fringe2}), these four
options for the reference angles $\theta_k$  would give exactly the
same amplitude $A_2$ but four different choices for the azimuth,
i.e. $-\theta_0$, $-\theta_0 +\theta^*$, $\theta_0 -\theta^*$ and
$\theta_0$. This basic ambiguity should be added to the standard
uncertainty in the azimuth that, due to the 2nd-harmonic nature of
the measurements, could always be changed by adding $\pm 180$
degrees \footnote{As an example, one can consider the azimuth for
Joos' picture 20. Depending on the choice of the reference angles
$\theta_k$, one finds $\theta_0 \sim $ $329^o$, $329^o + \theta^*$,
$31^o-\theta^*$, $31^o$ or $\theta_0 \sim$ $149^o$, $149^o +
\theta^*$, $211^o- \theta^*$, $211^o$.}. Therefore, since clearly
there is only one correct choice for the angles $\theta_k$, we have
preferred not to quote theoretical uncertainties on the azimuth and
just concentrate on the amplitudes. Their values are reported in
Table 7 and in Fig.10.

\begin{figure}[ht]
\begin{center}
\psfig{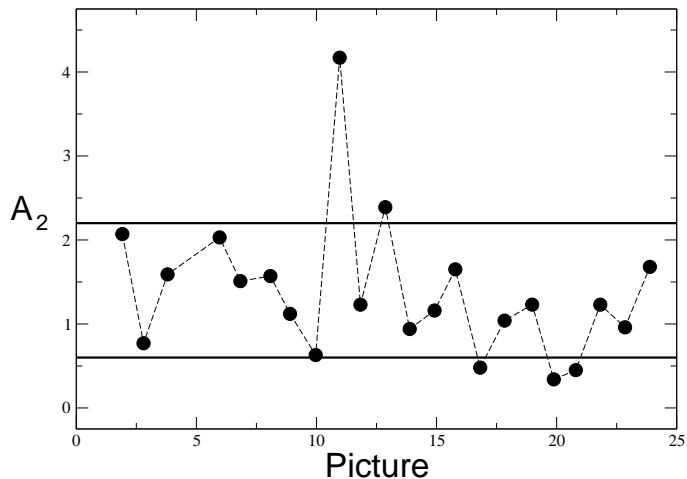}
\end{center}
\caption{ {\it Joos' 2nd-harmonic amplitudes, in units $10^{-3}$.
The vertical band between the two lines corresponds to the range
$(1.4 \pm 0.8)\cdot10^{-3}$. }} \label{Fig.10}
\end{figure}

\begin{table*}
\caption{\it The 2nd-harmonic amplitude obtained from the 22 Joos
pictures of our Fig.9. The uncertainty in the extraction of these
values is about $\pm 0.2 \cdot 10^{-3}$ (the size of the dots in
Fig.9). The mean amplitude over the 22 determinations is $\langle
A^{\rm joos}_2\rangle =1.4 \cdot 10^{-3}$. }
\begin{center}
\begin{tabular}{cl}
\hline Picture & $ A^{\rm joos}_2 [10^{-3}]~ $ \\ \hline
2  &  $2.05 $\\
3   &  $0.75$  \\
4 &  $1.60 $ \\
6  &  $2.00 $\\
7   &  $1.50$  \\
8 &  $1.55$  \\
9   &  $1.10 $ \\
10  &  $0.60$ \\
11  &  $4.15$ \\
12  &  $1.20  $   \\
13&  $2.35$  \\
14  &  $0.95$  \\
15  &  $1.15$ \\
16  &  $1.65$ \\
17  &  $0.50 $ \\
18  &  $1.05  $   \\
19&  $1.25 $ \\
20  &  $0.35 $  \\
21  &  $0.45 $ \\
22  &  $1.25 $ \\
23  &  $0.95 $\\
24  &  $1.65 $ \\
\hline
\end{tabular}
\end{center}
\end{table*}

By computing mean and variance of the individual values, we obtain
an average 2nd-harmonic amplitude \BE \langle A^{\rm joos}_2\rangle
=(1.4\pm 0.8)\cdot 10^{-3}\EE and a corresponding observable
velocity \BE \label{vobsjoos}
       v_{\rm obs} \sim 1.8^{+0.5}_{-0.6}~{\rm km/s}
\EE By correcting with the helium refractive index, Eqs.(\ref{vobs}) and (\ref{vobsjoos})
would then imply a true kinematical velocity $v \sim 217^{+66}_{-79} $ km/s.

However, this is only a first and very partial view of Joos' experiment. In
fact, we have compared Joos' amplitudes with theoretical models of
cosmic motion. To this end, one has first to transform the civil
times of Joos' measurements into sidereal times. For
the longitude 11.60 degrees of Jena, one finds that Joos'
observations correspond to a complete round in sidereal time in
which the value $\tau=0^o \equiv 360^o$ is very close to Joos' picture
20. Then, by using Eqs.(\ref{nassau1}) and (\ref{projection}), one
can use this input and compare with theoretical predictions for the
amplitude which, for the given latitude $\phi=50.94$ degrees of
Jena, depend on the right ascension $\alpha$ and the angular
declination $\gamma$. To this end, it is convenient to first
re-write the theoretical forms as \BE \label{amplitude101}
       A_2(t)\cos 2\theta_0(t)=2C(t)= {{2L ({\cal N}-1)
}\over{\lambda}}~ {{v^2_x(t)- v^2_y(t)  }
       \over{c^2}}\sim 2.6\cdot 10^{-3}~ {{v^2_x(t)- v^2_y(t)}
       \over{\rm (300~km/s)^2 }}  \EE and
\BE \label{amplitude102}       A_2(t)\sin 2\theta_0(t)=2S(t)= {{2L
({\cal N}-1) }\over{\lambda}} ~{{2v_x(t)~v_y(t)  }\over{c^2}} \sim
2.6\cdot 10^{-3} ~{{2v_x(t)~v_y(t)}\over{\rm (300~km/s)^2}} \EE
where we have used the numerical relation for Joos's experiment
${{L}\over{\lambda}} ~{{(30 {\rm km/s})^2 }\over{c^2}}\sim 0.375$
and the value of the helium refractive index. Then, by approximating
 $v_x(t) \sim \tilde v_x(t)$,  $v_y(t) \sim\tilde v_y(t)$ and
using Eq.(\ref{projection}) for the scalar combination $\tilde
v(t)\equiv \sqrt{ \tilde v^2_x(t) + \tilde v^2_y(t)}$, we have
fitted the amplitude data of Table 7 to the smooth form \BE{A}^{\rm
smooth}_2(t)= {\rm const}\cdot \sin^2 z(t)\EE where $\cos z(t)$ is
defined in Eq. (\ref{nassau1}). The results of the fit
\footnote{Actually, there is another degenerate minimum at $\alpha=
348^o\pm 30^o$ and $\gamma= 13^o\pm 14^o$ because $\sin^2 z(t)$
remains invariant under the simultaneous replacements $\alpha \to
\alpha + 180^o$ and $\gamma \to -\gamma$. However, due to the close
agreement with the CMB parameters we have concentrated on solution
(\ref{fitalpha}).} \BE \label{fitalpha} \alpha= 168^o\pm
30^o~~~~~~~~~~~~~~~~~\gamma= -13^o\pm 14^o\EE confirm that, as found
in connection with the Illingworth experiment, the Earth's motion
with respect to the CMB (which has $\alpha \sim  168^o$ and
$\gamma\sim -6^o$) could serve as a useful model to describe the
ether-drift data.

Still, in spite of the good agreement with the CMB $\alpha-$ and
$\gamma-$values obtained from the fit Eq.(\ref{fitalpha}), the
nature of the strong fluctuations in Fig.10 remains unclear. Apart
from this, there is also a sizeable discrepancy in the absolute
normalization of the amplitude. In fact, by assuming the standard
picture of smooth time modulations, the mean amplitude over all
sidereal times can trivially be obtained from the mean squared
velocity Eq.(\ref{projection})
 \BE \langle {\tilde v} ^2(t) \rangle= V^2
       \left(1- \sin^2\gamma\sin^2\phi
       - {{1}\over{2}} \cos^2\gamma\cos^2\phi \right) \EE
For the CMB and Jena, this gives $\sqrt{\langle \tilde {v}^2
\rangle} \sim 330$ km/s so that one would naively predict from
Eqs.(\ref{amplitude101}), (\ref{amplitude102}) \BE \label{atilde}
\langle A^{\rm smooth}_2(t)\rangle \sim  2.6\cdot 10^{-3}~ {{\langle
\tilde{v}^2(t)\rangle}\over{(300~ {\rm km/s})^2}}\sim 3.2\cdot
10^{-3} \EE to be compared with Joos' mean value $\langle A^{\rm
joos}_2\rangle= (1.4 \pm 0.8)\cdot 10^{-3}$. In the standard
picture, this experimental value leads to the previous estimate
$\sqrt{\langle \tilde{v}^2 \rangle} \sim 217$ km/s and {\it not} to
$\sqrt{\langle \tilde{v}^2 \rangle}  \sim 330$ km/s so that it is
necessary to change the theoretical model to try to make Joos'
experiment completely consistent with the Earth's motion with
respect to the CMB.

To try to solve this problem, and understand the origin of the
observed strong fluctuations, we have used the same model
Eqs.(\ref{vx}), (\ref{vy}) of Sect.5, to simulate stochastic
variations of the velocity field. As anticipated however, due to the
higher accuracy of the Joos experiment, we have modified the
theoretical framework. Namely, we have allowed the two random
parameters $x_n(1)$ and $x_n(2)$ to vary in the range $[-\tilde
v_x(t),\tilde v_x(t)]$ and the other two parameters $y_n(1)$ and
$y_n(2)$ to vary in the different range $[-\tilde v_y(t),\tilde
v_y(t)]$, where $\tilde v_x(t)$ and  $\tilde v_y(t)$ are defined in
Eqs.(\ref{nassau1})$-$(\ref{nassau3}). In this way, for each time
$t$, Eqs.(\ref{quadratic}) now become \BE \label{quadratict} \langle
x^2_n(i=1,2)\rangle_{\rm stat}={{{\tilde v^2_x(t)} }\over{3
~n^{2\eta}}} ~~~~~~~~~~~~~~~~~~~~~~~~\langle
y^2_n(i=1,2)\rangle_{\rm stat}={{{\tilde v^2_y(t)} }\over{3
~n^{2\eta}}}\EE It is understood that the latitude corresponds to
Joos' experiment while $V$, $\alpha$ and $\gamma$ describe the
Earth's motion with respect to the CMB. Notice that, in this model,
there will be a substantial reduction of the amplitude with respect
to its smooth prediction. To estimate the order of magnitude of the
reduction, one can perform a full statistical average (as for an
infinite number of measurements) and use Eqs.(\ref{quadratict}) in
Eqs.(\ref{amplitude101}), (\ref{amplitude102}) for our case
$\eta=1$. This gives \BE \label{reduction}\langle A_2(t)
\rangle_{\rm stat} \sim 2.6\cdot 10^{-3}~
{{\tilde{v}^2(t)}\over{(300~ {\rm km/s})^2}} ~{{1}\over{3}}
\sum^{\infty}_{n=1} {{1}\over{n^2}}= {{\pi^2}\over{18}}~A^{\rm
smooth}_2(t) \EE By also averaging over all sidereal times, for the
CMB and Jena, one would now predict a mean amplitude of about
$1.7\cdot 10^{-3}$ and not of $3.2\cdot 10^{-3}$.

\begin{table*}
\caption{\it The 2nd-harmonic amplitude obtained from a single
simulation of 22 instantaneous measurements performed at Joos' times. The
stochastic velocity components are controlled by the kinematical
parameters $(V,\alpha,\gamma)_{\rm CMB}$ as explained in the text.
The mean amplitude over the 22 determinations is $\langle A^{\rm
simul}_2\rangle =1.38 \cdot 10^{-3}$. }

\begin{center}
\begin{tabular}{cl}
\hline Picture & $ A^{\rm simul}_2 [10^{-3}]~ $
\\
\hline
2  &  $1.26$ \\
3  &  $3.50$ \\
4   &  $0.46$ \\
6 &  $0.34 $  \\
7  &  $2.71$ \\
8   &  $0.35 $  \\
9 &  $2.19 $  \\
10   &  $0.52$  \\
11  &  $5.24$ \\
12  &  $0.24$ \\
13  &  $1.19 $   \\
14&  $1.93 $  \\
15  &  $0.08$  \\
16  &  $1.52$ \\
17  &  $2.29$ \\
18  &  $0.24$ \\
19  &  $1.02$   \\
20&  $0.07 $  \\
21  &  $0.09$  \\
22  &  $2.18$ \\
23  &  $1.50$  \\
24  &  $1.52$ \\
\hline
\end{tabular}
\end{center}
\end{table*}
After having fixed all theoretical inputs, we have analyzed the
dependence of the numerical results on the remaining parameters of
the simulation, namely the number $N$ of Fourier modes (in the
available range  $N\lesssim 10^7$) and the integer number $s$ (the
`seed') which determines the random sequence. In particular, the
dependence on the latter is usually quoted as theoretical
uncertainty. For this reason, for Illingworth's experiment in Sect.5
we had produced several copies of the high-statistics simulation in
Fig.6 by quoting values for the standard deviations
Eq.(\ref{sigmas}) which take into account the observed
$s-$dependence of the results.

Here, we have started by doing something similar. However,
since it is not possible to consider at once all characteristics of a given
configuration, we have first concentrated on the simplest statistical indicator,
namely the mean amplitude $\langle A^{\rm simul}_2 \rangle$
obtained by averaging over all sidereal times.
Quite in general, this can be evaluated for a variety
of configurations which depend on the number $n$ of measurements
that one wants to simulate and the interval $\Delta t$ between two consecutive
measurements. For instance, Joos' experiment corresponds to $n=24$ (actually
$n=22$ since Joos finally deleted two observations) and $\Delta t \sim $ 3600 seconds.
At the same time,
the simulations become quite lengthy for large $N$, large $n$ and small
$\Delta t$. Therefore, we have first performed a scan of $s-$values for $N=10^4$ and then
studied a few $s$ by increasing $N$. To give an idea of the spread of
the central values, due to changes of the pair $(N,s)$,
we report below the approximate results of this analysis
for some choices of the pair $(n,\Delta t)$
\BE \langle A^{\rm simul}_2(n=24,\Delta t=3600~s) \rangle \sim (1.7 \pm 0.8)\cdot 10^{-3}\EE
\BE \langle A^{\rm simul}_2(n=1440,\Delta t=60~s) \rangle \sim (1.7 \pm 0.3)\cdot 10^{-3}\EE
\BE \langle A^{\rm simul}_2(n=240,\Delta t=3600~s) \rangle \sim (1.8 \pm 0.5)\cdot 10^{-3}\EE
As it might be expected, the average $\langle A^{\rm simul}_2 \rangle$
becomes more stable
by increasing the number of observations. Concerning the individual values
$A^{\rm simul}_2(t_i)$, with $i=1,..,n$, they have a large spread,
about $(1\div 4) \cdot 10^{-3}$. This is in agreement with the
 `fat-tailed' distributions of instantaneous values expected
in turbulent flows \cite{sreenivasan,beck} (compare with Fig. 6 in
Sect.5). However this other spread can be reduced by starting to
average the data in some interval of time $t_0$. In this case,  the
spread of the resulting average values $\langle A^{\rm
simul}_2(t_i)\rangle_{t_0}$ decreases as ${{1}\over{\sqrt{t_0}}}$.
We emphasize that, by performing extensive simulations, there are
occasionally very large spikes of the amplitude at some sidereal
times, of the order $(10\div 20)\cdot 10^{-3}$. The effect of these
spikes gets smoothed when averaging over many configurations but
their presence is characteristic of a stochastic-ether model. With a
standard attitude, where the ether drift is only expected to exhibit
smooth time modulations, the observation of such effects would
naturally be interpreted as a spurious disturbance (Joos' omitted
observations 1 and 5?).

\begin{figure}[ht]
\begin{center}
\psfig{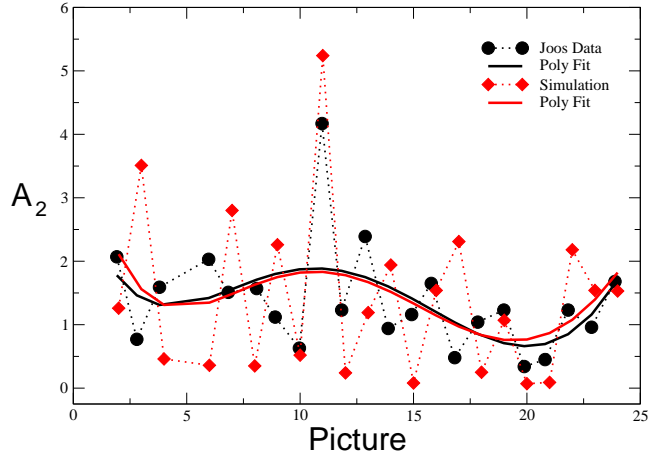}
\end{center}
\caption{ {\it Joos' experimental amplitudes in Table 7 are compared
with the single simulation of 22 measurements for fixed $(N,s)$ in
Table 8. By changing the pair $(N,s)$, the typical variation of each
simulated entry is $(1\div 4)\cdot10^{-3}$ depending on the sidereal
time. We also show two 5th-order polynomial fits to the two
different sets of values.  }} \label{Fig.11}
\end{figure}

After this preliminary study, we have then concentrated on the real
goal of our simulation, i.e. to compare with the {\it single} Joos
configuration of 22 entries in Table 7. To this end, one could first
try to look for the `best seed', or subset of seeds, which can
minimize the difference between the generated configurations and
Joos' data. This standard task, usually accomplished by minimizing a
chi-square, is difficult to implement here. In fact, it is
problematic to construct a function $\chi^2(s)$ and look for its
minima because a seed $s$ and the closest seeds $s \pm 1$ give often
vastly different configurations and chi-square. For this reason, we
have followed an empirical procedure by forming a grid and selecting
a set of seeds whose mean amplitude (for $n=24$ and $\Delta t=$ 3600
s) gets close to Joos's mean amplitude $\langle A^{\rm
joos}_2\rangle =1.4\cdot 10^{-3}$ for a large number $N$ of Fourier
modes. One of such seeds gave a sequence  $\langle A^{\rm
simul}_2\rangle=$1.66, 1.40, 1.08, 1.21 and 1.38 (in units
$10^{-3}$), for $N= 10^3$, $10^4$, $10^5$, $10^6$ and $5\cdot 10^6$
respectively, and the configuration with $N=5\cdot 10^6$ was finally
chosen to give an idea of the agreement one can achieve between data
and a single numerical simulation for fixed $(N,s)$. The simulated
values are reported in Table 8 and a graphical comparison with Joos'
data is shown in Fig. 11. We emphasize that one should not compare
each individual entry with the corresponding data since, by changing
$(N,s)$, the simulated instantaneous values vary typically of about
$(1\div 4) \cdot 10^{-3}$ depending on the sidereal time. Instead,
one should compare the overall trend of data and simulation. To this
end, we show two 5th-order polynomial fits to the two different sets
of values.

A more conventional comparison with the data consists in quoting for
the various 22 entries simulated average values and uncertainties.
\begin{table*}
\caption{\it The 2nd-harmonic amplitudes obtained by simulating the
averaging process over 10 hypothetical measurements performed, at
each Joos' time, on 10 consecutive days. The stochastic velocity
components are controlled by the kinematical parameters
$(V,\alpha,\gamma)_{\rm CMB}$ as explained in the text. The effect
of varying the pair $(N,s)$ has been approximated into a central
value and a symmetric error. The mean amplitude over the 22
determinations is $\langle A^{\rm simul}_2\rangle =1.8 \cdot
10^{-3}$.}

\begin{center}
\begin{tabular}{cl}
\hline Picture~~ & $ A^{\rm simul}_2 [10^{-3}]~~~~$
\\
\hline
2  &  $2.5\pm 1.0$ \\
3  &  $1.80\pm 0.85$\\
4   &  $1.95\pm 0.85$  \\
6 &  $1.90\pm 0.85 $ \\
7  &  $1.65\pm 0.90$ \\
8   &  $2.1\pm 1.0 $ \\
9 &  $2.0\pm 1.0 $  \\
10   &  $2.2\pm 1.2$  \\
11  &  $2.4\pm 1.4$ \\
12  &  $2.7\pm 1.6$ \\
13  &  $2.3\pm 1.5 $   \\
14&  $2.4 \pm 1.4 $  \\
15  &  $1.85 \pm 0.85$  \\
16  &  $1.70 \pm 0.75$ \\
17  &  $1.20 \pm 0.75$ \\
18  &  $1.20 \pm 0.70$ \\
19  &  $1.15 \pm 0.70$   \\
20&  $1.05 \pm 0.70 $  \\
21  &  $1.25 \pm 0.60$  \\
22  &  $1.55 \pm 0.60$ \\
23  &  $1.60\pm 0.80$  \\
24  &  $1.7 \pm 1.0$ \\
\hline
\end{tabular}
\end{center}
\end{table*}
To this end, we have considered the mean amplitudes  $\langle A^{\rm
simul}_2(t_i)\rangle$ defined by averaging, for each Joos' time
$t_i$, over 10 hypothetical measurements performed on 10 consecutive
days. For each $t_i$, the observed effect of varying  $(N,s)$ has
been summarized into a central value and a symmetric error. The
values are reported in Table 9 and the comparison with Joos'
amplitudes is shown in Fig.12.
\begin{figure}[ht]
\begin{center}
\psfig{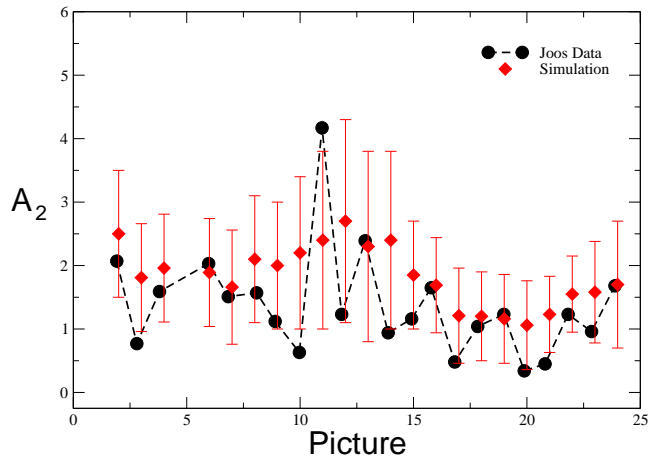}
\end{center}
\caption{ {\it Joos' experimental amplitudes in Table 7 are compared
with our simulation in Table 9.  }} \label{Fig.12}
\end{figure}

The spread of the various entries is larger at the sidereal times
where the projection at Jena of the cosmic Earth's velocity becomes
larger. The tendency of Joos' data to lie in the lower part of the
theoretical predictions in Table 9 mostly depends on our use of
symmetric errors. In fact, by comparing in some case with the
histograms of the basic generated configurations $A^{\rm
simul}_2(t_i)$, we have seen that our sampling method of $\langle
A^{\rm simul}_2(t_i)\rangle$, based on a grid of $(N,s)$ values,
typically underestimates the weight of the low-amplitude region in a
prediction at the $70 \%$ C.L. . This can also be checked by
considering the single simulation of Table 8 and counting the
sizeable fraction of amplitudes $A^{\rm simul}_2(t_i) \lesssim
0.5\cdot 10^{-3}$. For this reason, one could improve the evaluation
of the probability content. However, in view of the good agreement
already found in Fig.12 ($\chi^2=13/22$), we did not attempt to
carry out this more refined analysis.

In conclusion, after the first indication obtained from the fit
Eq.(\ref{fitalpha}), we believe that the link between Joos' data and
the Earth's motion with respect to the CMB gets reinforced by our
simulations. In fact, by inspection of Figs.11 and 12, the values of
the amplitudes and the characteristic scatter of the data are
correctly reproduced. In principle, there could be space for further
refinements by taking into account the Earth's orbital motion in the
input values for $V$, $\alpha$ and $\gamma$.

From this agreement with the data, we then deduce that the previous
value for the kinematical velocity $v \sim 217^{+66}_{-79} $ km/s,
obtained by simply correcting with the helium refractive index the
average observable velocity (\ref{vobsjoos}), has to be considerably
increased if one allows for stochastic variations of the velocity
field. In fact, the magnitude of the fluctuations in $v_x$ and $v_y$
is controlled by the same scalar parameter  $\tilde v(t)\equiv \sqrt{
\tilde v^2_x(t) + \tilde v^2_y(t)}$ of Eq.(\ref{projection}).
In view of the good agreement between data and our numerical
simulations, we conclude that Joos' data are consistent with a range
of kinematical velocity $v=330^{+40}_{-70}$ km/s which corresponds
to Eq.(\ref{projection}) for $\phi=50.94^o$, $V=370$ km/s,
$\alpha=168^o$ and $\gamma=-6^o$.

\section{Summary and conclusions}

The condensation of elementary quanta and their macroscopic
occupation of the same quantum state is the essential ingredient of
the degenerate vacuum of present-day elementary particle physics. In
this description, one introduces implicitly a reference frame
$\Sigma$, where the condensing quanta have ${\bf k}=0$, which
characterizes the physically realized form of relativity and could
play the role of preferred reference frame in a modern
re-formulation of the Lorentzian approach. To this end, we have
given in the Introduction some general theoretical arguments related
to the problematic notions of a non-zero vacuum energy and of an
exact Lorentz-invariant vacuum state. These arguments suggest the
possibility of a tiny vacuum energy-momentum flux, associated with
an Earth's absolute velocity $v$, which could affect {\it
differently} the various forms of matter. Namely, it could produce
small convective currents in a loosely bound system such as a gas or
dissipate mainly by heat conduction with no appreciable particle
flow in strongly bound systems as liquid or solid transparent media.
In the former case, by introducing the refractive index $ {\cal N}$
of the gas, convective currents of the gas molecules would produce a
small anisotropy, proportional to $({\cal N}-1) (v/c)^2$, of the
two-way velocity of light in agreement with the general structure
Eq.(\ref{legendre}) or with its particular limit Eq.(\ref{twoway1}).
Notice that this tiny anisotropy refers to the system $S'$ where the
container of the gas is at rest. In this sense, contrary to standard
Special Relativity, $S'$ might not define a true frame of rest. This
conceptual possibility can be objectively tested with a new series
of dedicated ether-drift experiments where two orthogonal optical
resonators are filled with various gaseous media by measuring the
fractional frequency shift $\Delta\nu/\nu$ between the two
resonators. By assuming the typical value $v\sim 300$ km/s of most
cosmic motions, one expects frequency shifts $\lesssim 10^{-10}$ for
gaseous helium and  $\lesssim 10^{-9}$ for air, which are well
within the present technology.

Given the heuristic nature of our approach, and to further motivate
the new series of dedicated experiments, we have tried to get a
first consistency check. In fact, by adopting Eq.(\ref{twoway1}),
the frequency shift between the optical resonators is governed by
the same classical formula for the {\it fringe shifts } in the old
ether-drift experiments with the only replacement \BE v^2 \to
2({\cal N}-1)v^2 \equiv v^2_{\rm obs}\EE In this way, where one
re-obtains the same classical formulas (with the only replacement $v
\to v_{\rm obs}$), testing the present scheme is very simple: one
should just check the consistency of the true kinematical $v'$s
obtained in different experiments.

In the old times, experiments were performed with interferometers
where light was propagating in gaseous media, air or helium at
atmospheric pressure, where $({\cal N}-1)$ is a very small number.
In this regime, the theoretical fringe shifts expected on the basis
of Eqs.(\ref{legendre}) and (\ref{twoway1}) are much smaller than
the classical prediction $(v/c)^2$. Another important aspect of
these classical experiments is that one was always expecting smooth
sinusoidal modulations of the data due to the Earth's rotation, see
Eqs. (\ref{fringe}), (\ref{amorse1}) and (\ref{amorse2}). As
emphasized in Sect.2, we now understand the logical gap missed so
far. The relation between the macroscopic Earth's motions (daily
rotation, annual orbital revolution,...) and the ether-drift
experiments depends on the physical nature of the vacuum. Assuming
Eqs.(\ref{amorse1}) and (\ref{amorse2}), to describe the effect of
the Earth's daily rotation, amounts to considering the vacuum as
some kind of fluid in a state of regular, laminar motion for which
global and local properties of the flow coincide. Instead, several
theoretical arguments (see e.g.
refs.\cite{wheeler1,migdal,ng2,ng3,troshkin,puthoff,tsankov,chaos,plafluid})
suggest that the physical vacuum might behave as a stochastic medium
similar to a turbulent fluid where large-scale and small-scale
motions are only {\it indirectly} related. In this case, there might
be non-trivial implications. For instance, due to the irregular
behaviour of turbulent flows, vectorial observables collected at the
same sidereal time might average to zero. However, this does not
mean that there is no ether-drift. More generally, the relevant
Earth's motion with respect to $\Sigma$ might well correspond to
that indicated by the anisotropy of the CMB,  but it becomes non
trivial to reconstruct the kinematical parameters from microscopic
measurements of the velocity of light in a laboratory.  These
arguments make more and more plausible that a genuine physical
phenomenon, much smaller than expected and characterized by
stochastic variations, might have been erroneously interpreted as an
instrumental artifact thus leading to the standard `null
interpretation' of the experiments reported in all textbooks.

Now, our analysis of Sects.3$-$8 shows that this traditional
interpretation is far from obvious. In fact, by using
Eqs.(\ref{twoway1}), (\ref{fringe0}) and (\ref{vobs}), the small
residuals point to an average velocity of about 300 km/s, as in most
cosmic motions. In this alternative interpretation, the indications
of the various experiments are summarized in our Table 10
\footnote{Other determinations of less accuracy could also be
included, as for the 1881 Michelson experiment in Potsdam
\cite{potsdam} or Tomaschek's starlight experiment \cite{tomaschek}
or the Piccard and Stahel experiment which was first performed in a
ballon \cite{piccard} and later \cite{piccard2} on the summit of Mt.
Rigi in Switzerland. These results were summarized in Table I of
ref.\cite{shankland} and by Miller \cite{miller}. In the 1881
Potsdam experiment the fringe shifts were in the range $0.002\div
0.007$ to be compared with an expected 2nd-harmonic of 0.02 for 30
km/s. This means observable velocities $(9\div 18)$ km/s which are
comparable and even larger than those of the 1887 experiment. In
Tomaschek's starlight experiment, fringe shifts were about 15 times
smaller than those classically expected for an Earth's velocity of
30 km/s. This gives $v_{\rm obs}\lesssim$ 7.7 km/s or $v\lesssim$
320 km/s. From Piccard and Stahel, in the most refined version of
Mt. Rigi, one gets an observable velocity $v_{\rm obs}\lesssim$ 1.5
km/s. Since their optical paths were enclosed in an evacuated
enclosure, this very low value can easily be reconciled with the
typical kinematical velocity $v\sim$ 300 km/s of the most accurate
experiments in Table 10. }.
\begin{table*}
\caption{\it The average velocity observed (or the limits placed) by
the classical ether-drift experiments in the alternative
interpretation of Eqs.(\ref{twoway1}), (\ref{fringe0}),
(\ref{vobs}).}
\begin{center}
\begin{tabular}{clll}
\hline Experiment &gas in the interferometer
&~~~~$v_{\rm obs}({\rm km/s})$ & ~~~~$v$({\rm km/s})\\
\hline
Michelson-Morley(1887)   & ~~~~~~~~~~ air & ~~~~ $8.4^{+1.5}_{-1.7}$&~~~$349^{+62}_{-70}$ \\
Morley-Miller(1902-1905)  & ~ ~~~~~~~~~air& ~~~~ $8.5\pm 1.5$ & ~~ $ 353 \pm 62$ \\
Kennedy(1926)  & ~~~~~~~~~~~helium &~~~~~~$<5 $  &  ~~~~$<600 $\\
Illingworth(1927) & ~~~~~~~~~~~helium & ~~~~~$3.1 \pm 1.0$  &~~~$370 \pm 120$ \\
Miller(1925-1926) & ~~~~~~~~~~~air & ~~~~~$8.4^{+1.9}_{-2.5}$ &~~~$349^{+79}_{-104}$ \\
Michelson-Pease-Pearson(1929)& ~~~~~~~~~~~air &~~~~~$4.5\pm... $  &  ~~~$ 185 \pm ... $\\
Joos(1930)  &~~~~~~~~~~~helium&~~~~$\ 1.8^{+0.5}_{-0.6} $  & ~~ $330^{+40}_{-70} $\\
\hline
\end{tabular}
\end{center}
\end{table*}
As a summary of our work, we emphasize the following points:

~~~i) an analysis of the individual sessions of the original
Michelson-Morley experiment, in agreement with Hicks \cite{hicks}
and Miller \cite{miller} (see our Figs. 1 and 2), gives no
justification to its standard null interpretation. As discussed in
Sect.3, this type of analysis is more reliable. In fact, averaging
directly the fringe displacements of different sessions requires two
additional assumptions, on the nature of the ether-drift as a smooth
periodic effect and on the absence of systematic errors introduced
by the re-adjustment of the mirrors on consecutive days, that in the
end may turn out to be wrong.

~~~ii) one gets consistent indications from the Michelson-Morley,
Morley-Miller, Miller and Illingworth-Kennedy experiments. In view
of this consistency, an interpretation of Miller's observations in
terms of a temperature gradient \cite{shankland} is only acceptable
provided this gradient represents a non-local effect as in our
picture where the ether-drift is the consequence of a fundamental
vacuum energy-momentum flow. We have also produced numerical
simulations of the Illingworth experiment in a simple statistically
isotropic and homogeneous turbulent-ether model. This represents a
zeroth-order approximation and is useful to illustrate basic
phenomenological features associated with the picture of the vacuum
as an underlying stochastic medium. In this scheme, Illingworth's
data are consistent with fluctuations of the velocity field whose
absolute scale is controlled by $\tilde v=V_{\rm CMB} \sim $370
km/s, the velocity of the Earth's motion with respect to the CMB.

~~~iii) on the other hand, there is some discrepancy with the
experiment performed by Michelson, Pease and Pearson (MPP). However,
as discussed in Sect.7, the uncertainty cannot be easily estimated
since only a single basic MPP observation is explicitly reported in
the literature. Therefore, since Miller's extensive observations
(see Fig.22 of ref.\cite{miller} and our Fig.8), within their
errors, gave fluctuations of the observable velocity in the wide
range 4$-$14 km/s, a single observation giving $v_{\rm obs} \sim$
4.5 km/s cannot be interpreted as a refutation. This becomes even
more true by noticing that the single session selected by Pease,
within a period of several months, was chosen to represent an
example of extremely small ether-drift effect.

~~~iv) some more details are needed to account for the Joos
observations. This experiment is particularly important since the
data were collected at steps of 1 hour to cover the full sidereal
day and were recorded by photocamera. For this reason, Joos'
experiment is not comparable with other experiments (e.g.
Michelson-Morley, Illingworth) where only observations at few
selected hours were performed and for which, in view of the strong
fluctuations of the azimuth, one can just quote the average
magnitude of the observed velocity. Moreover, differently from
Miller's, the amplitudes of all Joos's observations can be
reconstructed from the published articles \cite{joos,joos1}. For
these reasons, this experiment has deserved a more refined analysis
and is central for our work. As discussed in Sect.8, due to
uncertainties in the original data analysis, the standard 1.5 km/s
velocity value quoted for this experiment should be understood as an
order of magnitude estimate and not as a true upper limit. Instead,
our reported observable velocity $v_{\rm obs} \sim 1.8^{+0.5}_{-
0.6} $ km/s has been obtained from a direct analysis of Joos' fringe
shifts. From this value, to deduce a kinematical velocity, one still
needs the refractive index. The traditional view, motivated by
Miller's review article \cite{miller} and Joos's own statements in
ref.\cite{joos}, is that the experiment was performed in an
evacuated housing. In these conditions, it would be easy to
reconcile a large kinematical velocity $v \sim 350$ km/s with the
very small values of the observable velocity. On the other hand,
Swenson \cite{loyd2} explicitly reports that fringe shifts were
finally recorded with optical paths placed in a helium bath. Since
Joos' papers do not provide any definite clue on this aspect, we
have decided to follow Swenson's indications. In this case, by
simply correcting with the helium refractive index the result
$v_{\rm obs} \sim 1.8^{+0.5}_{-0.6} $ km/s, one would get a
kinematical velocity $v \sim 217^{+66}_{-79} $ km/s. However, as
discussed in detail in Sect.8, this is only a first partial view of
Joos' experiment. In fact, by fitting the experimental amplitudes in
Table 7 to various forms of cosmic motion (see Eq.(\ref{fitalpha}))
we have obtained angular parameters which are very close to those
that describe the CMB anisotropy (right ascension $\alpha_{\rm CMB}
\sim 168^o$ and angular declination $\gamma_{\rm CMB} \sim -6^o$).
Still, to get a complete agreement, one should explain the absolute
normalization of the amplitudes and the strong fluctuations of the
data. Thus we have improved our analysis by performing various
numerical simulations where the velocity components in the plane of
the interferometer $v_x(t)$ and $v_y(t)$, which determine the basic
functions $C(t)$ and $S(t)$ through Eqs.(\ref{amplitude10}) and the
fringe shifts through Eq.(\ref{fringe}), are not smooth functions
but are represented as turbulent fluctuations. Their Fourier
components in Eqs.(\ref{vx}) and (\ref{vy}) now vary within
time-dependent ranges Eqs.(\ref{nassau2})$-$(\ref{nassau3}),
$[-\tilde v_x(t),\tilde v_x(t)]$ and $[-\tilde v_y(t),\tilde
v_y(t)]$ respectively, controlled by the macroscopic parameters
$(V,\alpha,\gamma)_{\rm CMB}$. Taking into account these stochastic
fluctuations of the velocity field tends to increase the fitted
average Earth's velocity, see Eq.(\ref{reduction}), and can
reproduce correctly Joos' 2nd-harmonic amplitudes and the
characteristic scatter of the data, see Figs. 11 and 12. In view of
this consistency, we conclude that  the range $v= 330^{+40}_{-70}$
km/s (corresponding to Eq.(\ref{projection}) for CMB and Joos'
laboratory) is actually the most appropriate one.

The more refined analysis adopted for the Joos experiment provides
an explicit example of the previously mentioned non-trivial
ingredients that might be required to reconstruct the global Earth's
motion from microscopic measurements performed in a laboratory. For
this reason, the results reported in Table 10, besides providing an
impressive evidence for a light anisotropy proportional to $({\cal
N}-1) (v/c)^2$, with the realistic velocity values $v \sim$ 300 km/s
of most cosmic motions, could also represent the first experimental
indication for the Earth's motion with respect to the CMB. Due to
the importance of this result, and to provide the reader with all
elements of the analysis, we present in a second Appendix a brief
numerical simulation of one noon session of the Michelson-Morley
experiment. This has been performed in the same framework adopted
for the Joos experiment where the velocity components $v_x(t)$ and
$v_y(t)$ in the plane of the interferometer are represented as
turbulent fluctuations varying within time-dependent ranges
controlled by the macroscopic parameters $(V,\alpha,\gamma)_{\rm
CMB}$. We postpone to a future publication the non-trivial task of
performing a complete numerical simulation of the whole
Michelson-Morley experiment and of the Illingworth experiment (with
its 32 sessions and the associated sets of 20 rotations for each
session) where we'll also compare the various theoretical schemes
mentioned in Sect.5 to handle the stochastic components of the
velocity field.

We emphasize that the simulation reported in our second Appendix
corresponds to a single configuration whereas taking into account
more and more configurations is essential to properly estimate
theoretical uncertainties (as for the Joos experiment with the
results in our Table 9 and Fig.12). Nevertheless, even this very
small sample can provide interesting clues on the real data. For
instance, the strong scatter of the fringe shifts at the same
$\theta-$values in consecutive rotations and the good agreement with
the experimental azimuths obtained by accepting Hicks'
interpretation of the observations of July 8th (see Sect.3). In this
sense, this brief numerical analysis reinforces the picture of the
classical experiments emerging from our Table 10. According to the
usual view, the theoretical predictions, with a very low velocity
$v\sim$ 30 km/s, were much larger than the observed values. Instead,
in a modern view of the vacuum as a stochastic medium, theoretical
predictions, for the realistic velocities $v\sim$ 300 km/s of most
cosmic motions, are now well compatible and, sometimes, even smaller
than the actual outcome of the observations. This latter case simply
means that the experimental data were also affected by spurious
effects such as deformations induced by the rotation of the
apparatus or local thermal conditions. This gives a strong
motivation to repeat these crucial measurements with today's much
greater accuracy.

To this end, let us now briefly consider the modern ether-drift
experiments. As anticipated, in the modern experiments, the test of
the isotropy of the velocity of light consists in measuring the
relative frequency shift $\Delta \nu$ of two orthogonal optical
resonators \cite{applied,lammer}. Here, the analog of
Eq.(\ref{fringe0}), for a hypothetical physical part of the
frequency shift (after subtraction of all spurious effects), is \BE
\label{prediction}
              {{\Delta \nu^{\rm phys} (\theta) }\over{\nu_0}}=
{{\bar{c}_\gamma(\pi/2 +\theta)- \bar{c}_\gamma (\theta)} \over{c}}=
{\cal B}_{\rm medium} {{v^2 }\over{c^2}} \cos2(\theta-\theta_0) \EE
where $\theta_0$ is the direction of the ether-drift. This can be
interpreted within Eq.(\ref{twoway}) where \BE \label{rmsbb}
       |{\cal B}_{\rm medium}|\sim {\cal N}_{\rm medium}-1
\EE ${\cal N}_{\rm medium}$ being the refractive index of the
gaseous medium filling the optical resonators. Testing this
prediction, requires replacing the high vacuum usually adopted
within the optical resonators with a gaseous medium and studying the
substantially larger frequency shift introduced with respect to the
vacuum experiments.

As a rough check, a comparison was made \cite{pla,cimento} with the
results obtained by Jaseja et. al \cite{jaseja} in 1963 when looking
at the frequency shift of two orthogonal He-Ne masers placed on a
rotating platform. To this end, one has to preliminarily subtract  a
large systematic effect that was present in the data and interpreted
by the authors as probably due to magnetostriction in the Invar
spacers induced by the Earth's magnetic field. As suggested by the
same authors, this spurious effect, which was only affecting the
normalization of the experimental $\Delta \nu$, can be subtracted by
looking at the variations of the data. As discussed in
refs.\cite{pla,cimento}, the measured variations of a few kHz are
roughly consistent with the refractive index ${\cal N}_{\rm
He-Ne}\sim 1.00004$ and the typical variations of an Earth's
velocity as in Eq.(\ref{vearth}).

More recent experiments \cite{brillet}$-$\cite{schillernew} have
always been performed in a very high vacuum where, as emphasized in
the Introduction, the differences between Special Relativity and the
Lorentzian interpretation are at the limit of visibility. In fact,
in a perfect vacuum by definition ${\cal N}_{\rm vacuum}=1$ so that
${\cal B}_{\rm vacuum}$ will vanish \footnote{Throughout this paper
we have assumed the limit of a zero light anisotropy for experiments
performed in vacuum. However, as discussed at the end of Appendix I,
one could also consider the more general scenario where a metric of
the form \cite{ma1} $ g^{\mu\nu}= \eta^{\mu\nu} + \Delta^{\mu\nu}$
is introduced from the very beginning. In this case, the vacuum
behaves as a medium and light can spread with different velocities
for different directions. As an example, by adopting various
parameterizations for $\Delta^{\mu\nu}$, the non-zero one-way light
anisotropy reported by the GRAAL experiment \cite{graal} requires
typical values of the matrix elements $|\Delta^{\mu\nu}|=
10^{-13}\div 10^{-14}$ \cite{ma2}.In any case, as anticipated in
Sect.2, these genuine vacuum effects are much smaller than those
discussed in the present paper in connection with a gas refractive
index.}. Thus one should switch to the new generation of dedicated
ether-drift experiments in gaseous systems. Our conclusion is that
these new experiments should just confirm Joos' remarkable
observations of eighty years ago. \vskip 10 pt \centerline{\bf
Acknowledgments}

\noindent We thank Angelo Pagano for useful discussions.

\vfill\eject

\vfill\eject \centerline{\bf{\LARGE Appendix I}} \vskip 15 pt

To derive Eq.(\ref{twoway1}), one should start from
Eq.(\ref{legendre}) which describes light propagation in a gaseous
system in the presence of convective currents of the gas molecules
originating from a fundamental vacuum energy-momentum flow. Due to
these convective currents, isotropy of the velocity of light would
only hold if the solid container of the gas and the observer were
both at rest in  the particular reference frame $\Sigma$ where the
macroscopic condensation of quanta correspond to the state ${\bf
k}=0$. This introduces obvious differences with respect to the
standard analysis. For instance, let us compare with Jauch and
Watson \cite{jauch} who worked out the quantization of the
electromagnetic field in a moving medium of refractive index ${\cal
N}$. They noticed that the procedure introduces unavoidably a
preferred frame, the one where the photon energy does not depend on
the direction of propagation. Their conclusion, that this frame is
``usually taken as the system for which the medium is at rest",
reflects however the point of view of Special Relativity with {\it
no} preferred frame. Instead, one could consider a different
scenario where, at least in some limit, the angle-independence of
the photon energy might only hold for some special frame $\Sigma$.

To discuss this different case, let us first consider a dielectric
medium of refractive index ${\cal N}$ whose container is at rest in
$\Sigma$. For an observer at rest in this reference frame, light
propagation within the medium is isotropic and described by \BE
\label{masshell0}
       \pi_\mu \pi_\nu \gamma^{\mu\nu}= 0
\EE where \BE \label{isotropy} \gamma^{\mu\nu}={\rm diag}({\cal
N}^2,-1,-1,-1) \EE and $\pi_\mu$  denotes the light 4-momentum
vector for the $\Sigma$ observer. Let us now consider that the
container of the medium is moving with some velocity ${\bf V}$ with
respect to $\Sigma$ and {\it is at rest in some other frame $S'$}.
By analogy, light propagation within the medium for the observer in
$S'$ will be described by \BE \label{masshell}
       p_\mu p_\nu g^{\mu\nu}= 0 ,
\EE where $p_\mu\equiv (E/c,{\bf p})$ and $g^{\mu\nu}$ denote
respectively the light 4-momentum and the effective metric for $S'$.
On this basis, by introducing the $S'$ dimensionless velocity
4-vector $u^\mu\equiv(u^0,{\bf V}/c)$ (with $u_\mu u^\mu=1$), one
can define a transformation matrix
$A^{\mu}{_\nu}=A^{\mu}{_\nu}(u_\mu,{\cal N}) $ and express \BE
\label{transform0}
g^{\mu\nu}=A^{\mu}{_\sigma}A^{\nu}{_\rho}\gamma^{\sigma\rho} \EE In
this context, requiring the consistency of vacuum condensation with
Special Relativity corresponds to place all reference frames on the
same footing and assume $g^{\mu\nu}=\gamma^{\mu\nu}$ or
\BE\label{special}  A^{\mu}{_\nu}(u_\mu,{\cal N})=\delta^{\mu}_{\nu}
\EE This identification is independent of the physical nature of the
medium, being valid for gaseous systems as well as for liquid or
solid transparent media. In this sense Special Relativity, by
construction, cannot describe light propagation in the presence of a
vacuum energy-momentum flow which could affect the various forms of
matter differently.

Instead, consistently with the basic ambiguity in the interpretation
of relativity mentioned in the Introduction, and with Lorentz' point
of view \cite{electron} (``it seems natural not to assume at
starting that it can never make any difference whether a body moves
through the ether or not''), one could adopt different choices
without pretending to determine {\it a priori} the outcome of any
ether-drift experiment. Thus, by noticing that we have at our
disposal two matrices, namely $\delta^{\mu}_{\nu}$ and the Lorentz
transformation matrix $\Lambda^{\mu}{_\nu}$ associated with ${\bf
V}$, one could still maintain Eq.(\ref{special}) for strongly bound
systems, such as solid or liquid transparent media, where the small
energy flux generated by the motion with respect to $\Sigma$ should
mainly dissipate by heat conduction with no appreciable particle
flow and no light anisotropy in the rest frame of the medium. One
could instead identify \BE \label{continue}
A^{\mu}{_\nu}(u_\mu,{\cal N}=1)=\Lambda^{\mu}_{\nu} \EE to solve
non-trivially the equation $g^{\mu\nu}=\gamma^{\mu\nu}$ when ${\cal
N}=1$, i.e. when light propagates in vacuum and $\gamma^{\mu\nu}$
reduces to the Minkowski tensor $\eta^{\mu\nu}$. But then, by
continuity, it is conceivable that Eq.(\ref{continue}), up to
higher-order terms, can also describe the case ${\cal N}=1+
\epsilon$ of gaseous media. This choice provides a simple
interpretative model and a particular form of the more general
structure Eq.(\ref{legendre}) which corresponds to the
Robertson-Mansouri-Sexl (RMS) \cite{robertson,mansouri}
parametrization for the two-way velocity of light. Moreover, when
comparing with experiments with optical resonators, the resulting
frequency shift is governed by the same classical formula for the
fringe shifts in the old ether-drift experiments with the only
replacement $V^2 \to 2({\cal N}-1)V^2$. To see this, let us compute
$g^{\mu\nu}$ through the relation \BE \label{transform}
g^{\mu\nu}=\Lambda^{\mu}{_\sigma}\Lambda^{\nu}{_\rho}\gamma^{\sigma\rho}
\EE with $\gamma^{\mu\nu}$ as in Eq.(\ref{isotropy}). This gives the
effective metric for $S'$ \BE \label{effective} g^{\mu\nu}=
\eta^{\mu\nu} + \kappa u^\mu u^\nu \EE with
 \BE \label{kappagamma}
       \kappa={\cal N}^2 - 1 \EE  In this way,
Eq.(\ref{masshell}) gives a photon energy ($u^2_0=1 + {\bf
V}^2/c^2$) \BE \label{watson}
       E(| {\bf{p}}| , \theta)= c~{{ -\kappa u_0 \zeta
       + \sqrt{ |{\bf{p}}|^2(1+\kappa u^2_0) -
       \kappa \zeta^2 }}\over{ 1 + \kappa u^2_0}}
\EE with
\BE
       \zeta={\bf{p}}\cdot{{{\bf{V}}}\over{c}}= |{\bf{p}}|\beta \cos\theta ,
\EE where $\beta={{|{\bf{V}}|}\over{c}}$ and
$\theta\equiv\theta_{\rm lab}$ indicates the angle defined, in the
laboratory $S'$ frame, between the photon momentum and ${\bf{V}}$.
By using the above relation, one gets the one-way velocity of light
\begin{eqnarray}
       & &{{E(| {\bf{p}}| , \theta)}\over{|{\bf{p}}|}}=
       c_\gamma(\theta)= c~{{ -\kappa \beta  \sqrt{1+\beta ^2} \cos\theta
+ \sqrt{ 1+ \kappa+ \kappa \beta^2 \sin^2\theta} }
       \over{1+ \kappa(1+\beta^2)}} .
\end{eqnarray}
or to ${\cal O}(\kappa)$ and ${\cal O}(\beta^2)$
\BE \label{oneway}
       c_\gamma(\theta) \sim {{c} \over{{\cal N}}}~\left[
       1- \kappa \beta \cos\theta -
       {{\kappa}\over{2}} \beta^2(1+\cos^2\theta)\right]
\EE
From this one can compute the two-way velocity
\begin{eqnarray}
\label{twoway}
       \bar{c}_\gamma(\theta)&=&
       {{ 2  c_\gamma(\theta) c_\gamma(\pi + \theta) }\over{
       c_\gamma(\theta) + c_\gamma(\pi + \theta) }} \nonumber \\
       &\sim& {{c} \over{{\cal N}}}~\left[1-\beta^2\left(\kappa -
       {{\kappa}\over{2}} \sin^2\theta\right) \right]
\end{eqnarray}
which, as anticipated, is a special form of the more general
Eq.(\ref{legendre}). We can then define the RMS anisotropy parameter
${\cal B}$ \footnote{There is a subtle difference between our
Eqs.(\ref{oneway}) and(\ref{twoway}) and the corresponding Eqs. (6)
and (10) of ref.~\cite{pla} that has to do with the relativistic
aberration of the angles. Namely, in ref.\cite{pla}, with the
(wrong) motivation that the anisotropy is ${\cal O}(\beta^2)$, no
attention was paid to the precise definition of the angle between
the Earth's velocity and the direction of the photon momentum. Thus
the two-way velocity of light in the $S'$ frame was parameterized in
terms of the angle $\theta\equiv\theta_\Sigma$ as seen in the
$\Sigma$ frame. This can be explicitly checked by replacing in our
Eqs.~(\ref{oneway}) and(\ref{twoway}) the aberration relation
$\cos \theta_{\rm lab}=(-\beta + \cos\theta_\Sigma)/
       (1-\beta\cos\theta_\Sigma)$
or equivalently by replacing $\cos \theta_{\Sigma}=(\beta +
\cos\theta_{\rm lab})/ (1+\beta\cos\theta_{\rm lab})$ in Eqs. (6)
and (10) of ref.~\cite{pla}. However, the apparatus is at rest in
the laboratory frame, so that the correct orthogonality condition of
two optical cavities at angles $\theta$ and $\pi/2 + \theta$ is
expressed in terms of $\theta=\theta_{\rm lab}$ and not in terms of
$\theta=\theta_{\Sigma}$. This trivial remark produces however a
non-trivial difference in the value of the anisotropy parameter. In
fact, the correct resulting $|{\cal B}|$ Eq. (\ref{rmsb}) is now
smaller by a factor of 3 than the one computed in ref.\cite{pla} by
adopting the wrong definition of orthogonality in terms of
$\theta=\theta_{\Sigma}$.}
\BE \label{rms}
       {{\bar{c}_\gamma(\pi/2 +\theta)- \bar{c}_\gamma (\theta)} \over
       {\langle \bar{c}_\gamma \rangle }} \sim
       {\cal B}{{v^2 }\over{c^2}} \cos2(\theta-\theta_0) \EE
where the pair $(v,\theta_0)$ describes the projection of ${\bf{V}}$
onto the relevant plane and
\BE \label{rmsb}
       |{\cal B}|\sim {{\kappa}\over{2}}\sim ({\cal N}-1)
\EE
From the previous analysis, by replacing the two-way velocity in
Eq.(\ref{deltaT}), one finally obtains the observable velocity \BE
v^2_{\rm obs} \sim 2|{\cal B}| v^2 \sim 2({\cal N}-1)v^2 \EE to be
used in Eq.(\ref{fringe0}). In this way, where one re-obtains the
classical formulas with the only replacement $v \to v_{\rm obs}$,
testing the present scheme requires to check the consistency of the
kinematical $v'$s obtained in different experiments.

Before concluding this Appendix, we emphasize that throughout this
paper we have assumed the limit of a zero light anisotropy for
experiments performed in vacuum. In fact, the effective metric
Eq.(\ref{effective}) reduces to the Minkowski tensor $\eta^{\mu\nu}$
in the limit ${\cal N}\to 1$. Admittedly, this might represent a
restrictive scenario and one could also consider the more general
case where a metric of the form \cite{ma1} \BE \label{effective1}
g^{\mu\nu}= \eta^{\mu\nu} + \Delta^{\mu\nu}\EE is introduced from
the very beginning  in extensions of the Standard Model. In this
sense, once Eq.(\ref{effective1}) is adopted, the vacuum behaves as
a medium and the dispersion relations that describe light and
particle propagation can have several solutions. For instance, light
will spread with different velocities in different directions as
with anisotropic media in optics. Therefore, by adopting various
parameterizations  for $\Delta^{\mu\nu}$, one can restrict its size
by comparing with measurements of the one- and two-way velocity of
light. As an example, the one-way light anisotropy reported by the
GRAAL experiment \cite{graal} requires typical values of the matrix
elements $|\Delta^{\mu\nu}|= 10^{-13}\div 10^{-14}$ \cite{ma2}. In
any case, as anticipated in Sect.2, these genuine vacuum effects are
much smaller than those discussed in the present paper in connection
with a gas refractive index.

\vfill\eject

\centerline{\bf{\LARGE Appendix II}} \vskip 15 pt

In this second Appendix we'll report the results of a single
simulation of an individual noon session of the Michelson-Morley
experiment. This will be performed within the same stochastic-ether
model described in Sect.8 for the Joos experiment. For sake of
clarity, we recapitulate the various steps so that an interested
reader can also run his own simulations.

One should first express the functions $C(t)$ and $S(t)$ as in
Eqs.(\ref{amplitude10}) and model the two velocity components
$v_x(t)$ and $v_y(t)$ as in Eqs.(\ref{vx}) and (\ref{vy}). A basic
input value is the sidereal time of the observation. This has to be
inserted, together with the CMB kinematical parameters $V_{\rm
CMB}\sim $ 370 km/s, $\alpha_{\rm CMB}\sim 168^o$, $\gamma_{\rm
CMB}\sim -6^o$, in Eqs.(\ref{nassau1})$-$(\ref{nassau3}) to fix the
boundaries $[-\tilde v_x(t),\tilde v_x(t)]$ and $[-\tilde
v_y(t),\tilde v_y(t)]$ respectively for the random parameters
$x_n(i=1,2)$ and $y_n(i=1,2)$ entering Eqs.(\ref{vx}) and
(\ref{vy}). In the end, with the simulated $C(t)$ and $S(t)$, one
should form the fringe shift combination \BE \label{convolution}
{{\Delta \lambda(\theta)}\over{\lambda}} \equiv 2C(t)\cos2\theta
+2S(t) \sin 2\theta = A_2(t) \cos2(\theta-\theta_0(t))\EE

As recalled in Sect.3, an individual session of the Michelson-Morley
experiment consisted of 6 rotations. Each complete rotation of the
interferometer took 6 minutes and the consecutive readings of the
fringe shifts were performed every 22.5 degrees. Therefore, two
consecutive readings differed by 22.5 seconds. In these conditions,
a numerical simulation of a single rotation consists in generating
16 pairs $\left[C(t), S(t)\right]$  at steps of 22.5 seconds.

As the central time of the observations, we have chosen 12 A. M. of
July 10, 1887 which, for Cleveland, corresponds to a sidereal time
$\tau \sim 102^o$. To select the parameters of the simulation, we
have compared with the traditional analysis of the experiment where
one performs a fit to the fringe shifts obtained by averaging the
results of the various experimental sessions. In our case, averaging
the data of the three noon sessions in our Table 1 gives a
2nd-harmonic amplitude \BE A^{\rm fit}_2({\rm average~ data-noon})
\sim 0.012 \EE We have thus considered the exact amplitude \BE
A^{\rm exact}_2(t)= 2 \sqrt{ C^2(t) +S^2(t)} \EE and selected a
particular configuration whose global average over the 6 turns gives
 $\langle A^{\rm exact}_2(t)\rangle\sim$ 0.012. Of course, this
condition can be realized by a very large number of configurations.
These can produce very different fringe shifts at the same
$\theta-$values and sizeable variations of the fitted amplitude and
 azimuth. Taking into account these variations is essential to
perform a full numerical simulation and estimate theoretical
uncertainties (as done for the Joos experiment with Table 9 and
Fig.12). However, our intention here is just to give an idea of the
agreement one can achieve between data and a single numerical
simulation. We thus postpone to a future publication a complete
analysis of the whole Michelson-Morley experiment and of the
Illingworth experiment (with its 32 sessions and the associated sets
of 20 rotations for each session) where we'll also compare the
various theoretical schemes mentioned in Sect.5 to handle the
stochastic components of the velocity field.

The results of our single simulation for $\left[2C(t), 2S(t)\right]$
are reported in Tables 11 and 12 while the combinations ${{\Delta
\lambda(\theta)}\over{\lambda}}$ Eq.(\ref{convolution}) are reported
in Table 13, for $\theta={{i-1}\over{16}}~2\pi$ together with the
results of 2-parameter fits to the simulated data. Notice the strong
scatter of the simulated data at the same $\theta-$values. Of
course, to compare with the {\it real} data, one should first take
the even combination Eq.(\ref{even}) of the entries in Table 1 which
otherwise also contain odd-harmonic terms.

We conclude this brief analysis by emphasizing the importance of
Hicks' observation (see Sect.3) concerning the fringe shifts from
the session of July 8th. By accepting his interpretation, the
experimental azimuths from the three noon sessions of July 8th, 9th
and 11th, respectively $\theta^{\rm EXP}_0\sim $ 357, 285 and 317
degrees, would become $\theta^{\rm EXP}_0\sim $ 267, 285 and 317
degrees and thus be in rather good agreement with the simulated
azimuths reported in Table 13.

\begin{table*} \caption{\it The coefficients 2C(t) Eqs.(\ref{amplitude10})
from a single simulation of 6 rotations in one noon session of the
Michelson-Morley experiment. The stochastic components of $v_x(t)$
and $v_y(t)$ in Eqs.(\ref{vx}) and (\ref{vy}) are controlled by the
kinematical parameters $(V,\alpha,\gamma)_{\rm CMB}$ as explained in
the text.}
\begin{center}
\begin{tabular}{cllllll}
\hline i ~~     &~~~~    1 &~~~~   2 & ~~~~  3
       &  ~~~~  4 & ~~~~  5 & ~~~~  6     \\
\hline
1 ~~     & $-0.023$ & $+0.002$ &$-0.024$& $+0.001$& $-0.004$& $-0.003$ \\
2 ~~     & $+0.003$ & $-0.011$ &$-0.000$& $-0.006$& $-0.021$& $-0.034$ \\
3 ~~     & $-0.001$ & $-0.001$ &$+0.000$& $-0.009$& $+0.002$& $+0.007$ \\
4 ~~     & $-0.002$ & $+0.003$ &$-0.008$& $+0.002$& $-0.060$& $-0.030$ \\
5 ~~     & $+0.002$ & $-0.017$ &$+0.002$& $-0.001$& $+0.003$& $-0.008$ \\
6 ~~     & $-0.007$ & $-0.006$ &$-0.059$& $-0.013$& $-0.008$& $-0.047$ \\
7 ~~     & $-0.020$ & $-0.001$ &$-0.019$& $-0.000$& $-0.003$& $-0.003$ \\
8 ~~     & $-0.011$ & $-0.001$ &$-0.011$& $-0.002$& $-0.026$& $+0.001$ \\
9 ~~     & $-0.015$ & $-0.000$ &$-0.008$& $-0.001$& $-0.008$& $-0.022$ \\
10~~     & $-0.037$ & $-0.005$ &$+0.000$& $-0.002$& $-0.003$& $+0.003$ \\
11~~     & $+0.003$ & $-0.022$ &$-0.015$& $-0.005$& $-0.003$& $+0.002$ \\
12~~     & $-0.002$ & $-0.049$ &$-0.023$& $-0.016$& $-0.009$& $-0.006$ \\
13~~     & $-0.001$ & $+0.002$ &$+0.000$& $+0.001$& $+0.003$& $-0.001$ \\
14~~     & $+0.003$ & $-0.003$ &$+0.003$& $-0.023$& $-0.001$& $-0.019$ \\
15~~     & $-0.012$ & $-0.034$ &$-0.013$& $-0.001$& $-0.001$& $-0.011$ \\
16~~     & $-0.004$ & $-0.017$ &$-0.004$& $+0.002$& $-0.010$& $-0.010$ \\
\hline
\end{tabular}
\end{center}
\end{table*}

\begin{table*} \caption{\it The coefficients 2S(t) Eqs.(\ref{amplitude10})
from a single simulation of 6 rotations in one noon session of the
Michelson-Morley experiment. The stochastic components of $v_x(t)$
and $v_y(t)$ in Eqs.(\ref{vx}) and (\ref{vy}) are controlled by the
kinematical parameters $(V,\alpha,\gamma)_{\rm CMB}$ as explained in
the text.}
\begin{center}
\begin{tabular}{cllllll}
\hline i ~~     &~~~~    1 &~~~~   2 & ~~~~  3
       &  ~~~~  4 & ~~~~  5 & ~~~~  6     \\
\hline
1 ~~     & $+0.011$ & $+0.001$ &$+0.011$& $+0.000$& $+0.003$& $+0.001$ \\
2 ~~     & $+0.004$ & $-0.003$ &$-0.001$& $-0.008$& $+0.007$& $-0.009$ \\
3 ~~     & $+0.003$ & $-0.005$ &$+0.000$& $-0.007$& $+0.007$& $+0.000$ \\
4 ~~     & $+0.001$ & $-0.007$ &$-0.003$& $-0.001$& $+0.001$& $+0.016$ \\
5 ~~     & $+0.002$ & $-0.007$ &$+0.000$& $+0.001$& $-0.000$& $-0.003$ \\
6 ~~     & $-0.004$ & $+0.010$ &$+0.001$& $-0.001$& $+0.006$& $+0.009$ \\
7 ~~     & $-0.017$ & $+0.002$ &$+0.005$& $-0.000$& $-0.003$& $+0.002$ \\
8 ~~     & $+0.011$ & $-0.002$ &$-0.012$& $-0.001$& $+0.010$& $-0.000$ \\
9 ~~     & $+0.011$ & $+0.001$ &$+0.008$& $+0.011$& $+0.005$& $-0.011$ \\
10~~     & $+0.016$ & $+0.005$ &$-0.001$& $+0.001$& $-0.002$& $+0.001$ \\
11~~     & $+0.000$ & $-0.013$ &$+0.015$& $+0.005$& $-0.002$& $-0.001$ \\
12~~     & $-0.001$ & $-0.022$ &$+0.004$& $-0.003$& $-0.000$& $+0.002$ \\
13~~     & $-0.001$ & $-0.001$ &$+0.001$& $+0.000$& $-0.004$& $+0.001$ \\
14~~     & $-0.001$ & $+0.002$ &$+0.002$& $+0.018$& $+0.001$& $+0.009$ \\
15~~     & $-0.012$ & $+0.018$ &$-0.002$& $+0.001$& $-0.004$& $+0.007$ \\
16~~     & $+0.002$ & $+0.008$ &$+0.001$& $+0.002$& $-0.005$& $-0.006$ \\
\hline
\end{tabular}
\end{center}
\end{table*}

\begin{table*} \caption{\it The fringe shifts ${{\Delta
\lambda(\theta)}\over{\lambda}}$ Eq.(\ref{convolution}) for the
single simulation of one noon session of the Michelson-Morley
experiment reported in Tables 11 and 12. The angular values are
defined as $\theta={{i-1}\over{16}}~2\pi$. The variance of the
averages is about $\pm 0.004$ for the amplitude and about $\pm 11^o$
for the azimuth.}
\begin{center}
\begin{tabular}{clllllll}
\hline i ~~     &~~~~    1 &~~~~   2 & ~~~~  3
       &  ~~~~  4 & ~~~~  5 & ~~~~  6  & average   \\
\hline
1 ~~     & $-0.023$ & $+0.002$ &$-0.024$& $+0.001$& $-0.004$& $-0.003$&$-0.009$ \\
2 ~~     & $+0.005$ & $-0.010$ &$-0.001$& $-0.010$& $-0.010$& $-0.031$&$-0.010$ \\
3 ~~     & $+0.003$ & $-0.005$ &$+0.000$& $-0.007$& $+0.007$& $+0.000$&$-0.000$ \\
4 ~~     & $+0.002$ & $-0.007$ &$+0.004$& $-0.002$& $+0.043$& $+0.033$&$+0.012$ \\
5 ~~     & $-0.002$ & $+0.017$ &$-0.002$& $+0.001$& $-0.003$& $+0.008$&$+0.003$ \\
6 ~~     & $+0.008$ & $-0.003$ &$+0.041$& $+0.010$& $+0.002$& $+0.027$&$+0.014$ \\
7 ~~     & $+0.017$ & $-0.002$ &$-0.005$& $+0.000$& $+0.003$& $-0.002$&$+0.002$ \\
8 ~~     & $-0.016$ & $+0.001$ &$+0.001$& $-0.001$& $-0.025$& $+0.001$&$-0.007$ \\
9 ~~     & $-0.015$ & $-0.000$ &$-0.008$& $-0.001$& $-0.008$& $-0.022$&$-0.009$ \\
10~~     & $-0.015$ & $-0.000$ &$-0.000$& $-0.001$& $-0.003$& $+0.003$&$-0.003$ \\
11~~     & $+0.000$ & $-0.013$ &$+0.015$& $+0.005$& $-0.002$& $-0.001$&$+0.001$ \\
12~~     & $+0.000$ & $+0.019$ &$+0.020$& $+0.009$& $+0.006$& $+0.006$&$+0.010$ \\
13~~     & $+0.001$ & $-0.002$ &$-0.000$& $-0.001$& $-0.003$& $+0.001$&$-0.001$ \\
14~~     & $-0.001$ & $+0.001$ &$-0.004$& $+0.004$& $-0.000$& $+0.007$&$+0.001$ \\
15~~     & $+0.012$ & $-0.018$ &$+0.002$& $-0.001$& $+0.004$& $-0.007$&$-0.001$ \\
16~~     & $-0.005$ & $-0.018$ &$-0.003$& $-0.000$& $-0.004$& $-0.003$&$-0.005$ \\
\hline
$A^{\rm fit}_2$~  &~~0.009 &~~0.005 &~~0.009 &~~0.003 &~~0.011 &~~0.013 &~~0.008\\
$\theta^{\rm fit}_0$~ &~~~$279^o$ &~~~$259^o$ &~~~$266^o$ &~~~$285^o$ &~~~$255^o$ &~~~$272^o$ &~~~$269^o$\\
\hline
\end{tabular}
\end{center}
\end{table*}


\begin{thebibliography} {99}
\bibitem{einstein}
A. Einstein, Ann. der Physik, {\bf 17} (1905) 891.
\bibitem{lorentz}
H. A. Lorentz,  Proceedings of the Academy of Sciences of Amsterdam,
{\bf 6}, 1904.
\bibitem{poincare}
H. Poincar\'e, La Science et l'Hypothese, Flammarion, Paris 1902; C.
R. Acad. Sci. Paris {\bf 140} (1905) 1504.
\bibitem{electron}
H. A. Lorentz, The Theory of Electrons, Leipzig 1909, B. G. Teubner
Ed.
 \bibitem{bell}
J. S. Bell, How to teach special relativity, in Speakable and
unspeakable in quantum mechanics, Cambridge University Press 1987,
pag. 67.
 \bibitem{brown}
H. R. Brown and O. Pooley, The origin of the space-time metric:
Bell's Lorentzian pedagogy and its significance in general
relativity, in `Physics meets Philosophy at the Planck Scale', C.
Callender and N. Hugget Eds., Cambridge University Press 2000
(arXiv:gr-qc/9908048).
\bibitem{pla}
M. Consoli and E. Costanzo, Phys. Lett. {\bf A333} (2004) 355.
\bibitem{annals}
S. Liberati, S. Sonego and M. Visser, Ann. Phys. {\bf 298} (2002)
167.
\bibitem{scarani}
V. Scarani et al., Phys. Lett. {\bf A276} (2000) 1.
\bibitem{volo}
G. E. Volovik, Phys. Rep. {\bf 351} (2001) 195.
\bibitem{volo1}
G. E. Volovik, Found.Phys. {\bf 33} (2003) 349.
\bibitem{chadha} S. Chadha and H. B. Nielsen, Nucl. Phys. {\bf 217}
(1983) 125.
\bibitem{nielsen}
C. D. Froggatt and H. B. Nielsen, Origin of Symmetries, World
Scientific, 1991.
\bibitem{wheeler1}
J. A. Wheeler, in Relativity, Groups and Topology, B. S. DeWitt and
C. M. DeWitt Eds., Gordon and Breach New York 1963, p. 315.
\bibitem{migdal}
A. A. Migdal, Int. J. Mod. Phys. {\bf A9} (1994) 1197.
\bibitem{ng2}
V. Jejjala, D. Minic, Y. J. Ng and C. H. Tze, Int. J. Mod. Phys.
{\bf D19} (2010) 2311.
\bibitem{ng3}
Y. J. Ng, Various Facets of Spacetime Foam, in Proceedings of the
Third Conference on Time and Matter, Budva, Montenegro 2010,
arXiv:1102.4109 [gr-qc].\bibitem{amelino1} G. Amelino-Camelia,
Nature {\bf 418} (2002) 34.
\bibitem{amelino2}
G. Amelino-Camelia, Int.J.Mod.Phys. {\bf D11} (2002) 35.
\bibitem{amelino3}
G. Amelino-Camelia, Symmetry {\bf 2} (2010) 230.
\bibitem{kleinert}
P. Jizba and H. Kleinert, Phys. Rev. {\bf D82} (2010) 085016.
\bibitem{thooft}
G. 't Hooft, In Search of the Ultimate Building Blocks, Cambridge
Univ. Press, 1997.
\bibitem{mech}
M. Consoli and P.M. Stevenson, Int. J. Mod. Phys. {\bf A15} (2000)
133, hep-ph/9905427.
\bibitem{pagano}
M. Consoli, A. Pagano and L. Pappalardo, Phys. Lett. {\bf A318},
(2003) 292.
\bibitem{epjc}
M. Consoli and E. Costanzo, Eur. Phys. Journ. {\bf C54} (2008) 285.
\bibitem{dedicated}
M. Consoli and E. Costanzo, Eur. Phys. Journ. {\bf C55} (2008) 469.
\bibitem{cpt}
See, for instance, R. F. Streater and  A. S. Wightman, PCT, Spin and
Statistics, and all that, W. A. Benjamin, New York 1964.
\bibitem{zeldovich}
Y. B. Zeldovich , Sov. Phys. Usp. {\bf 11}, 381 (1968).
\bibitem{weinberg}
S. Weinberg , Rev. Mod. Phys. {\bf 61}, 1 (1989).
\bibitem{barcelo1}
C. Barcelo, S. Liberati and M. Visser, Class. Quantum Grav. {\bf
18}, 3595 (2001).
\bibitem{barcelo2}
M. Visser, C. Barcelo and S. Liberati, Gen. Rel. Grav. {\bf 34},
1719 (2002).
\bibitem{ultraweak}
M. Consoli, Class. Quantum Grav. {\bf 26}, 225008 (2009).
\bibitem{rule}
R. P. Feynman, in Superstrings: A Theory of Everything ?, P. C. W.
Davies and J. Brown Eds., Cambridge University Press, 1997, pag.
201.
\bibitem{eckart}
C. Eckart, Phys. Rev. {\bf 58}, 919 (1940).
\bibitem{applied}
For a comprehensive review, see H. M\"uller et al., Appl. Phys. B
{\bf 77}, 719 (2003).
\bibitem{lammer}
Special Relativity, J. Ehlers and C. L\"ammerzahl Eds., Lectures
Notes in Physics, Springer, New York 2006.
\bibitem{hughes}
V. W. Hughes, H. G. Robinson, and V. Beltran-Lopez, Phys. Rev. Lett.
{\bf 4}, 342 (1960).
\bibitem{drever}
R. W. P. Drever, Phil. Mag. {\bf 6}, 683 (1961).
\bibitem{will}
C. M. Will, The Confrontation between General Relativity and
Experiment, arXiv:gr-qc/0510072.
\bibitem{bornwolf}
M. Born and E. Wolf, Principles of Optics, Cambridge University
Press, Cambridge 1999.
\bibitem{weber}
V. C. Ballenegger and T. A. Weber, Am. J. Phys.{\bf 67}, 599 (1999).
\bibitem{robertson}
H. P. Robertson, Rev. Mod. Phys. {\bf 21}, 378 (1949).
\bibitem{mansouri}
R. M. Mansouri and R. U. Sexl, Gen. Rel. Grav. {\bf 8}, 497 (1977).
\bibitem{brillet}
A. Brillet and J. L. Hall, Phys. Rev. Lett. {\bf 42} (1979) 549.
\bibitem{muller}
H. M\"uller et al.,  Phys. Rev. Lett. {\bf 91} (2003) 020401.
\bibitem{peters}
S. Herrmann, et al., Phys. Rev. Lett. {\bf 95} (2005) 150401.
\bibitem{schiller}
P. Antonini et al.,  Phys. Rev. {\bf A71} (2005) 050101(R).
\bibitem{crossed}
Ch. Eisele et al., Opt. Comm. {\bf 281}, 1189 (2008).
\bibitem{newberlin}
S. Herrmann, et al., Phys.Rev. D {\bf 80}, 105011 (2009).
\bibitem{schillernew}
Ch. Eisele, A. Newski  and S. Schiller, Phys. Rev. Lett. {\bf 103},
090401 (2009).
\bibitem{cimento}
M. Consoli and E. Costanzo, N. Cim. {\bf 119B} (2004) 393.
\bibitem{fox}
J. Shamir and R. Fox, N. Cim. {\bf 62B} (1969) 258.
\bibitem{guerra}
R. De Abreu and V. Guerra, Relativity-Einstein's Lost Frame, 2005,
Extra]muros[ Publ.
\bibitem{holger}
H. M\"uller, Phys. Rev. {\bf D71}, 045004 (2005).
\bibitem{nassau}
J. J. Nassau and P. M. Morse, Astrophys. Journ. {\bf 65} (1927) 73.
\bibitem{troshkin}
O. V. Troshkin, Physica {\bf A168} (1990) 881.
\bibitem{puthoff}
H. E. Puthoff, Linearized turbulent flow as an analog model for
linearized General Relativity, arXiv:0808.3404 [physics.gen-ph].
\bibitem{tsankov}
T. D. Tsankov, Classical Electrodynamics and the Turbulent Aether
Hypothesis, Preprint February 2009.
\bibitem{chaos}
M. Consoli, A. Pluchino and A. Rapisarda,  Chaos, Solitons and
Fractals {\bf 44}, 1089 (2011).
\bibitem{plafluid}
M. Consoli, Phys. Lett. {\bf A 376 }, 3377 (2012).
\bibitem{michelson}
A. A. Michelson and E. W. Morley, Am. J. Sci. {\bf 34} (1887) 333.
\bibitem{righi}
An incomplete list of references includes: W. Sutherland, Phil. Mag.
{\bf 45} (1898) 23; A. Righi, N. Cimento {\bf XVI} (1918) 213; {\it
ibidem} {\bf XIX} (1920) 141; {\it ibidem} {\bf XXI} (1921) 187; G.
Dalla Noce, N. Cimento {\bf XXIV} (1922) 17. Righi's theory was also
re-analyzed by P. Di Mauro, S. Notarrigo and A. Pagano, Quaderni di
Storia della Fisica, {\bf 2} (1997) 101.
\bibitem{conference}
A. A. Michelson, et al., Astrophys. Journ. {\bf 68} (1928) pag.
341-402.
\bibitem{kennedy}
R. J. Kennedy, Phys. Rev. {\bf 47} (1935) 965.
\bibitem{miller}
D. C. Miller, Rev. Mod. Phys. {\bf 5} (1933) 203.
\bibitem{hicks}
W. M. Hicks, Phil. Mag. {\bf 3} (1902) 9.
\bibitem{born}
M. Born,  Einstein's Theory of Relativity, Dover Publ., New York,
1962.
\bibitem{shankland}
R. S. Shankland et al., Rev. Mod. Phys., {\bf 27}, (1955) 167.
\bibitem{handshy}
M. A. Handshy, Am. J. of Phys. {\bf 50} (1982) 987.
\bibitem{morley}
E. W. Morley and D. C. Miller, Phil. Mag. {\bf 9} (1905) 680.
\bibitem{illingworth}
K. K. Illingworth, Phys. Rev. {\bf 30} (1927) 692.
\bibitem{munera}
H. A. M\'unera, APEIRON {\bf 5} (1998) 37.
\bibitem{kolmo}
A. N. Kolmogorov, Dokl. Akad. Nauk SSSR {\bf 30} (1940) 4; English
translation in Proc. R. Soc. {\bf A 434} (1991) 9.
\bibitem{landau}
L. D. Landau and E. M. Lifshitz, Fluid Mechanics, Pergamon Press
1959, Chapt. III.
\bibitem{fung}
J. C. H. Fung et al., J. Fluid Mech. {\bf 236}, 281 (1992).
\bibitem{sreenivasan}
K. R. Sreenivasan, Rev. Mod. Phys. {\bf 71}, Centenary Volume 1999,
S383.
\bibitem{beck}
C. Beck, Phys. Rev. Lett. {\bf 98}, 064502 (2007).
\bibitem{tsallis}
C. Tsallis, Introduction to Nonextensive Statistical Mechanics,
Springer, 2009.
\bibitem{demeo}
J. DeMeo, Dayton C. Miller Revisited, in Should the Laws of
Gravitation be Reconsidered?, H. A. M\'unera Ed., APEIRON Montreal
2011, pp. 285-315.
\bibitem{roberts}
T. Roberts, An Explanation of Dayton Miller's Anomalous "Ether
Drift" Result, arXiv:physics/0608238.
\bibitem{minuit}
F. James, MINUIT: Function minimization and error analysis,
CERN Computing and Networks Division, Long Writeup D506, Geneva 1994.
\bibitem{laue}
M. von Laue, Handbuch der Experimentalphysik, Vol. XVIII (1926) 95.
\bibitem{thirring}
H. Thirring, Zeit. Physik {\bf 35} (1926) 723; Nature {\bf 118}
(1926) 81.
\bibitem{mpp}
A. A. Michelson, F. G. Pease and F. Pearson, Nature, {\bf 123},
(1929) 88
\bibitem{mpp2}
A. A. Michelson, F. G. Pease and F. Pearson, J. Opt. Soc. Am. {\bf
18} (1929) 181.
\bibitem{pease}
F. G. Pease, Publ. of the Astr. Soc. of the Pacific, {\bf XLII}, 197
(1930).
\bibitem{joos}
G. Joos, Ann. d. Physik {\bf 7} (1930) 385.
\bibitem{joos1}
G. Joos, Naturwiss. {\bf 38} (1931) 784.
\bibitem{joos2}
G. Joos, D. Miller, Letters to the Editor, Phys. Rev. {\bf 45}, 114
(1934).
\bibitem{loyd2}
Loyd S. Swenson Jr., Journ. for the History of Astronomy, {\bf 1}, 56 (1970).
\bibitem{potsdam}
A. A. Michelson, Am. J. Sci. {\bf 22}, 120 (1881).
\bibitem{tomaschek}
R. Tomaschek, Ann. d. Physik, {\bf 73}, 105 (1924).
\bibitem{piccard}
A. Piccard and E. Stahel, Compt. Rend. {\bf 183}, 420 (1926);
Naturwiss. {\bf 14}, 935 (1926).
\bibitem{piccard2}
A. Piccard and E. Stahel, Compt. Rend. {\bf 185}, 1198 (1927);
Naturwiss. {\bf 16}, 25 (1928).
\bibitem{jaseja}
T. S. Jaseja, et al.,  Phys. Rev. {\bf 133} (1964) A1221.
\bibitem{jauch}
J. M. Jauch and K. M. Watson, Phys. Rev. {\bf 74}, 950 (1948).
\bibitem{ma1}
Zhou L. L. and B. -Q. Ma, Mod. Phys. Lett. {\bf A25}, 2489 (2010).
\bibitem{graal}
V. G. Gurzadyan et al., arXiv:1004.2867 [physics.acc-ph], and
references quoted therein, in Proc. 12th M. Grossmann Meeting on
General Relativity, p.1495, World Sci. (2012).
\bibitem{ma2}
Zhou L. L. and B. -Q. Ma, Astropart. Phys.  {\bf 36}, 37 (2012),
 arXiv:1009.1675 [hep-ph].



\end{thebibliography}
\end{document}